\documentclass{article}[12pt]
\usepackage[a4paper,margin=1.25in,footskip=0.25in]{geometry}

\usepackage{color,soul}

\usepackage{afterpage}
\usepackage[final]{pdfpages}
\usepackage{titlesec}
\usepackage[hidelinks]{hyperref}
\usepackage{setspace}
\usepackage{caption}
\usepackage{subcaption}
\usepackage{rotating}
\usepackage{tabu}
\usepackage{tabularx}
\usepackage[width=.9\textwidth]{caption}
\usepackage[labelfont=bf,skip=0pt]{caption}
\usepackage{arydshln}
\usepackage[multiple]{footmisc}
\usepackage{wrapfig}

\usepackage[english]{babel}
\usepackage{natbib}
\usepackage{amssymb}

\usepackage{amsmath}
\usepackage{mathtools}
\usepackage{breqn}
\usepackage{geometry}
\usepackage{amsmath}

\usepackage{nicematrix}
\usepackage{multicol}
\usepackage{multirow}
\usepackage{float}  

\usepackage{amsmath}
\usepackage{mathtools}
\usepackage{tabularx}

\usepackage{lscape}
\usepackage{pdflscape}

\usepackage{geometry}
\usepackage{lipsum}

\usepackage{multirow}
\usepackage{threeparttable}

	\usepackage{booktabs,siunitx}
	
\newcolumntype{P}[1]{>{\centering\arraybackslash}p{#1}}
\newcolumntype{M}[1]{>{\centering\arraybackslash}m{#1}}
\usepackage{booktabs}

\usepackage{tikz}
\usetikzlibrary{positioning}
\usetikzlibrary{shapes,arrows,arrows,positioning,fit}
\usetikzlibrary{calc}

\usepackage{amssymb, amsfonts}

\usepackage[nottoc,notlot,notlof]{tocbibind}

\usepackage[toc,page]{appendix}
\usepackage{appendix}

\usepackage{eurosym}

\usepackage{subcaption}

\usepackage{listings}

\title{\textbf{Jawboning the Fed: Do Markets Listen When Administrations Talk?}\thanks{The authors would like to thank Gabriel P\'erez Quir\'os, Giovanni Ricco, Massimo Giovannini, and John Caruana for the valuable feedbacks. Any remaining errors are the authors' own. The views expressed in this paper are those of the authors and do not necessarily reflect the views of the Central Bank of Malta.\\ %Correspondence: \texttt{publicrelations@centralbankmalta.org}.
}}
\vspace{5cm}
\author{
Michela Baretti\thanks{Michela Baretti is a PhD candidate at the Universidad de Alicante}
\and
Umberto Collodel\thanks{Umberto Collodel is a Senior Economist in the Monetary Policy \& Eurosystem Department of the Central Bank of Malta.}
\and
Jacopo Zacché\thanks{Jacopo Zacché is a Senior Research Economist in the Modelling Office, within the Economic Research Department of the Central Bank of Malta.}
}
\vspace{2cm}
\date{}

\begin{document}

%\maketitle
%\begin{figure}
 %   \begin{center}
  %  \noindent\includegraphics[width=0.8\textwidth]{figures/cbm_logo.jpg}
   % \end{center}
%\end{figure}

\maketitle
\vspace{2cm}
\begin{center}
    10/2026
\end{center}

\clearpage
\null\vspace{\fill}
\begin{abstract}

 \noindent        
\normalsize{
We provide the first systematic evidence on whether financial markets price political pressure on the Federal Reserve as a signal of future monetary policy.  Using approximately four decades of Reuters newswire articles from 1988 to 2025, classified by an open-weight large language model ensemble into dovish and hawkish pressure, we estimate the daily
response of interest rate expectations across the term structure, the US dollar, inflation compensation, and market volatility.  Dovish pressure lowers near-term rate expectations by roughly one basis point per pressure event---equivalent to a four-percent increase in the implied
probability of a rate cut at the next FOMC meeting---with the effect decaying monotonically and vanishing beyond the three-month horizon. Hawkish pressure has no detectable effect. Furthermore, excluding the Trump
presidencies does not weaken the estimate, ruling out the possibility that the result is confined to an idiosyncratic recent episode. The response is accompanied by dollar depreciation, with stable inflation compensation and market volatility. We interpret the pattern as markets pricing anticipated monetary accommodation, rather than a credibility tax on the nominal anchor.}

 \vspace{1cm}
\end{abstract}
\vspace{\fill}
\clearpage

%%%%%%%%%%%%%%%%%%%%%%%%%%%%%%%%%%%%%%
%Table of Contents
%%%%%%%%%%%%%%%%%%%%%%%%%%%%%%%%%%%%%%
\bigskip 
\tableofcontents
\newpage

\newpage

%%%%%%%%%%%%%%%%%%%%%%%%%%%%%%%%%%%%%%
%Introduction
%%%%%%%%%%%%%%%%%%%%%%%%%%%%%%%%%%%%%%

\section{Introduction}
\label{sec:intro}
% =============================================================================
% Section: Introduction
% =============================================================================
Central bank independence presumes a clean boundary between politics and monetary policy. In practice, that boundary has always been porous: from Reagan-era tensions with Paul Volcker over the severity of disinflation to the Bush Senior administration's pressure on Greenspan ahead of the 1992 election, elected officials have routinely and publicly attempted to influence the Federal Reserve.\footnote{See \citet{volcker2018keeping} for a thorough discussion.} The topic has attracted renewed interest in the wake of President Trump's confrontational public campaign against the Fed and its Chair, Jerome Powell, including explicit threats to remove him before the expiry of his mandate. 

Whether financial markets price this pressure bears directly on the institutional design of monetary policy: were independence fully credible, pressure would carry no authority over the instrument, and the expected path of the policy rate would not move. Systematic repricing is therefore informative in its own right, and potentially consequential. If central banks condition on the market-based expectations they observe, as recent evidence suggests \citep{cieslak2021}, repricing need not remain confined to asset prices: it can feed through to the policy rate itself, so that pressure shapes outcomes without any deliberation being swayed.

We provide a measure of public political pressure on the Fed by collecting approximately four decades of Reuters newswire articles, from 1988 through 2025, filtered to retain articles that simultaneously mention a U.S. elected or appointed official and the Federal Reserve in the context of monetary policy. Each event is classified using an ensemble of open-weight large language models into dovish or hawkish pressure. The resulting series spans eight U.S. presidential administrations, from Reagan's final years through Trump's second term, and constitutes, to the best of our knowledge, the first daily multi-decade measure of public political pressure on the Federal Reserve. This measure is complementary to \citet{Drechsel2026PoliticalPressureFed}, who constructs an index of \emph{private} pressure from archival FOMC and White House records: we instead measure \emph{public} pressure, which is what markets observe and can price.

We exploit daily variation in our measure to estimate the effect of political pressure on interest rate expectations across the full term structure, the US dollar, inflation compensation, and market volatility. Identification rests on daily variation in pressure conditional on scheduled FOMC announcements, slow-moving business-cycle controls, and the \citet{bauer_swanson_2023} orthogonalisation variables, which together are designed to absorb the Fed's own information set and mitigate concern that concurrent macroeconomic news are driving the results.

The key finding is that dovish political pressure on the Fed is associated with a significant and robust decline in near-term interest rate expectations, consistent with markets pricing in a more accommodative monetary policy path. Hawkish pressure has no detectable effect. The response is concentrated at the short end of the yield curve: dovish pressure shifts rate expectations significantly at maturities of up to one quarter ahead, with the effect fading and becoming statistically indistinguishable from zero at longer horizons. This result is not a Trump-era artifact: the effect is larger in magnitude and more precisely estimated outside the Trump presidencies than within them, establishing political pressure as a persistent feature of the post-1988 U.S. monetary policy landscape rather than the idiosyncrasy of a single high-intensity administration. Beyond interest rates, dovish pressure days are accompanied by a depreciation of the USD, offering corroboration from a second asset class consistent with the same accommodative repricing.

Market-based inflation expectations and financial volatility measures, by contrast, do not respond to political pressure. This is consistent with markets retaining confidence in the Fed's ability to maintain its inflation target even under political pressure, and helps distinguish a path-of-policy channel from a credibility channel.

Finally, dovish pressure days are days on which futures move closer to the eventual policy outturn: the absolute pricing error falls by 0.5–0.8 basis points per event, 15–25\% of a typical daily fluctuation, and roughly three times that on days when markets actually priced in the pressure. This pattern is consistent with accommodation operating through the market signal rather than the statement itself.

A natural concern is that politicians press for easing precisely when conditions call for it, so pressure may simply proxy for observable economic conditions rather than anticipated accommodation. Although we proxy extensively for domestic conditions, no set of daily controls can fully eliminate this concern. We address this endogenous confound directly by interacting pressure with the alignment of the Board of Governors with the administration in power: if pressure only proxies for economic conditions, the response should not depend on Board alignment; if it reflects accommodation, it should, instead, be stronger when the Board is aligned. We find that the response is significantly larger when the Board is aligned, difficult to reconcile with a purely informational account.

The rest of the paper is organized as follows. Section~\ref{sec:literature} reviews the related literature. Section~\ref{sec:measurement} describes the construction of our political pressure indicator. Section \ref{sec:data} explains our dependent variables. Section~\ref{sec:strategy} presents the empirical strategy and reports the main results. Section~\ref{sec:robustness} presents robustness checks. Section~\ref{sec:conclusion} concludes.

\section{Related Literature}
\label{sec:literature}
This paper connects to a rich literature on central bank independence
and political pressure.  The theoretical case for independence runs
from the inflation-bias models of \citet{barro_gordon_1983} and
\citet{rogoff_1985} through the cross-country empirical evidence of
\citet{alesina_summers_1993} and subsequent work documenting that
statutory independence correlates robustly with lower and less
volatile inflation across advanced and developing economies.  Testing
whether independence is challenged in practice, however, requires a
measure of pressure itself.  The pioneering contribution is
\citet{Havrilesky1995Pressures}, who constructs an index of
executive-branch signalling toward the Federal Reserve by manually
coding \textit{Wall Street Journal} articles over 1964--1991 as $\pm
1$ for pressure toward ease or restraint.  Regressing weekly changes
in the federal funds rate on cumulated pressure counts, he finds that
a one-unit increase in the index is associated with roughly a
five-basis-point adjustment in the policy rate.  The methodology was
subsequently applied to other central banks: comparable indices have
been constructed for the Bundesbank
\citep{maier_sturm_dehaan_2000}, the Czech National Bank
\citep{Gersl2006PoliticalPressureCNB}, and the early ECB
\citep{MaierBezoen2004BashingSupporting}, with broadly similar
conclusions that political signalling correlates with subsequent
policy moves.  In a related line of work, \citet{ehrmann2011}
assemble a database of politicians' statements about ECB monetary
policy across the twelve founding EMU members, but use it to
characterise the \emph{composition} of political preferences rather
than their effect on market or policy outcomes. More recently, \citet{binder2021} extends the measurement exercise
to a much broader sample, constructing a monthly dataset of
political pressure events on 118 central banks using news archives
and documenting that dovish pressure is associated with higher subsequent
inflation outcomes.

A parallel high-frequency literature has focused more narrowly on
the Trump administrations, shifting the outcome of interest from
realised inflation to the response of financial markets.
\citet{Bianchi2023} find that markets responded to Trump's monetary
policy tweets during his first term by lowering interest rate
expectations alongside rising equity prices and a depreciating
dollar, consistent with markets pricing in monetary easing.
\citet{Liu2023} further show that Trump's tweets raised exchange
rate volatility, particularly when accompanied by negative
sentiment.  For the second administration, preliminary evidence from
\citet{FrankovicKarau2026PoliticalPressureFed} documents a
qualitatively different pattern: while markets continue to price in
lower short-term interest rates following political interventions,
these episodes simultaneously trigger a repricing of US risk,
reflected in higher implied volatility, rising gold prices, a
depreciating dollar, and declining equity and oil valuations.\footnote{A growing literature documents political pressure on
central banks outside the United States, including the Bank of Japan
\citep{chien_cole_lustig_2023} and several emerging-market central
banks \citep{gurkaynak_kisacikoglu_lee_2023, ioannidou_et_al_2025,
bolhuis_mano_thorell_2026}.}

Despite this progress, two limitations remain.  The first, and most
consequential, is that the existing high-frequency literature is
episodic: studies are confined to specific administrations rather
than providing a systematic measure of political pressure over a
long historical sample.  This is largely a methodological
constraint---manual coding of press articles is labour-intensive,
making multi-decade classification infeasible without automation.
The second, closely related, is that the pressure measures employed are narrow in
scope, typically capturing signals from a single actor (the
President), which leaves the broader universe of executive and
congressional pressure unexplored.  We address both limitations by
constructing a four-decade measure of public political pressure in the US across multiple actors over 1988--2025, classified by an open-weight LLM
ensemble. We then exploit daily variation in this measure to
estimate its effect on the full term structure of interest rate
expectations, the USD exchange rate, and market-based measures of
medium- and long-run inflation expectations. The sample
length allows to respond directly to the impact question: the estimated
market response averages over episode-specific idiosyncrasies rather
than reflecting the peculiarities of any one administration.  Because the pipeline extracts the identity of the pressuring actor, the
communication channel, and the target, we can also study heterogeneity along
each of these dimensions---asking, for instance, whether markets
respond differently to presidential pressure than to statements by
other executives.  The methodology is fully extensible: any central bank
subject to documented public pressure can be analysed using the
same pipeline, opening the door to systematic cross-country
comparison of how political interference with monetary policy is
transmitted to financial markets.

\section{Measuring Political Pressure}
\label{sec:measurement}
\subsection{News Corpus}
\label{subsec:corpus}

The primary data source is Reuters newswires, accessed through the 
Factiva Dow Jones service and covering the period from 1988 through 
end-2025.\footnote{Reuters newswire coverage on Factiva begins in 1988, 
which determines the start date of our sample.} We rely on a single 
source rather than combining multiple outlets, as in 
\citet{caldara_iacoviello_2022}, in order to identify unique political 
pressure events and assess their relevance without conflating event 
occurrence with media salience. This approach avoids using the volume 
of news coverage as a proxy for importance—a concern that arises when 
the same event is counted multiple times across outlets. Reuters is 
particularly well suited for this purpose: it provides comprehensive, 
timely, and editorially consistent coverage of major political and 
economic developments over an extended time horizon, making it a 
natural textual source for consistent event classification over an extended sample.

The first step in constructing the dataset is to collect all articles 
containing at least one keyword from each of three categories of 
interest: political actors, the Federal Reserve, and monetary policy, a standard procedure in textual analysis since the pioneering work of \citet{baker_bloom_davis_2016}). 
We adopt a deliberately broad filtering strategy based on expansive 
keyword classes (Table~\ref{tab:keywords}): this 
approach prioritizes recall over precision, allowing for the inclusion 
of false positives in order to minimize the risk of excluding relevant 
articles (false negatives). False positives are subsequently filtered 
out in the LLM classification pipeline. By construction, our measure 
captures direct public political pressure on the Fed: every 
article in the sample explicitly references both a political actor and Federal Reserve monetary policy. We therefore abstract from indirect 
pressure operating through politicians' stated preferences over 
macroeconomic outcomes, which need not invoke the Fed by name.\footnote{To ensure robustness of our approach to the choice of keywords, we run an alternative Factiva search with an expanded set, capturing the evolving monetary framework of the Federal Reserve over time: although we identify a larger pool of articles, results remain virtually unchanged (see Section \ref{subsec:expandend_keywords}).}

\begin{table}[h!]
\centering
\caption{Keyword Categories for Article Filtering}
\label{tab:keywords}
\small
\begin{tabular}{p{3.2cm} p{10.5cm}}
\toprule
\textbf{Category} & \textbf{Keywords} \\
\midrule
Political actors &
\textit{president, White House, administration, Treasury Secretary, 
Treasury Department, Senate Banking, Banking Committee, House Banking, 
economic advisor, Council of Economic Advisors, senator, senators, 
representative, representatives} \\
\addlinespace
Federal Reserve &
\textit{Federal Reserve, Fed, FOMC, Fed Chair, Fed Chairman, Fed 
governor, Fed governors, Fed president, Fed presidents, Fed official, 
Fed officials, Fed policymaker, Fed policymakers} \\
\addlinespace
Monetary policy &
\textit{interest rate, interest rates, monetary policy, rate cut, rate 
cuts, rate hike, rate hikes, fed policy, federal funds, fed funds rate, 
inflation, price stability, money supply, money growth} \\
\bottomrule
\end{tabular}
\end{table}

\subsection{LLM Classification Pipeline}
\label{subsec:llm}

Starting from a broad collection of articles requires a sequence of 
refinement steps to isolate the subset of truly relevant observations. 
To this end, we develop a three-stage LLM pipeline that transforms the 
raw corpus into a precise set of classified pressure events. The use 
of LLMs is motivated by the need to automate a classification task 
that would otherwise require extensive manual labor, making large-scale, 
multi-decade analysis infeasible. Moreover, LLMs bring two specific 
advantages over simpler rule-based or keyword approaches: they are 
particularly suited for inferring the direction of a communication—
distinguishing, for instance, a politician pressuring the Fed from the 
other way around—and for extracting structured metadata from 
unstructured text, such as the identity of the pressuring actor, the 
target, and the communication channel, without requiring hand-crafted 
extraction rules.

The main challenge in using LLMs for replication purposes is ensuring 
that identical inputs and prompts yield identical outputs. We address 
this by using open-weight models with temperature set to zero, which 
allow full control over the inference environment.\footnote{Open-weight models are models whose trained parameters (weights) are publicly released, allowing researchers to run and evaluate them locally and control fully the experimental conditions.} Open-weight models 
may, however, come at some cost in classification accuracy relative to 
proprietary alternatives. To mitigate this, each stage of the pipeline 
is run in parallel across three LLMs—Qwen, Llama, and 
DeepSeek\footnote{Specifically, Qwen (\texttt{qwen3-235b-a22b-2507}), 
Meta Llama (\texttt{llama-4-maverick}), and DeepSeek 
(\texttt{deepseek-v3-0324}).}—and the final set of events retains 
only those observations for which at least two of the three models 
agree. Beyond improving accuracy, the majority-voting architecture 
reduces classification variance and partially decorrelates measurement 
error: because the three models differ in architecture, training data, 
and alignment procedure, their error patterns are plausibly independent, 
so idiosyncratic biases of one model are unlikely to be ratified by 
the other two.\footnote{Reducing the systematic component of measurement error is pivotal for inference as it reduces the probability of having a non-classical 
measurement error.}

\begin{figure}[h!]
    \centering
    \caption{Three-Stage LLM Classification Pipeline}
    \label{fig:pipeline}
    \includegraphics[width=0.95\textwidth]{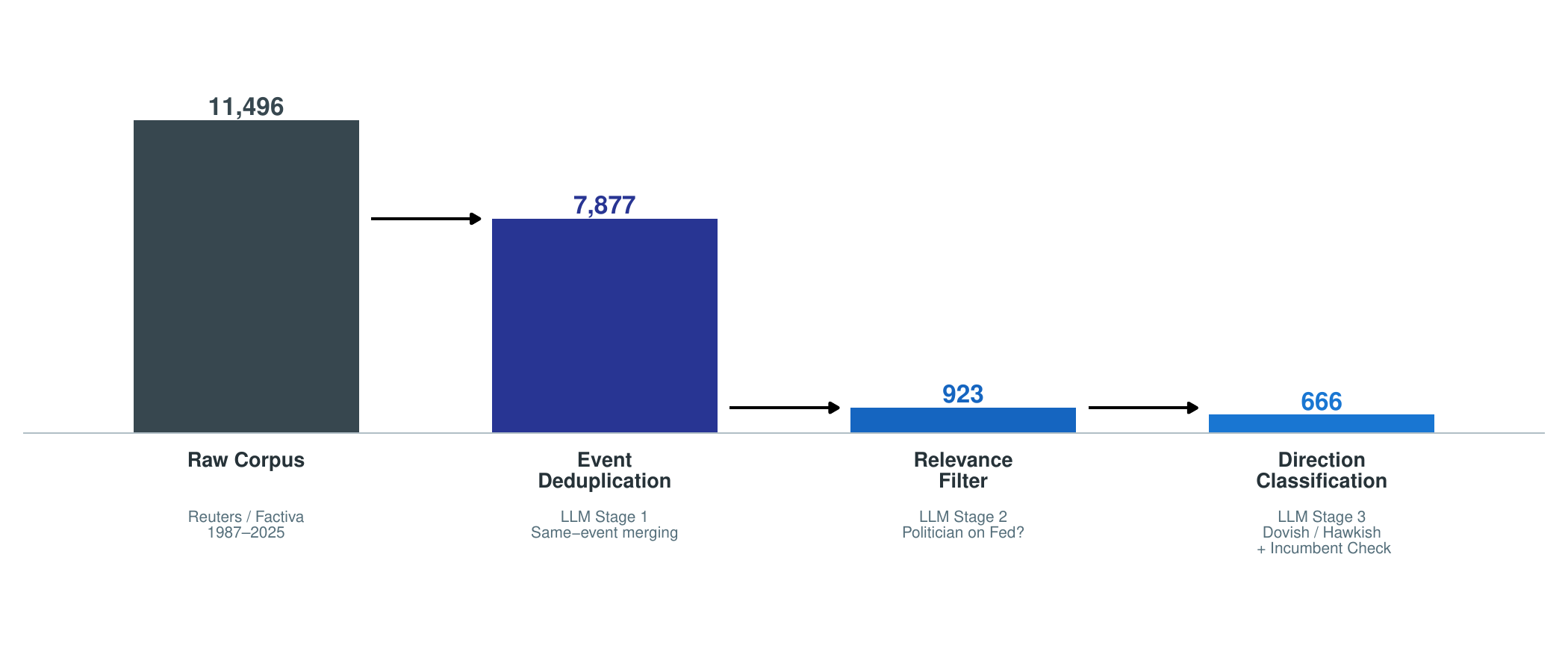} 
    \vspace{0.2em}
\footnotesize{\begin{flushleft}
\textbf{Note}: The figure illustrates the three-stage LLM pipeline 
used to classify political pressure events. Each stage is run in 
parallel across Qwen (\texttt{qwen3-235b-a22b-2507}), Meta Llama 
(\texttt{llama-4-maverick}), and DeepSeek 
(\texttt{deepseek-chat-v3-0324}), and only observations for which at 
least two of the three models agree are retained.
\end{flushleft}}
\end{figure}

The pipeline proceeds in three stages, illustrated in 
Figure~\ref{fig:pipeline}. The first stage consolidates multiple 
journalistic entries into distinct event units. The unit of analysis 
is the political pressure event—a discrete act of public interference 
with the Fed—not the article reporting it. Treating each article as a 
separate observation would conflate the intensity of media coverage 
with the frequency of pressure events, reintroducing precisely the 
problem that motivates relying on a single source. Even within a single 
source, the same event may be reported multiple times as new 
information becomes available; it is therefore essential to group 
articles referring to the same underlying event. This initial step 
reduces the raw corpus from approximately 11,696 articles to 7,707 
unique events.\footnote{Technically, same day articles share the same numeric identifier: we pass each couple to our LLM jury and at the end resolve transitive inconsistencies.}

The second stage identifies whether an event involves a U.S. politician 
commenting on the Federal Reserve, excluding cases in which the 
direction of communication is reversed or where the content reflects 
journalistic interpretation rather than direct political statement. 
Two exclusion criteria are applied. First, the direction of 
communication must run from a political actor toward the Fed. Second, 
the content must reflect an attributable statement by a named political 
actor rather than editorial framing or analyst commentary embedded in 
the article. This stage acts as the primary precision filter, reducing 
the sample to 923 events. In addition to applying these criteria, we 
extract structured metadata for each event: the name and position of 
the politician, the object of the statement, and the communication 
channel through which it is conveyed, as summarized in 
\autoref{fig:sankey}.\footnote{The extraction of these features is 
performed using Qwen.} These metadata serve two purposes: they enable 
the heterogeneity analysis in Section~\ref{app:heterogeneity_actor}, and 
they feed directly into the quality check below.

Since we are interested in the market effects of pressure exerted by 
actors currently holding positions of institutional authority, we 
restrict the sample to statements made by incumbent officeholders, 
excluding candidates and former officials. Markets are more likely to 
reprice assets in response to pressure from actors with current 
decision-making power over fiscal policy, legislative agenda, or 
executive appointments. We verify Reuters' incumbent classification 
by cross-checking extracted politician names and positions against 
official records from the U.S. Congress, the White House, and the 
Treasury.\footnote{This procedure confirms the reliability of Reuters' 
labeling in the vast majority of cases. Across the full sample, we 
identify only 14 instances of misclassification, all involving 
Treasury Secretary nominees reported as incumbents prior to Senate confirmation. We discard these observations as the actors are still not part of the executive.}

In the final stage, the three LLMs classify each remaining event along 
two dimensions. First, they determine whether the statement constitutes 
genuine political pressure—that is, whether it goes beyond mere 
commentary and represents an attempt to influence the conduct of 
monetary policy. Second, conditional on being classified as pressure, 
they determine its direction: dovish, if the politician advocates for 
easier monetary conditions such as rate cuts or more stimulus; or 
hawkish, if the politician pushes for tighter policy or more aggressive 
action against inflation.\footnote{The classification does not require 
an explicit directional statement. When pressure is implicit, the model 
infers direction from the economic context embedded in the article—the 
prevailing inflation or unemployment environment and the recent stance 
of monetary policy. Explicit statements take precedence over contextual 
inference whenever both are present.}\footnote{This context-sensitive 
inference illustrates a key advantage of LLMs over bag-of-words or 
dictionary-based classifiers. A tweet calling for the Fed Chair to be 
fired contains no rate-relevant keywords, yet an LLM correctly infers 
dovish pressure by recognising that removing the FOMC Chair during a 
tightening cycle is an implicit demand for easier policy.} This step 
further reduces the sample to 666 events, of which 579 are classified 
as dovish and 87 as hawkish—a pronounced asymmetry reflecting the 
well-documented bias of political actors toward easier monetary 
conditions, consistent with \citet{barro_gordon_1983}.\footnote{Full text of the three prompts in Appendix \ref{sec:app:prompts}.}

\subsection{Political Pressure Measure}
\label{subsec:measures}

\begin{figure}[h!]
    \centering
    \caption{Political Pressure over Time}
    \label{fig:pressure_timeseries}
    \begin{subfigure}{\textwidth}
        \centering
                \caption{Dovish pressure}
        \label{fig:pressure_ts_dovish}
        \includegraphics[width=\linewidth]{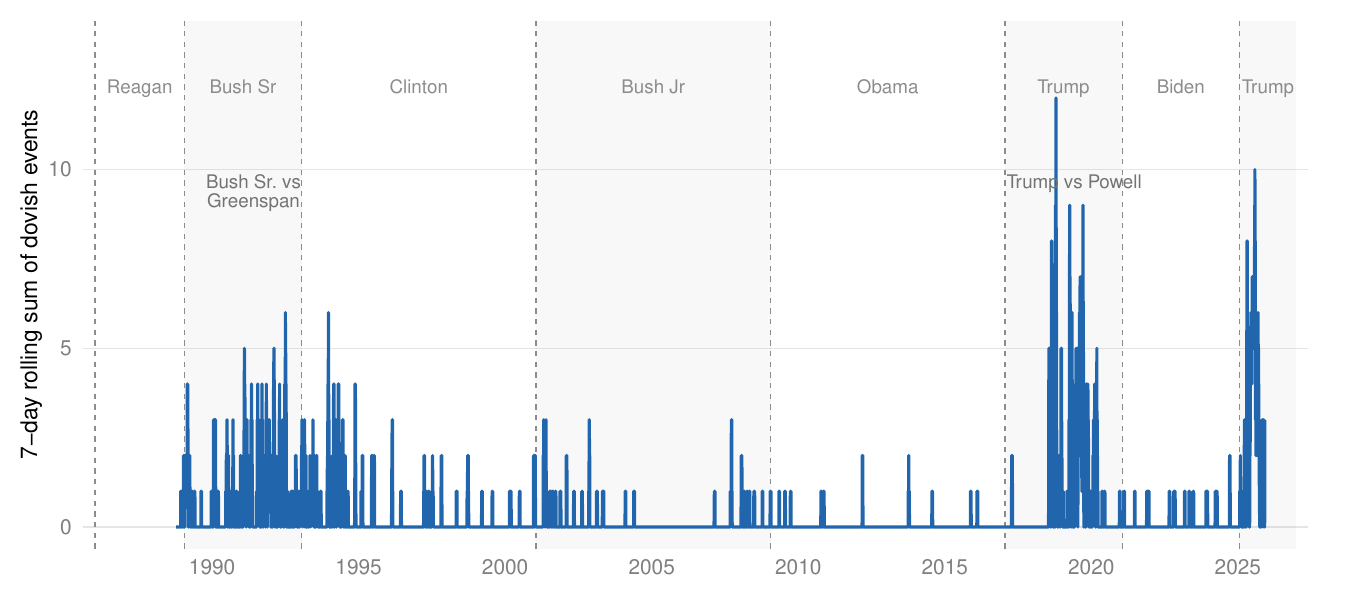}
    \end{subfigure}

    \vspace{0.5em}

    \begin{subfigure}{\textwidth}
        \centering
        \caption{Hawkish pressure}
        \label{fig:pressure_ts_hawkish}
        \includegraphics[width=\linewidth]{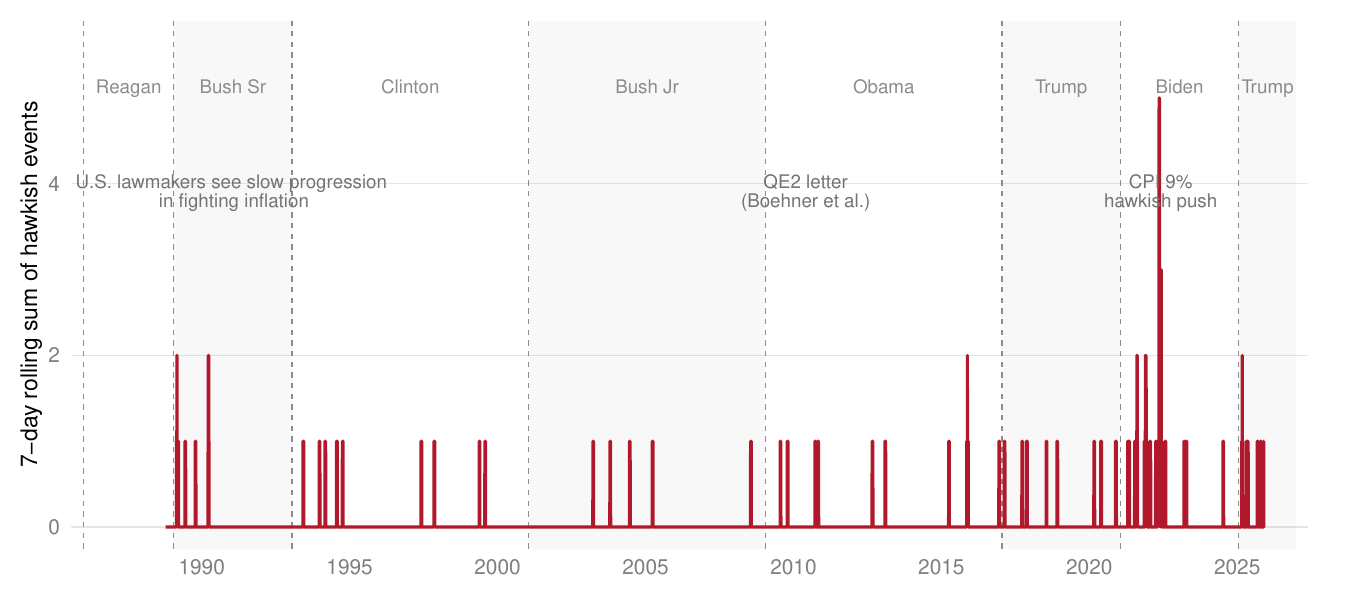}
    \end{subfigure}

    \vspace{0.2em}
    \footnotesize{\begin{flushleft}
    \textbf{Note:} Dashed lines indicate change in presidency.
    \end{flushleft}}
\end{figure}

Classified articles are then aggregated to the daily frequency to construct 
the main treatment variables used in the regression analysis. For each 
trading day $t$, we count the number of articles classified as dovish 
and hawkish political pressure, yielding $n_{\text{dovish},t}$ and 
$n_{\text{hawkish},t}$, respectively. Both variables are non-negative 
integers that equal zero on days with no qualifying articles.\footnote{The 
choice of daily rather than intraday frequency is motivated by metadata availability: precise publication timestamps are available in the 
newswires only from mid-1997 onward, so moving to higher frequencies 
would sacrifice the first decade of the sample—a period that includes 
some of the most intense pressure episodes in our data.}
\autoref{fig:pressure_timeseries} shows the temporal evolution of 
our indicators.

Several patterns stand out. First, dovish pressure is the dominant 
form throughout the sample, consistent with the split 
documented in the previous section. Second, pressure is episodic 
rather than continuous, clustering around specific confrontations 
between the executive and the Fed. The most prominent dovish episodes 
are the Trump--Fed conflict during the first, and the second Trump administration. The Bush Sr.--Greenspan tensions of 
the early 1990s are also clearly visible, with President Bush publicly 
and repeatedly criticizing Greenspan for keeping rates too high ahead 
of the 1992 election—a conflict widely seen as having contributed to 
Bush's electoral defeat and to a lasting norm against overt 
presidential commentary on monetary policy. Third, hawkish pressure 
is rare but notable: the most intense hawkish episode coincides with 
the 2022 inflation surge, when CPI reached 9\% and several legislators 
publicly pushed for more aggressive tightening. A secondary cluster 
appears around the 2010 QE2 announcement, when a group of Republican 
lawmakers wrote an open letter urging the Fed to abandon its asset 
purchase program.\footnote{In November 2010, Speaker Boehner, Majority 
Leader Cantor, Senate Majority Leader McConnell, and Whip Kyl sent an 
open letter to Chairman Bernanke urging the Fed to ``resist further 
extraordinary intervention in the U.S. economy''.}

\begin{figure}[h!]
    \centering
    \caption{Comparison with Existing Benchmarks}
    \label{fig:tweets_vs_ours}
    \begin{subfigure}[b]{0.95\textwidth}
        \centering
        \includegraphics[width=\textwidth]{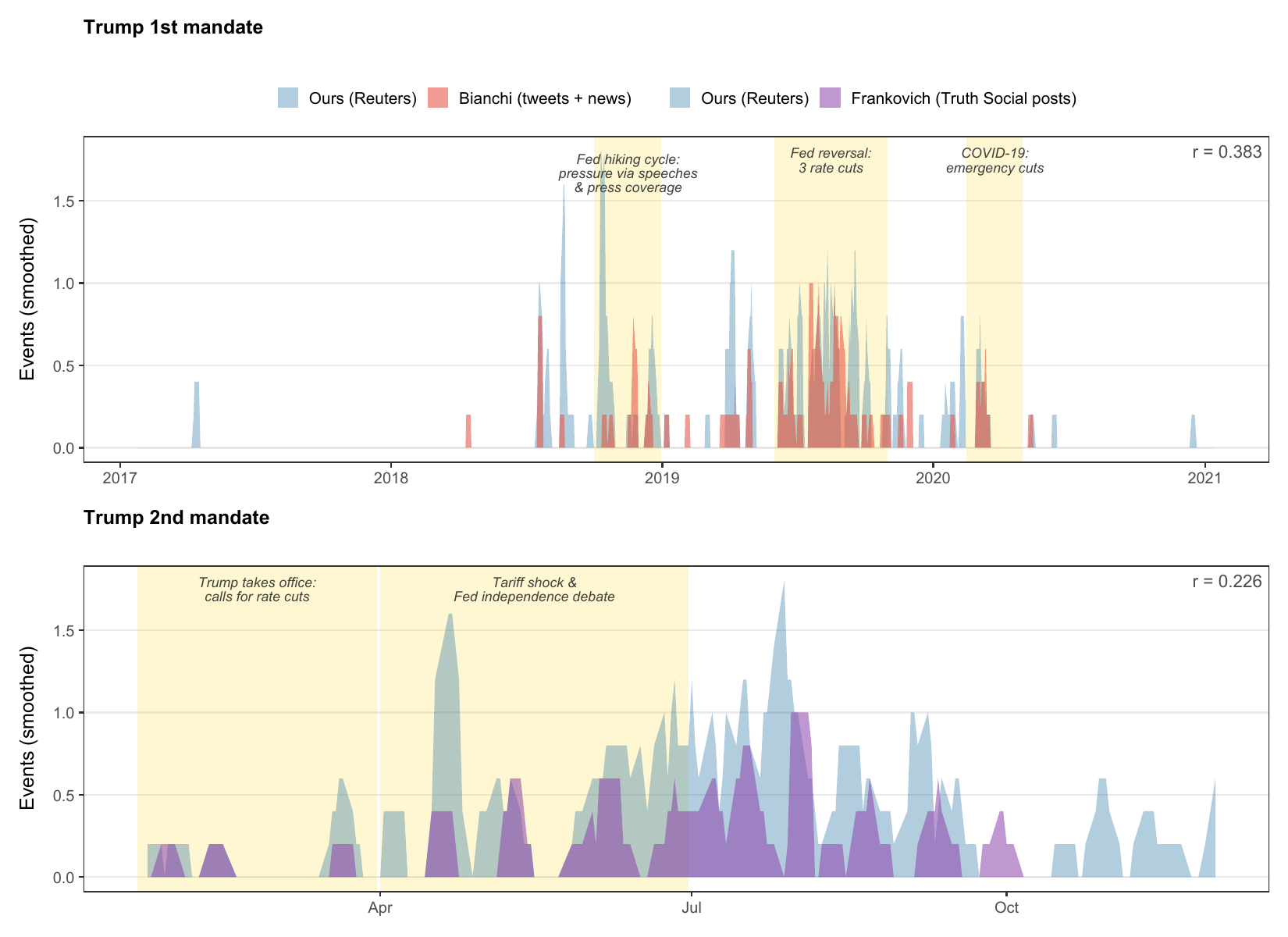}
    \end{subfigure}
    \vspace{0.2em}
    \footnotesize{\begin{flushleft}
    \textbf{Note}: All series are smoothed using a five-day centred rolling mean. Yellow shading denotes periods of heightened pressure 
    identified in the narrative record. Pearson correlation in the top-right corner.
    \end{flushleft}}
\end{figure}

We next benchmark our measure against two existing high-frequency measures for the Trump administrations: \citet{Bianchi2023}, who construct a 
daily pressure measure based on Trump's monetary policy tweets combined 
with Bloomberg news coverage during the first administration, and 
\citet{FrankovicKarau2026PoliticalPressureFed}, who use Truth Social 
posts during the second term. \autoref{fig:tweets_vs_ours} reports the 
comparison at the daily frequency, smoothed by a five-day centred 
rolling mean.

The two measures co-move broadly, correctly identifying the same major 
pressure episodes: the rate-hiking cycle of 2018--2019, the Fed 
reversal and subsequent rate cuts, the COVID-19 emergency cuts, and 
the tariff shock and Fed independence debate of 2025. The daily 
correlation between our measure and \citet{Bianchi2023} is 0.40, and 
0.23 against \citet{FrankovicKarau2026PoliticalPressureFed}. These 
moderate correlations are expected: \citet{Bianchi2023} and 
\citet{FrankovicKarau2026PoliticalPressureFed} restrict attention to 
a single actor and a single channel—presidential social media—whereas 
our measure aggregates pressure from the full spectrum of political 
actors and communication channels.

Tables~\ref{tab_confusion_bianchi} and~\ref{tab_confusion_frankovich} in Appendix A
formalize the overlap analysis. Against \citet{Bianchi2023}, our 
measure identifies pressure on 106 trading days versus 57 for the 
tweet-based measure, with 36 days in common. Against 
\citet{FrankovicKarau2026PoliticalPressureFed}, we identify 70 days 
versus 31, agreeing on 22. In both cases, the dominant off-diagonal 
cell consists of days flagged by us but not by the benchmark, 
confirming that the gap reflects the broader scope of our source. The 
small number of days captured by the benchmarks but not by us—21 and 
9 respectively—are consistent with social media posts that did not 
generate Reuters newswire coverage, plausibly because they were deemed 
insufficiently newsworthy.

Taken together, our measure advances the existing literature along two 
dimensions: it extends the time horizon to a systematic four-decade 
record, and it captures the full universe of actors and channels 
through which political pressure on the Fed is publicly exerted—
recovering a substantially larger set of pressure events while 
remaining well-anchored to the same historical episodes identified by existing measures.

\section{Data}
\label{sec:data}
% =============================================================================
% Section: Data & Measurement
% =============================================================================

\subsection{Dependent Variables}
\label{subsec:dep_var}

\subsubsection{Federal Funds Futures}

The primary dependent variables in this study are federal funds futures (hereafter, \emph{ffc}) traded on the Chicago Mercantile Exchange (CME), at different maturities. Ffc are contracts that settle on the average effective  federal funds rate over a given expiration month, quoted as 100 minus that expected rate. For an expiring contract, the last trading day corresponds to the last business day in the delivery month of the futures contract. The
corresponding daily federal funds overnight rate is provided by the Federal Reserve Bank of
New York. For each contract maturing $i$ months ahead, the futures rate can be decomposed into the expected federal funds target, a term capturing the discrepancy between the target and the effective rate, and a bias term reflecting risk premia. Under the standard assumption that political pressure events do not systematically affect these latter two components within a short window— an assumption shared by the high-frequency monetary policy literature since \citet{gurk_et_al_2005}— changes in futures prices over a narrow window identify the market's revision in interest rate expectations. Contracts at successively longer maturities allow us to trace this revision across the term structure, from the current month out to the twelve consecutive contract, providing a comprehensive picture of how political pressure shifts expectations not only about the immediate policy rate but also about the medium-run path of monetary policy.

We use daily settlement (or, where unavailable, closing) prices, sourced from LSEG Workspace. In the LSGE dataset, however, continuous federal funds futures series are available only for a limited historical window. To extend the series backward, we roll individual futures according to the following procedure: for each maturity $N$, we select the contract expiring exactly $N+1$ months ahead and stitch them together over time. Under this convention,
\emph{ffc1} corresponds to the current-month contract, \emph{ffc2} to the one-month-ahead contract, and so on.\footnote{Because these contracts reflect the average effective federal funds rate over the delivery month, price discontinuities arise when transitioning from one contract to the next. We account for these mechanical jumps by including a dummy variable for the beginning of month in our successive estimations to ensure that they do not affect the results.} In addition, since the front-month futures price represents a monthly average of the federal funds rate over the entire settlement month, a raw change in the futures price understates the true policy surprise when the pressure event occurs mid-month. To see this, consider a month with 30 days and a pressure event on day 20: the futures price only reflects the expected rate for the remaining 10 days of the month, so a one-basis-point revision in the expected overnight rate translates into only a $10/30$ basis point move in the futures price. Scaling by the inverse of this fraction recovers the true revision in the expected rate level. Following \citet{kuttner_2001}, we implement this adjustment formally as: 
\begin{equation}
\text{surprise}_t = \frac{D}{D - d}\cdot\Delta ffc^{1}_t,
\label{eq:kuttner}
\end{equation}

where $D$ denotes the total number of days in the settlement month 
and $d$ is the current calendar day. To avoid the scaling factor 
becoming unstable as the settlement date approaches---when the 
denominator $D - d$ shrinks toward zero---we cap the adjustment 
for the final three calendar days of each settlement period.\footnote{This correction applies exclusively to the front-month 
contract (\texttt{FFc1}), whose settlement price averages the 
federal funds rate over the current calendar month. For contracts 
at longer maturities, the settlement month lies entirely in the 
future and the futures price already reflects the full expected 
rate for that month, so no timing adjustment is required.}

\subsubsection{Treasury Futures}
\label{subsec:treasury_futures}

Moving to longer maturities, though, trading volumes in federal funds futures, generally become increasingly thin at the beginning of the sample, implying a progressive shortening of the usable time span for a given data quality threshold. To maintain a comparable number of observations across maturities and, crucially, to extend the analysis to the long end of the yield curve—where inflation expectations and term premium channels of political pressure could arise— we complement the federal funds futures with Treasury-based instruments. Specifically, we include the 3-month Treasury Bill yield from FRED as a benchmark for the short end of the curve, and add Treasury futures contracts for the 2-year (TU), 5-year (FV), 10-year (TY), and 30-year (US) maturities, sourced from the CME via LSEG Workspace.  

Treasury futures are standardized exchange-traded contracts that obligate the holder to buy or sell a notional basket of U.S. Treasury securities at a predetermined price on a future delivery date.\footnote{Rather than referencing a single bond, each contract specifies an eligible basket of Treasury securities within a given maturity range—for instance, the 10-year contract (TY) accepts delivery of Treasury notes with remaining maturities between 6.5 and 10 years—and uses a conversion factor system to make bonds of different coupons and maturities economically equivalent for delivery purposes.}\footnote{We prefer Treasury futures over on-the-run Treasury 
yields for two reasons. First, futures markets are among the most 
liquid financial markets in the world: the 10-year Treasury futures 
contract alone regularly trades several hundred thousand contracts 
per day, with bid-ask spreads of a single tick ($1/64$ of a point), 
making them less susceptible to microstructure noise than cash bond 
markets. Second, futures prices are not contaminated by the 
specialness premium that affects on-the-run Treasury yields in the 
repo market---whereby a bond in high demand as collateral trades at 
an artificially depressed yield relative to its fair value---providing 
a cleaner measure of interest rate expectations at standardized 
maturities along the yield curve \citep{duffie_1996, 
krishnamurthy_2002}.}

For these contracts, we construct continuous series by splicing quarterly-expiring contracts. Trading volumes decline near expiry as market participants roll into the next contract; rather than dropping these observations, we include a dummy variable in the regression that equals one on days for which $\log(\textit{volume})$ falls more than one rolling standard deviation below its 63-day rolling mean, absorbing any expiry-related noise without reducing the sample (Figure \ref{fig:roll_vol_decade}). Price returns on Treasury futures are expressed in basis points, with the sign convention that a positive return corresponds to a price increase and therefore a \emph{decline} in the underlying yield.

This cross-instrument coverage allows us to characterize the 
complete yield-curve response to political pressure on the Federal 
Reserve, from the near-term policy rate embedded in the front-month 
federal funds futures contract to the very long-run expectations captured by the 30-year Treasury futures.

\section{Empirical Strategy and Results}
\label{sec:strategy}
% =============================================================================
% Section: Empirical Strategy
% =============================================================================
Building on the data and classification pipeline described in Section~\ref{sec:measurement} and ~\ref{sec:data}, this section formalises the regression framework used to estimate the effect of political pressure on rate expectations. The core specification regresses the daily change in each rate instrument on the contemporaneous count of political-pressure articles, two lags of the dependent variable, a vector of dummies, and a vector of controls. Formally:

\begin{equation}
  \Delta r^{m}_t 
  = \alpha 
  + \beta\, \text{PolPres}_t 
  + \sum_{k=1}^{2}\gamma_k\, \Delta r^{m}_{t-k} 
  + \mathbf{D}_t' \delta
  + \mathbf{Z}_t' \lambda
  + \varepsilon_t
  \label{eq:baseline}
\end{equation}

where $\Delta r^{m}_t$ is the daily change in the rate instrument at maturity $m$, the variable $\text{PolPres}_t$ denotes the daily count of news articles classified as dovish or hawkish political pressure, and the coefficient $\beta$ therefore measures the Conditional Average Treatment Effect (CATE) in the rate instrument associated with one additional political-pressure article on day $t$.\footnote{We drop pressure events falling on weekends or bank holidays, since markets are closed and no same-day price change exists to identify a response. Reassigning them to the following trading session would mechanically load their effect onto that day's coefficient, confounding it with pressure genuinely originating then.} The two lags $\Delta r^{(m)}_{t-1}$ and $\Delta r^{(m)}_{t-2}$ absorb short-run autocorrelation that is common in daily financial series, and $\varepsilon_t$ is the regression residual. The regressions are estimated at the daily frequency, which implies that rate fluctuations on a given day may reflect a combination of political pressure and other contemporaneous factors. Isolating the effect of political pressure from these confounders requires a rich set of controls, which we organize into two groups.

The first group ($\mathbf{D}_t$) consists of indicator variables. We include a beginning-of-month dummy ($\text{BOM}_t$) to absorb the mechanical 
price discontinuities that arise when rolling federal funds futures contracts at month turns, and an FOMC announcement dummy to capture the large movements in rate expectations that occur on policy decision days and that are by construction unrelated to political pressure. For Treasury futures, we substitute the $\text{BOM}_t$ dummy with the low-volume dummy described in section \ref{subsec:treasury_futures}.

The second group collects time-varying covariates ($\mathbf{Z}_t$) that proxy for the macroeconomic and financial environment in which pressure occurs. As discussed above, the core concern is that political pressure and futures returns both correlate with the state of the economy. To absorb business-cycle variation and control for the state of the Federal Reserve's dual mandate, we include the inflation gap—defined as PCE in annual percentage changes minus the 2\% target—and the unemployment gap relative to the CBO estimate of the natural rate.\footnote{The FOMC announced an explicit 2\% objective only in its January 2012 Statement on Longer-Run Goals and Monetary Policy Strategy; for most of our sample no formal target existed.  The pre-2012 implicit objective was in any case close—participants' stated preferences clustered near $1.5\%$ before converging to 2\% by 2009, with long-run expectations anchored around 2\%.}\footnote{Both inflation and unemployment data were downloaded from FRED.} These are slow-moving variables that capture the macroeconomic environment in which political pressure occurs, and that directly correspond to the two objectives the Fed is mandated to pursue. Including them ensures that any estimated effect of political pressure on rate expectations is not simply picking up the mechanical response of markets to a deteriorating economic outlook that politicians and investors observe simultaneously. To further control for higher-frequency predictable variation in rate expectations, we include a subset of the controls proposed by \citet{bauer_swanson_2023}, who show that a range of publicly observable variables - percentage change in the oil price from three months (65 trading days), percentage change in the S\&P 500 stock market index from three months  (65 trading days), nonfarm payroll release surprises, and Treasury yield skewness \citep{chernov2024}— can predict monetary policy surprises ahead of FOMC meetings, implying that naive high-frequency regressions may confound genuine shocks with forecastable variation in the policy rate.\footnote{We retain only those Bauer--Swanson controls available at daily frequency over our full 1988--2025 sample}

The equation is estimated by OLS, and all standard errors are computed using Newey-West HAC corrections. Regarding the sample period, we consider two specifications: \emph{full sample} and \emph{excl.\ Trump}. The \emph{full sample} covers all trading days over 1988--2025. The \emph{Excl.\ Trump} specification excludes the first Trump term (20 January 2017 to 20 January 2021) and the post-inauguration period of the second term (from 20 January 2025 onwards). This split allows us to assess whether the unusually intense political pressure during these periods drives the results relative to more typical episodes.

To examine the persistence of political pressure shocks, we estimate a series of local projections following \cite{jorda2025}. To preserve the exogeneity of the shock at each forecast horizon, the control variables are constructed over the corresponding horizon rather than fixed at time t. Specifically, for flow variables, we use their cumulative values over the interval [t,t+h]. In contrast, for the inflation gap and unemployment gap—which are stock variables observed at a monthly frequency and carried forward to daily observations—we use their average value over the same horizon rather than their cumulative sum.
For each horizon $h = 0, 1, ..., H_MAX$, we estimate
\begin{equation}
\label{eq:lp}
\sum_{s=t}^{t+h} \Delta r_s
    = \alpha_h
    + \beta_h \, \textit{PolPres}_t
    + \sum_{k=1}^{2}\gamma_k\, \Delta r^{m}_{t-k} 
    + \mathbf{F}_{t,h}' \delta_h
    + \mathbf{S}_{t,h}' \lambda_h
    + \varepsilon_{t+h}
\end{equation}
where $F_{t,h}$ denotes the vector of flow controls, constructed as cumulative values over the horizon [t,t+h], and $S_{t,h}$ denotes the vector of stock controls, constructed as horizon averages over the same interval. In our application, $S_{t,h}$ consists of the inflation gap and unemployment gap, while $F_{t,h}$ includes the remaining control variables.

\subsection{Baseline Results}
\begin{figure}[H]
\centering
\caption{Impact of dovish and hawkish pressure on federal funds futures}
\label{fig:coef_drivers_ffc}
\includegraphics[width=0.9\linewidth]{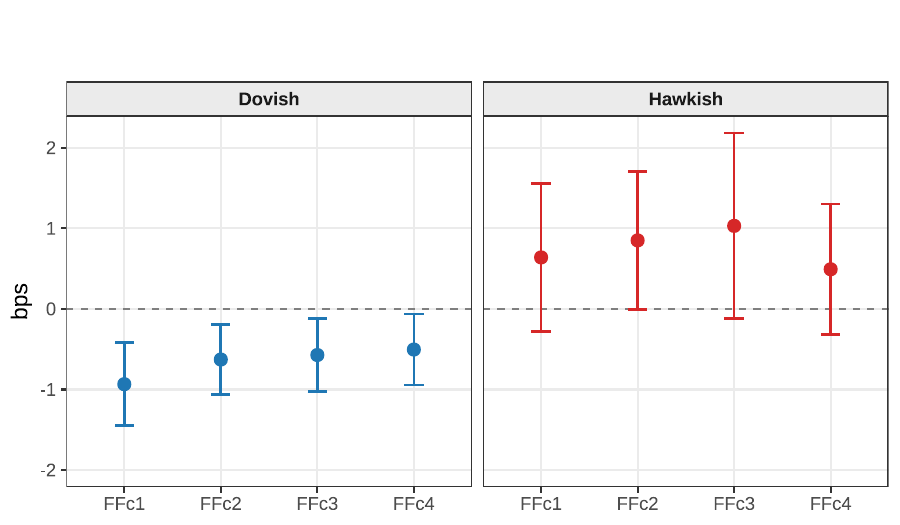}
  \footnotesize{\begin{flushleft}
\textbf{Note:} The figure shows OLS estimates of $\hat{\beta}$ from equation \eqref{eq:baseline} estimated separately for dovish and hawkish pressure across federal funds futures contracts. The dependent variable is the daily change in the implied rate (in basis points). Confidence intervals are constructed at the 90\% level using Newey--West HAC standard errors.\end{flushleft}}
\end{figure}

We first estimate the baseline specification on the full sample to assess whether the impact of political pressure on rate expectations varies by direction and maturity.
The central finding is that dovish political pressure is associated with a statistically significant decline in near-term federal funds rate expectations, whereas hawkish pressure yields no comparable effect, as shown in \autoref{fig:coef_drivers_ffc}.

For dovish pressure, the estimated effect is largest at the front of the curve: an additional pressure event is associated with a 0.91 bps decline in the front-month contract.\footnote{We focus initially on \emph{ffc1}--\emph{ffc4} as these are the most liquid contracts throughout the sample period; trading volume drops substantially beyond the four-month horizon in the earlier portion of the sample.} The effect decays monotonically along the near-term segment of the term structure, falling to 0.62 bps for one-month ahead, 0.57 bps for two-months ahead, and 0.50 bps for three-months ahead. This attenuation pattern is consistent with a market revision of expectations that only affects short horizons. To interpret the economic magnitude of the estimated effects, note that the standard Federal Reserve policy move is 25 bps. Under a simple two-outcome scenario in which the Fed either holds rates unchanged or cuts by 25 bps, a futures price decline of 0.25 bps corresponds to a 1 percentage point increase in the implied probability of a cut. On this metric, each additional dovish political pressure raises the probability of a near-term cut by approximately 3.7 percentage points at the front-month horizon, declining to 2.5, 2.3, and 2.0 percentage points for the subsequent maturities. While modest relative to the usual tightening and loosening cycles ($\pm 200\text{--}400\,\text{bp}$), these effects are per-event averages; therefore, the cumulative signal embedded in sustained political pressure campaigns can be meaningfully larger.\footnote{The regressions include two lags of the dependent variable (Equation \ref{eq:baseline}), so the long-run multiplier is $1/(1-\hat{\gamma}_1-\hat{\gamma}_2)$. Across contracts, the sum of AR coefficients ranges from 0.07 to 0.11, implying cumulative effects 8--12\% larger than the contemporaneous estimates reported in the text.}

Estimates for hawkish pressure are positive across all maturities and borderline significant at conventional levels for the second and third contract, though the wide confidence intervals — reflecting the relative scarcity of hawkish episodes, which account for roughly one-sixth of observed dovish pressure days — preclude sharp inference. The point estimates suggest a symmetric market response in direction, but the asymmetry in episode frequency prevents a definitive comparison of magnitudes.

The baseline coefficients are accompanied by statistically
significant estimates on most of the controls, confirming that the specification captures the daily determinants of futures-price variation.  The
political pressure coefficient is estimated against, rather than
in place of, this conditional information set (Table \ref{tab_drivers_dovish_ffc}). The inflation gap enters positively and the unemployment gap negatively across
all four maturities, each significant at conventional levels,
reproducing the sign pattern of a standard Taylor-type response:
futures price in a higher near-term policy rate when inflation runs
above target and a lower rate when unemployment runs above its
natural level.  The FOMC announcement dummy is negative and
significant throughout, with magnitudes between $-0.010$ and
$-0.012$ basis points, consistent with the average FOMC meeting in
our sample being dovish relative to prior expectations.  The
Nonfarm Payroll release dummy is also negative and strongly
significant at the front of the curve--- indicating that scheduled labour-market
releases in the sample coincided on average with downward
revisions in the expected policy path.  The yield-curve skewness
measure (ISK) enters positively and is highly significant at every
maturity, with magnitudes rising from roughly 0.0035 at FFc1 to
0.007 at FFc4, which is consistent with a standard term-premium
channel in which richer convexity in the yield distribution
translates into higher forward rates further out the curve. Analogous results hold for the hawkish regressions (Table \ref{tab_drivers_hawkish_ffc}),

\begin{table}[H]
   \caption{\label{tab_drivers_dovish_ffc} Dovish Political Pressure and Fed Funds Futures}
   \centering
   \footnotesize
   \setlength{\tabcolsep}{4pt}
   \renewcommand{\arraystretch}{0.95}
   \begin{tabular}{lcccc}
      \tabularnewline \midrule \midrule
      Dependent Variables:                   & FFc1 (Kuttner)          & FFc2                    & FFc3                         & FFc4\\  
      Model:                                    & (1)                          & (2)                         & (3)                     & (4)\\  
      \midrule
      \emph{Variables}\\
      Constant                                  & -0.0004                      & -0.0010$^{**}$              & -0.0011$^{**}$          & -0.0013$^{**}$\\   
                                                & (0.0007)                     & (0.0005)                    & (0.0005)                & (0.0005)\\   
      Dovish pressure ($n_{\text{dovish}}$)     & -0.0091$^{***}$              & -0.0062$^{**}$              & -0.0057$^{**}$          & -0.0050$^{**}$\\   
                                                & (0.0033)                     & (0.0027)                    & (0.0027)                & (0.0025)\\   
      Oil price growth (65-day \% return, \%)   & $5.36\times 10^{-5}$$^{**}$  & $3.26\times 10^{-5}$$^{*}$  & $2.42\times 10^{-5}$    & $2.14\times 10^{-5}$\\    
                                                & ($2.64\times 10^{-5}$)       & ($1.85\times 10^{-5}$)      & ($1.87\times 10^{-5}$)  & ($1.87\times 10^{-5}$)\\    
      S\&P~500 return (65-day \% return, \%)    & 0.0002$^{*}$                 & 0.0001$^{*}$                & 0.0002$^{**}$           & 0.0002$^{***}$\\   
                                                & (0.0001)                     & ($7.12\times 10^{-5}$)      & ($6.9\times 10^{-5}$)   & ($6.06\times 10^{-5}$)\\    
      NFP announcement                          & -0.0094$^{***}$              & -0.0062$^{***}$             & -0.0077$^{***}$         & -0.0063$^{**}$\\   
                                                & (0.0033)                     & (0.0022)                    & (0.0025)                & (0.0030)\\   
      FOMC announcement                         & -0.0102$^{**}$               & -0.0112$^{***}$             & -0.0119$^{***}$         & -0.0124$^{***}$\\   
                                                & (0.0047)                     & (0.0038)                    & (0.0039)                & (0.0036)\\   
      Yield skewness (ISK)                      & 0.0035$^{**}$                & 0.0046$^{***}$              & 0.0058$^{***}$          & 0.0070$^{***}$\\   
                                                & (0.0016)                     & (0.0015)                    & (0.0015)                & (0.0014)\\   
      Inflation gap                             & 0.0011$^{*}$                 & 0.0008$^{*}$                & 0.0008$^{*}$            & 0.0008$^{**}$\\   
                                                & (0.0006)                     & (0.0004)                    & (0.0004)                & (0.0004)\\   
      Unemployment gap                          & -0.0008$^{**}$               & -0.0005$^{**}$              & -0.0005$^{**}$          & -0.0006$^{***}$\\   
                                                & (0.0003)                     & (0.0002)                    & (0.0002)                & (0.0002)\\   
      
    \midrule
      \emph{Fit statistics}\\
      Observations                              & 8,963                        & 8,976                       & 8,972                   & 8,972\\  
      R$^2$                                     & 0.018                        & 0.034                       & 0.040                   & 0.032\\  
      \midrule \midrule
      \multicolumn{5}{l}{\emph{}}\\
   \end{tabular}
   
   \par \raggedright 
   \footnotesize \textbf{Note:}
   All columns use the full sample. Coefficients on FF futures variables multiplied by 100 are expressed in basis points. All regressions include: beginning-of-month dummy and two lags of the dependent variable (omitted from display). HAC (Newey-West) standard errors in parentheses. *, **, *** denote significance at the 10\%, 5\%, 1\% level.
\end{table}

\begin{figure}[H]
    \centering
    \caption{Impact of dovish political pressure across maturities}
    \label{fig:baseline_all_term}
    \includegraphics[width=\linewidth]{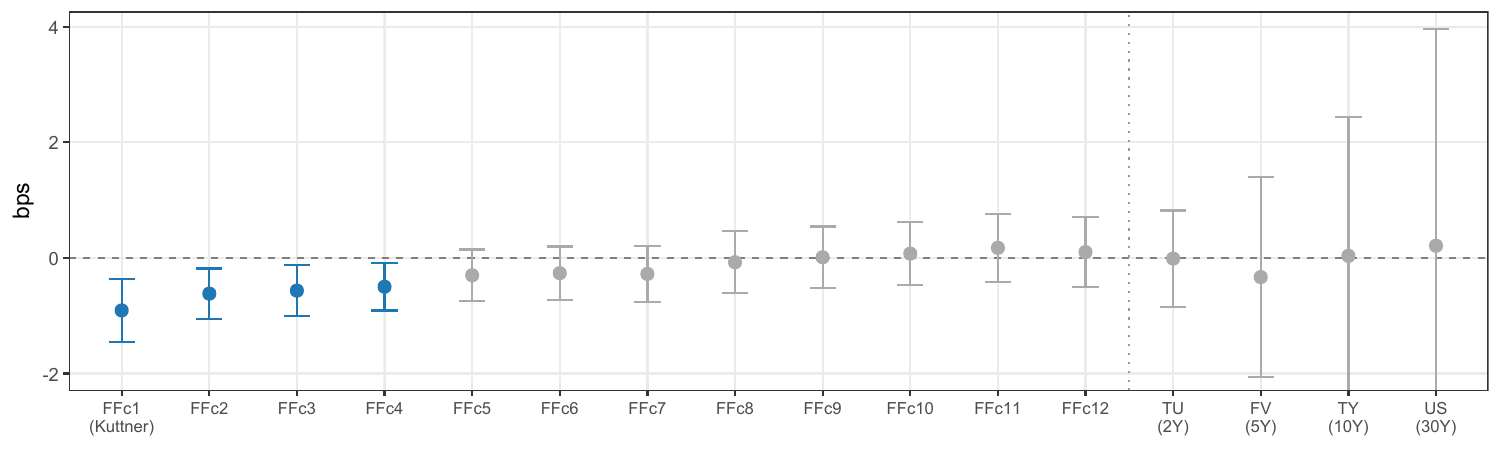}
    \footnotesize{\begin{flushleft}
    \textbf{Note:} The figure shows OLS estimates of $\hat{\beta}$ from equation \eqref{eq:baseline} for dovish pressure across the full maturity spectrum, from near-term federal funds futures (\emph{ffc1}--\emph{ffc12}) to Treasury futures (TU, FV, TY, US). Blue markers denote estimates significant at the 10\% level; grey markers denote insignificant estimates. The dependent variable is the daily change in the implied rate or yield (in basis points). A dotted vertical line separates federal funds futures from Treasury futures. Confidence intervals are constructed at the 90\% level using Newey--West HAC standard errors.\end{flushleft}}
\end{figure}

To assess whether the effect extends beyond the near-term horizon, we estimate the baseline specification across the full term structure of federal funds futures contracts, supplementing them with Treasury futures to gauge the medium- and longer-term impact. Figure \ref{fig:baseline_all_term} shows the results.

The statistically significant estimates cluster at the near end of the curve — \emph{ffc1} through \emph{ffc4} — consistent with dovish pressure primarily shifting rate expectations over the next three months. The effect remains borderline significant through \emph{ffc7}, with point estimates consistently negative and confidence intervals that only marginally encompass zero, before dissipating entirely at longer horizons.

For Treasury futures, point estimates are near zero or positive across all four contracts, with confidence intervals that widen substantially with maturity. This indicates that dovish political pressure does not lead markets to revise longer-term rate expectations, consistent with the view that political interference is perceived as a transitory phenomenon that leaves the medium- and long-run policy stance unaffected. Moreover, the absence of any upward pressure on longer-term yields argues against a risk-premium channel in which political interference de-anchors inflation expectations and pushes up nominal yields.

\begin{figure}[H]
    \centering
    \caption{Cumulative Responses of Federal Funds Futures to Dovish Political Pressure}
    \label{fig:LP_baseline}
    \includegraphics[width=\linewidth]{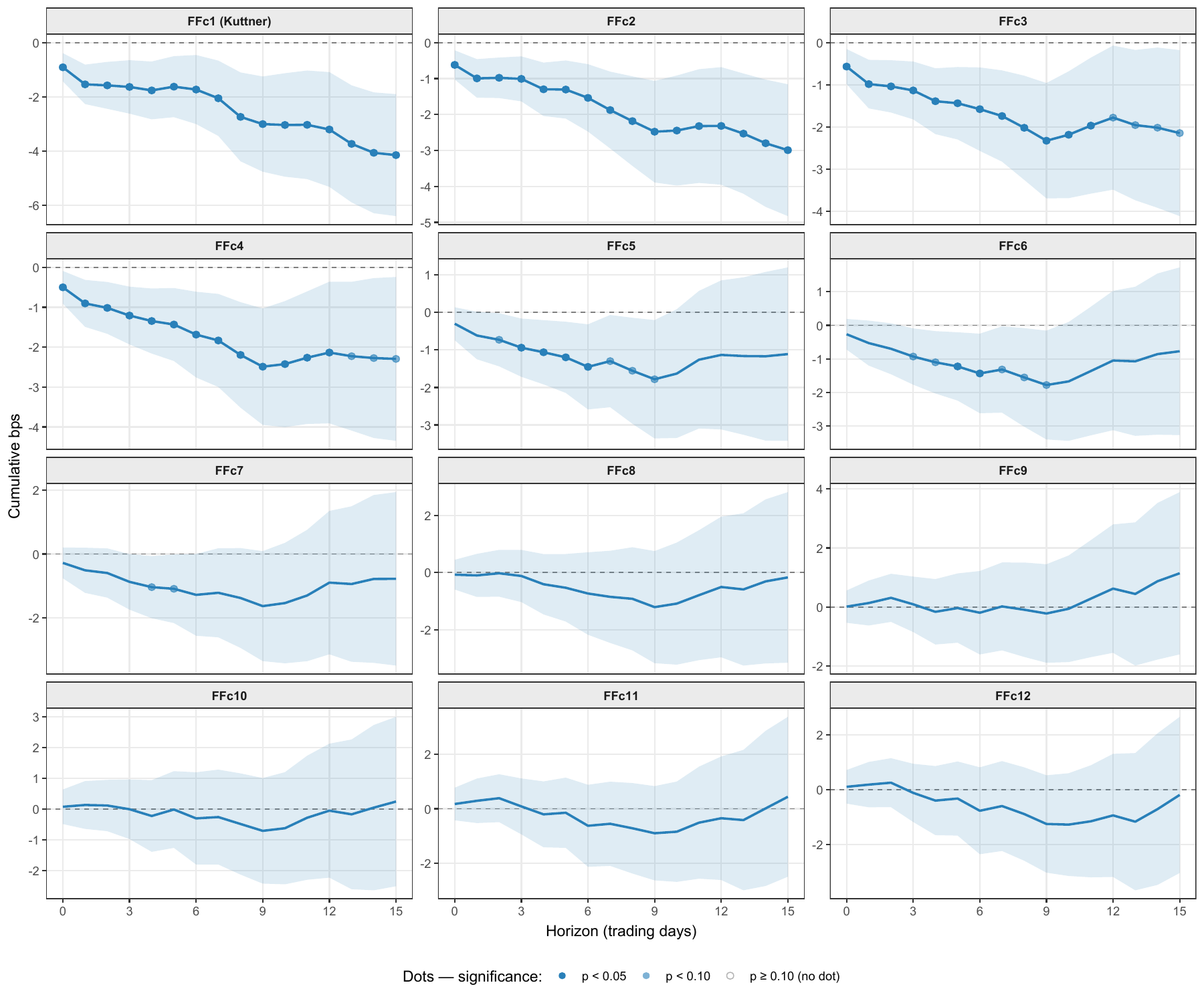}
    \footnotesize{
    \begin{flushleft}
    \textbf{Note:} The figure reports cumulative impulse-response functions of Federal Funds futures contracts (FFc1--FFc12) to a one-unit increase in dovish political pressure, estimated using the local projection framework of \citet{jorda2025} from equation \ref{eq:lp}. Shaded areas denote 90\% confidence intervals.  Horizon $h=0$ corresponds to the contemporaneous OLS estimate. 
    \end{flushleft}
    }
\end{figure}

The persistence of political pressure shocks is assessed through the cumulative impulse response functions obtained from the local projections, reported in Figure \ref{fig:LP_baseline}. By construction, the impact responses (h=0) coincide with the baseline OLS estimates. Since the analysis focuses on cumulative impulse responses, the estimated effects are not expected to converge back to zero, but rather capture the cumulative evolution of the response over the forecast horizon. Overall, the local projection results confirm the findings from the baseline specification. The effects of political pressure are stronger, more statistically significant, and more persistent for shorter-maturity federal funds futures contracts, while they gradually weaken for longer maturities. Interestingly, contracts with maturities five and six months exhibit a delayed response, with the cumulative effect becoming statistically significant only after approximately two to three trading days.

The main caveat of this analysis is that the baseline coefficient is a reduced-form total effect: dovish pressure may simply proxy for a deteriorating outlook that markets and politicians observe simultaneously, in which case the estimate reflects anticipated easing driven by the macroeconomic state rather than by pressure itself. Although our daily identification and rich set of controls should substantially mitigate this concern, we cannot entirely rule it out.

To further test for the exogeneity of our result, we introduce an indicator variable representing party alignment by the FOMC board: this indicator is equal to 1 when more than 50\% of the Federal Reserve Board of Governors is of the same party of the executive and 0 otherwise.\footnote{To classify board members we use the party of the appointing president. Although FOMC members are supposed to act independently once appointed, this is a commonly used shortcut in the literature: the seminal paper by \cite{chappell1993partisan} argues that presidential influence on the Fed operates largely through the appointment channel, and codes FOMC members by appointing-president party.} If dovish pressure merely proxied for a weak economy, its effect would not depend on Board composition — a weak economy calls for the same easing whoever holds the Governor seats. Board alignment, moreover, is fixed well before any single pressure event, so it cannot be a stand-in for the day's macroeconomic news. A response that strengthens under alignment therefore reflects the Fed's anticipated reaction function rather than shared information about the economy.

Table \ref{tab:interaction_board_zlb} shows the results of this test.

\begin{table}[htbp]\centering
\caption{FOMC Board Party Alignment and Political Pressure}
\label{tab:interaction_board_zlb}
\small
\begin{tabular}{lrrrr}
\hline\hline
Interaction Term & FFc1 & FFc2 & FFc3 & FFc4 \\
\hline
Board Majority (Pres. Party) & -1.56** (0.63) & -1.47*** (0.53) & -1.53*** (0.54) & -1.47*** (0.49) \\
\hline\hline
\multicolumn{5}{p{1\linewidth}}{\footnotesize \textbf{Note:} The table reports the coefficient on the interaction term $n_{\text{dovish},t} \times \text{BoardMajority}_t$ from the baseline specification in Equation \eqref{eq:baseline}, augmented with the interaction and the uninteracted FOMC Board-majority indicator. The dependent variable is the daily change in the implied rate (in basis points) for federal funds futures contracts FFc1--FFc4. $\text{BoardMajority}_t$ equals 1 when the sitting President's party holds a majority of the Federal Reserve Board of Governors on day $t$, and 0 otherwise; Board composition is coded from official appointment. The reported coefficient is the additional response to one dovish pressure event on aligned-Board days, over and above the unconditional $n_{\text{dovish}}$ effect; the baseline $n_{\text{dovish}}$ coefficients and the full control set (inflation gap, unemployment gap, Bauer--Swanson controls, FOMC and beginning-of-month dummies, and two lags of the dependent variable) are included. Estimation is by OLS over the full 1988--2025 sample. HAC (Newey--West) standard errors in parentheses. *, **, *** denote significance at the 10\%, 5\%, and 1\% level.} \\
\end{tabular}
\end{table}

The interaction is negative and significant at every near-term maturity (Table \ref{tab:interaction_board_zlb}). Under an aligned Board the incremental effect of about --1.5 bps per event exceeds the unconditional coefficient, with the response more than doubling. This is the pattern the informational reading cannot produce: a deteriorating outlook calls for the same easing regardless of Board composition, so a response that scales with alignment must operate through the Fed's anticipated reaction function rather than through shared macroeconomic news. The reduced-form estimate is thus not merely a repricing of the outlook — markets price a higher probability of accommodation precisely when the appointment configuration makes it more feasible. We read the interaction not as a structural compliance parameter but as a comparative static on the market's prior over accommodation.

\subsection{A Trump story?}
\label{subsec:a_trump_story}

\begin{figure}[H]
    \centering
    \caption{Dovish political pressure: full sample vs.\ excluding Trump periods}
    \label{fig:full_vs_no-Trump}
    \includegraphics[width=\linewidth]{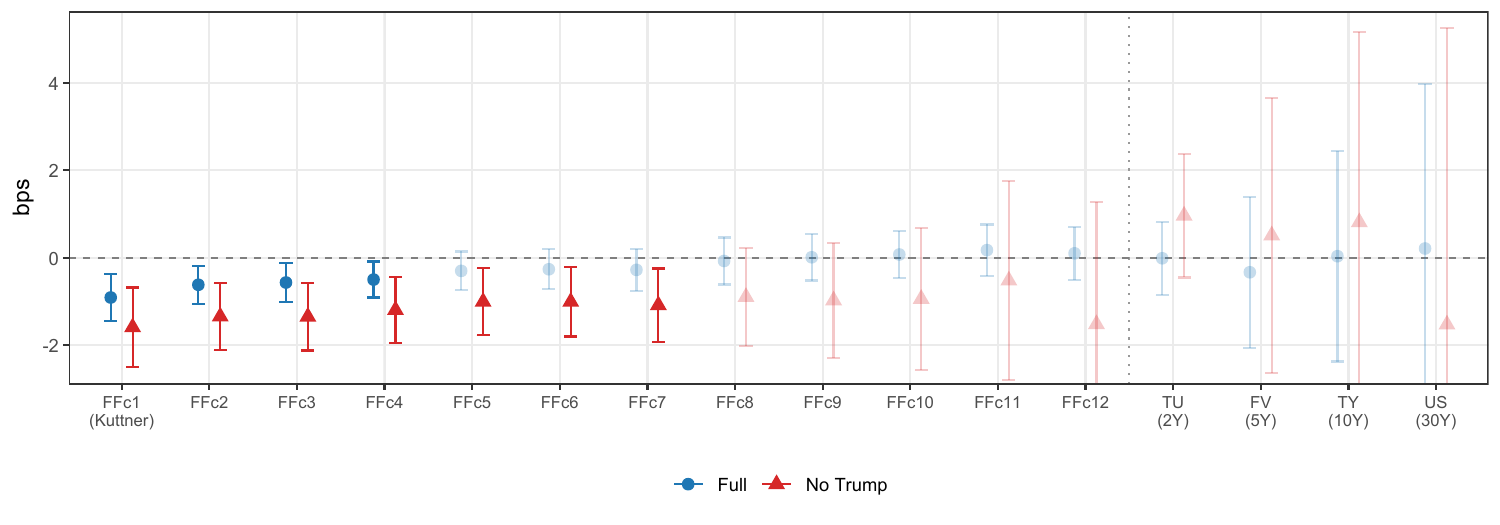}
    \footnotesize{\begin{flushleft}
    \textbf{Note:} The figure shows OLS estimates of $\hat{\beta}$ from equation \eqref{eq:baseline} for dovish pressure, estimated on the full sample and on the subsample excluding the Trump presidencies. The dependent variable is the daily change in the implied rate (in basis points). Confidence intervals are constructed at the 90\% level using Newey--West HAC standard errors.
    \end{flushleft}}
\end{figure}

A key question is whether the baseline results are driven solely by Trump presidencies. As shown in \autoref{fig:pressure_timeseries}, the intensity of political pressure increased markedly during both mandates, with the pattern shifting from isolated episodes to a more persistent and continuous process. To answer this question, we re-estimate the baseline specification on a restricted sample excluding the Trump periods. Results are reported in \autoref{fig:full_vs_no-Trump}; estimates for hawkish pressure remain statistically insignificant throughout, hence we exclude them.

The sign of the effect is stable across samples, while its magnitude is approximately 50\% larger outside the Trump period at all maturities. More specifically, in the non-Trump subsample, near-term rate expectations for the front-month contract decline by 1.58 bps per additional pressure event, with the effect attenuating gradually to 1.37 bps for \emph{ffc2}, 1.37 bps for \emph{ffc3}, and 1.24 bps for \emph{ffc4}.

These findings allow us to strongly reject the argument that the baseline results are driven by the Trump presidencies. If anything, the effect is larger and more consistently significant outside that period, confirming that the underlying mechanism operates throughout the full 1988--2025 sample and is not attributable to a single high-intensity episode. Notably, despite the reduction in sample size and the associated loss of precision, the non-Trump estimates remain statistically significant up to the six-month horizon. 

One natural interpretation of the attenuated Trump-period estimates is that markets progressively discount political pressure as its frequency rises. When pressure is a rare signal, each additional article carries substantial informational weight and is more likely to be perceived as a credible cue. When it becomes a near-daily occurrence, however, the marginal impact on rate expectations diminishes: markets may come to treat persistent pressure as background noise rather than as new information about the future path of monetary policy. An alternative interpretation is that the true effects differs across presidential administrations, reflecting heterogeneity in the perceived credibility of the threat or in the Fed's willingness to accommodate. Under this view, the attenuation is not a function of frequency but of the political context itself. Distinguishing between these two mechanisms —saturation versus structural heterogeneity — is not straightforward in a reduced-form framework, and we treat the split-sample evidence as suggestive rather than conclusive. However, we provide evidence consistent with the saturation mechanism in Appendix~\ref{subsec:pressure_intensity}. Regardless of which interpretation one favors, the key takeaway is unambiguous: the baseline results are not driven by the Trump presidencies, as the effect is stronger, more precisely estimated, and significant across a wider range of maturities precisely in the sample that excludes them.

\subsection{Beyond Interest Rates}
\label{subsec:beyond_rates}

The evidence thus far establishes that dovish political pressure
shifts near-term interest rate expectations, with the effect
concentrated at the short-end of the yield curve.  We now ask whether
the same pressure episodes move other financial markets, and what the
joint cross-asset response reveals about the transmission channel.

Two mechanisms can rationalise the baseline rate response.  The
first is a \emph{monetary channel}: pressure is read as news
about the near-term path of monetary policy, prompting markets to
reprice short-term rate expectations without any implication for
central bank credibility.  The second is a \emph{credibility
channel}: persistent or intense pressure raises the probability,
in the eyes of market participants, that the Fed's independence is
compromised, with implications that extend beyond the short end to
longer-horizon inflation expectations and to risk premia more
broadly.  The two channels are observationally equivalent at short
maturities---both predict a dovish repricing of FFc1 through
FFc4---but they diverge sharply across other instruments.  Under the
information channel, longer-horizon inflation expectations remain
anchored and financial market volatility does not respond; under the
credibility channel, long-run inflation compensation rises, implied
volatility increases in both equity (VIX) and bond (MOVE) markets,
and the dollar depreciates on a broader basis than can be accounted
for by the rate shift alone.  We examine each of these implications
in turn.

\paragraph{Credibility measures.}

Tables \ref{tab_drivers_inflexp_dovish} and ~\ref{tab_drivers_volatility_dovish} in the Appendix report the estimated response two inflation compensation
measures (2-year and 10-year inflation swap rates) and two volatility indices (VIX, MOVE) to dovish
political pressure, both with and without the Trump presidencies.
None of the eight dovish coefficients is statistically
distinguishable from zero, and the pattern is informative beyond
statistical insignificance.

The point estimates on inflation compensation are, if anything,
mildly \emph{negative} across all four specifications---the opposite
of the sign the credibility channel would predict.  A deterioration
in perceived Fed independence should raise long-run inflation
expectations as markets price a higher probability of accommodation
beyond what the Fed's mandate would imply; we see the reverse, with
small negative point estimates at both the two- and ten-year
horizons, in both the full sample and the subsample excluding the
Trump presidencies.  The market-implied inflation anchor is
therefore robust to dovish pressure: markets read these episodes as
shifting the near-term policy path without revising their
assessment of the Fed's long-run commitment to price stability.

The volatility estimates tell a consistent story.  Dovish pressure
does not generate a systematic increase in implied volatility
across equity or bond markets: the coefficients are small in
magnitude and inconsistent in sign across the full and no-Trump
subsamples, a pattern more consistent with noise around zero than
with a repricing of institutional risk.  If markets interpreted
political pressure as a credibility-undermining event, implied
volatility should rise across the board---in equities as investors
reprice policy uncertainty, and in bonds as the distribution of
future rates widens around its conditional mean.  We find no such
evidence.\footnote{The inflation swap and volatility regressions are
estimated on shorter samples than the federal funds futures
regressions (roughly 4,600--5,400 observations versus 9,000),
reflecting the later inception of these markets: the VIX began
trading in 1990, inflation swaps in the early 2000s, and the MOVE
index from 2002.  Statistical power is correspondingly lower.
The point estimates reported here are, however, economically
small relative to the unconditional daily variation in each
series: the dovish coefficient in the 2-year inflation swap
regression implies a response of less than one-twentieth of a
standard deviation per additional pressure article, an order of
magnitude smaller than the corresponding FFc1 effect.}

\paragraph{Exchange rates.}
\label{par:fx}
While the credibility channel appears inactive, the short-rate
expectations channel still predicts a concrete implication for
foreign exchange: under uncovered interest parity, a decline in
near-term US rate expectations should---holding foreign rates
constant---generate a depreciation of the US dollar.  To test this,
we re-estimate the baseline with the dependent variable replaced by
the daily log return on the Dollar Index (DXY) and the EUR/USD spot
rate.  Because the EUR/USD is a bilateral exchange rate, identifying
the US leg of the response requires conditioning on contemporaneous
euro-area monetary developments: a daily movement in EUR/USD reflects
both US and euro-area interest rate expectations, and without a
euro-leg control the estimated coefficient would conflate the two.
We therefore augment the baseline with the daily change in the
two-year German Bund yield, $\Delta r^{\text{Bund,2Y}}_t$, which
absorbs high-frequency movements in euro-area short-rate
expectations.\footnote{The Bund is available throughout our 1987--2025 sample,
whereas ESTR OIS is available only from 2019 and EONIA OIS from
1999; using a spliced control would introduce a structural break
in the euro-leg series that the Bund avoids.}  Formally, for $y_t \in \{\Delta\log \text{DXY}_t,\;
\Delta\log \text{EUR/USD}_t\}$ we estimate

\begin{equation}
\label{eq:fx_spec}
y_t = \alpha
    + \beta \, n_{\text{dovish},t}
    + \phi \, \Delta r^{\text{Bund,2Y}}_t
    + \sum_{l=1}^{2} \delta_l \, y_{t-l}
    + \boldsymbol{\gamma}' \mathbf{X}_t
    + \varepsilon_t,
\end{equation}

\noindent where $\mathbf{X}_t$ collects the baseline controls.  Under the UIP prediction,conditioning on the Bund yield isolates the dollar-leg response toUS political pressure: $\hat\beta < 0$ for the DXY and $\hat\beta > 0$for the EUR/USD would indicate that dovish pressure on the Fed
depreciates the dollar relative to a foreign currency whose own
short-rate expectations are held fixed by the control.

Table~\ref{tab_drivers_fx_dovish} confirms the prediction.
Dovish pressure depreciates the dollar on both measures, with the
effect stronger and more precisely estimated in the sample excluding
the Trump administrations.  In the full sample, the EUR/USD
appreciates by 4.5 basis points per additional dovish
article (significant at 10\%) and the DXY declines by 2.6 basis
points (not significant); excluding the Trump windows, the effects
rise to 10.87 and $-7.5$ basis points respectively, both significant.\footnote{The full-sample DXY coefficient is less precisely estimated
than its EUR/USD counterpart.  This is consistent with the composition
of the DXY, which weights five non-euro currencies---JPY, GBP, CAD,
SEK, and CHF---whose own short-rate expectations are not controlled
for in Equation~\eqref{eq:fx_spec}.  The EUR/USD specification is, by
construction, the cleaner test: it is bilateral, and the Bund control
isolates the dollar-leg response by absorbing contemporaneous
euro-leg movements.  The non-Trump subsample, in which pressure
events are more isolated and more informative at the margin, is
precise enough to deliver significance at 5\% on both measures
despite the DXY's noisier construction.}

The Bund control is itself quantitatively large and highly
significant in the EUR/USD regression---a daily rise of 1~percentage
point in the 2Y Bund yield is associated with a 0.53\% appreciation
of the euro against the dollar, exactly the sign predicted by
UIP---validating the identification strategy: the dovish coefficient
is estimated on variation in EUR/USD that remains after euro-area
short-rate news is absorbed. The non-Trump amplification mirrors the pattern documented for the FFc coefficient in Section~\ref{subsec:a_trump_story}: dovish pressure
in the non-Trump sample is less clustered, more informative at the
margin, and therefore more fully priced in both the rate and FX
margins.

Appendix Figure \ref{fig:lp_fx_appendix} reports the cumulative exchange-rate responses estimated using the local projection framework. In the full sample, the cumulative response of the DXY is not statistically different from zero at any horizon, despite negative point estimates. By contrast, the EUR/USD exhibits a positive and statistically significant response over the first few trading days following the shock, although statistical significance fades at longer horizons. The evidence is considerably stronger when the Trump presidencies are excluded. In that subsample, dovish political pressure is associated with a statistically significant cumulative depreciation of the U.S. dollar, reflected in both a decline in the DXY and an appreciation of the EUR/USD. These dynamic responses are consistent with the baseline OLS evidence and indicate that the exchange-rate effects are primarily driven by the non-Trump sample.

\section{Pressure and the Actual Policy Rate}
\label{sec:fed_response}
The main results establish that dovish political     pressure shifts market-implied rate expectations. A distinct question is whether those shifts are eventually validated by actual Fed    policy. Following
  \citet{Bianchi2023}, we construct the ex-post      
  pricing error of federal funds futures:
  \begin{equation}
      \mathrm{FE}_{jt} \;=\; r_{jt} -
  \mathrm{FEDFUNDS}_{d(j,t)},
  \end{equation}
  where $r_{jt}$ is the FFc$j$ implied rate on day   
  $t$ and
  $\mathrm{FEDFUNDS}_{d(j,t)}$ is the monthly   
  average effective FFR
  for the delivery month, the contract's settlement  
  value. We then
  estimate, for each maturity $j$,
  \begin{equation}
      \Delta|\mathrm{FE}|_{jt} \;=\; \gamma\,        
  n_{\mathrm{dovish},t} +
      \boldsymbol{\beta}'\mathbf{X}_t,
      \label{eq:fed_response}
  \end{equation}
  where $\Delta|\mathrm{FE}|_{jt} =
  |\mathrm{FE}_{jt}| -
  |\mathrm{FE}_{j,t-1}|$ and $\mathbf{X}_t$ contains 
   the same controls of Section~\ref{sec:strategy}. $\gamma < 0$
   implies that futures moved closer to the eventual policy outturn on days with more dovish pressure --- consistent with the Fed  eventually moving in the direction of political pressure. Standard errors are clustered
   by date.

We estimate two variants that answer distinct
  economic questions. The
  \emph{unconditional} specification uses              $n_{\mathrm{dovish},t}$ on all
  days and asks: when politicians press for easier policy, does the Fed eventually cut by more than macro conditions alone   would predict? The \emph{signed-conditioned} specification asks, instead, a narrower question: among
  days when political pressure is accompanied by a   
  market reaction ---
  that is, when investors respond by pricing in a    
  higher probability of a
  rate cut ($\Delta r_{jt} < 0$) --- does the Fed    
  then follow through
  with actual policy easing? Formally, we replace    
  $n_{\mathrm{dovish},t}$
  with $n_{\mathrm{dovish},t}^{\mathrm{eff}} \equiv  
  n_{\mathrm{dovish},t}
  \cdot \mathbf{1}[\Delta r_{jt}<0]$. The two        
  estimands thus differ in
  their implicit model of the transmission mechanism.
   A negative $\hat{\gamma}$ in the unconditional specification 
  is consistent with a
  direct channel from political pressure to Fed      
  behaviour. A negative
  $\hat{\gamma}$ in the signed-conditioned
  specification additionally
  requires that the Fed respond to the market signal 
  generated by political pressure, rather than to the political  statement itself ---
  a pattern related to the market-dependent monetary 
  policy in the spirit of \citet{cieslak2021}, whereby the Fed
  incorporates financial market
  information into its policy deliberations.

\begin{figure}[H]
    \centering
    \caption{Pressure and Changes in Forecasting Error}
    \label{fig:forecasting_error}
    \includegraphics[width=0.85\linewidth]{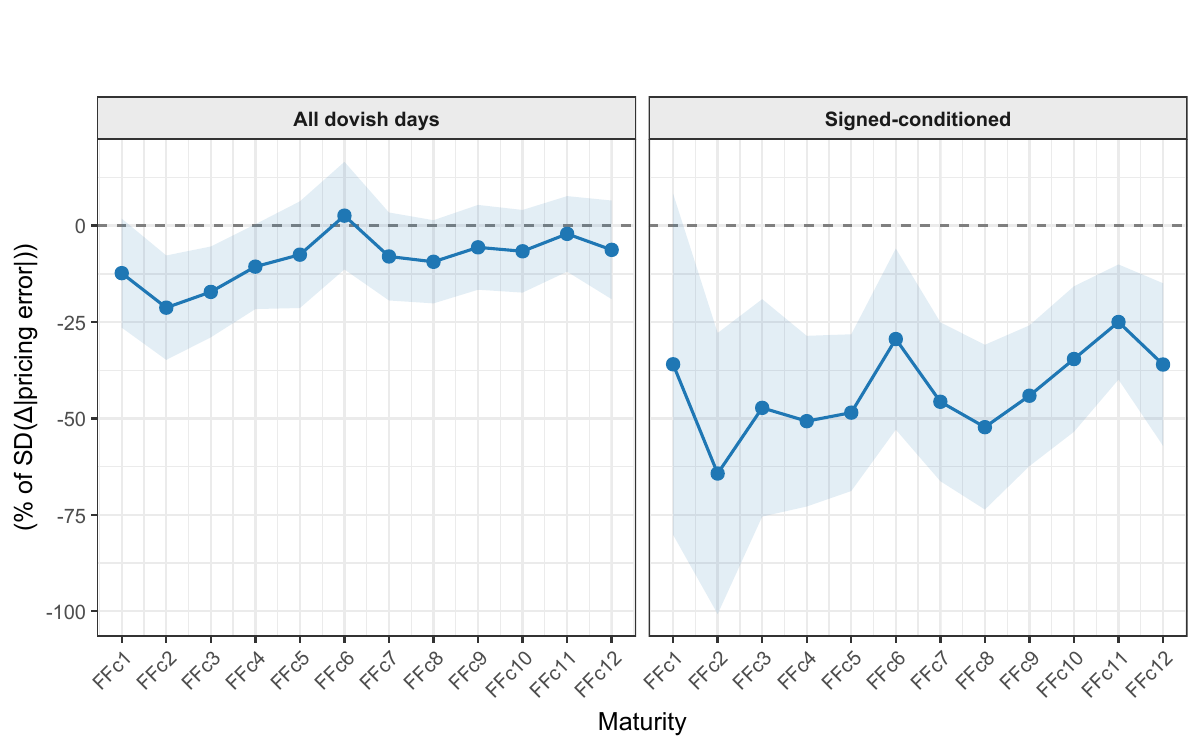}
    \footnotesize
    \begin{flushleft}
    \textbf{Note:} The figure plots $\hat{\gamma}$
   from equation~\eqref{eq:fed_response} across       
  maturities FFc1--FFc12, normalised by the within-maturity standard deviation of $\Delta|\mathrm{FE}|_{jt}$. The left panel uses
   $n_{\mathrm{dovish},t}$
      on all days (unconditional); the right panel   
  uses
      $n_{\mathrm{dovish},t}^{\mathrm{eff}} \equiv   
  n_{\mathrm{dovish},t}
      \cdot \mathbf{1}[\Delta r_{jt}<0]$, which      
  retains the pressure count
      only on days when the futures rate also fell   
  (signed-conditioned).
      Controls follow the drivers specification in   
      Section~\ref{sec:strategy}. Standard errors are 
  clustered by date;
      shaded bands denote 95\% confidence intervals.
      \end{flushleft}
\end{figure}

Figure~\ref{fig:forecasting_error} reports the
  results.
  
  The
  unconditional estimates are negative and             significant across all
  maturities, amounting to 0.5--0.8 basis points per   additional dovish                                    article and flat across the maturity spectrum ---    suggesting that                                     political pressure shifts the entire expected rate 
  path, not just the
  near-term horizon. Normalised by the
  within-maturity standard deviation
  of $\Delta|\mathrm{FE}|_{jt}$, this corresponds to 
  15--25\% of a
  typical daily fluctuation in the pricing error. The
   signed-conditioned
  estimates are roughly three times larger, reaching 
  35--65\% of the
  daily standard deviation. The three-fold amplification     
  between the two
  specifications implies that the observable
  convergence between market
  expectations and eventual policy is concentrated on
   days when political
  pressure and market co-movement coincide,
  consistent with the Fed
  responding to the market signal generated by       
  pressure rather than to
  the political statement itself.

Two caveats apply. First, politicians tend to call
  for rate cuts                                        precisely when the economy is weak --- the same      conditions that                                      independently lead the Fed to ease. Even with our extensive set of controls, some of the estimated
  $\hat{\gamma}$ may
  reflect this common driver rather than a causal    
  effect of pressure on
  Fed behaviour; the results should be read as       
  suggestive evidence of
  accommodation rather than as a causal claim.       
  Second, the
  signed-conditioned estimand selects days on which  
  the market reacted
  to political pressure by pricing in a higher       
  probability of a cut.
  Since the main results show that
  $n_{\mathrm{dovish},t}$ itself moves
  futures prices, this conditioning tends to retain  
  the largest and most
  market-relevant pressure events, occurring in      
  environments already
  tilted toward easing. The signed-conditioned       
  estimate therefore
  captures the upper end of the distribution of      
  pressure episodes rather
  than the average effect.

\section{Robustness}
\label{sec:robustness}
% =============================================================================
% Section: Robustness
% =============================================================================
This section evaluates the LLM jury classification results against a human benchmark and subjects the baseline estimates of Figure \ref{fig:baseline_all_term} to two sets of robustness checks. The human benchmark allows us to assess the accuracy of the classifier and verify that the directional pressure labels identified by the LLM are broadly consistent with human judgment. We then examine the robustness of the baseline estimates. The motivation is that the daily frequency, while unusually high by macroeconomic standards, remains coarse relative to the speed at which financial markets incorporate information: a trading day may contain scheduled macroeconomic releases, FOMC communications, and other non-political news, any of which may co-move with political pressure and contaminate the baseline coefficient. First, we augment the control set to reduce the scope for omitted variables that simultaneously drive interest-rate expectations and political pressure. Second, we implement a placebo test that replaces the dovish-pressure count in the baseline regression with the daily count of articles that the classifier flagged as mentioning the Federal Reserve without exerting directional pressure. A null result would indicate that the baseline estimate reflects the directional pressure signal rather than mechanical features of the classification pipeline or Federal Reserve-related news coverage more broadly.

\subsection{Human Validation}
\label{subsec:validation}

We validate the LLM classification against three independent human raters, blind to the model's output, on a stratified sample of 250 events (100 dovish, 75 hawkish, 75 not-relevant), roughly one-third of the classified sample.\footnote{Random sampling would under-represent hawkish and not-relevant events given their lower base rate in the corpus, hence the stratification.} We validate only the final classification step rather than each stage of the pipeline: the direction label is the object that enters the estimating equation directly, so end-to-end validation captures any error introduced upstream without requiring it to be attributed to a specific stage. For complete details on the structure of the exercise, see section \ref{subsec:human_validation} in the Appendix.

We report the results of the validation in Table \ref{tab:human_llm_validation}. 

\begin{threeparttable}
  \centering
  \caption{Human-LLM Classification Validation}
  \label{tab:human_llm_validation}
  \small
  \begin{tabular}{llrr}
  \hline\hline
  \multicolumn{4}{l}{\textit{Panel A: Headline agreement}} \\
  \hline
  Comparison & $n$ & Statistic & Value \\
  \hline
   Human consensus vs.\ LLM majority & 240 & Accuracy (weighted) & 81\% \\
  Human consensus vs.\ LLM majority & 240 & Accuracy (unweighted) & 78.8\%\% \\
  Human jury (within) & 225 & Fleiss' $\kappa$ & 0.49 \\
  LLM jury (within) & 225 & Fleiss' $\kappa$ & 0.74 \\
  Human consensus vs.\ LLM majority & 240 & Cohen's $\kappa$ & 0.68 \\
  \hline
  \multicolumn{4}{l}{\textit{Panel B: Confusion matrix (human consensus vs.\ LLM majority)}} \\
  \hline
  Human / LLM & DOVISH & HAWKISH & NOT_RELEVANT \\
  \hline
  DOVISH & 77 & 1 & 2 \\
  HAWKISH & 0 & 47 & 3 \\
  NOT_RELEVANT & 20 & 25 & 65 \\
  \hline
  \multicolumn{4}{l}{\textit{Panel C: Individual-level pairwise agreement (no consensus conditioning)}} \\
  \hline
  Rater & Model & $n$ & \% agreement \\
  \hline
  rater_1 & qwen & 225 & 72.0\% \\
  rater_1 & llama & 225 & 68.0\% \\
  rater_1 & deepseek & 225 & 68.0\% \\
  rater_2 & qwen & 225 & 65.8\% \\
  rater_2 & llama & 225 & 62.7\% \\
  rater_2 & deepseek & 225 & 65.8\% \\
  rater_3 & qwen & 225 & 75.1\% \\
  rater_3 & llama & 225 & 70.2\% \\
  rater_3 & deepseek & 225 & 71.1\% \\
  \hline\hline
  \end{tabular}
  \begin{tablenotes}
\item \footnotesize \textbf{Note:} Panel A's within-jury rows and Panel C use the
$n=225$ articles with a complete rating from all 3 human raters and a non-missing vote from all
3 LLM models, so that both juries' Fleiss' $\kappa$ are computed over an identical set of articles
for a like-for-like comparison. Panel A's cross-jury row and Panel B instead use the larger $n=250$
sample of articles with complete human ratings and an assigned LLM majority label -- a label is
assigned once at least 2 of the 3 models (\texttt{qwen3-235b-a22b-2507}, \texttt{llama-4-maverick},
\texttt{deepseek-chat-v3-0324}) agree, even if the third model's vote is missing. Of these 250,
10 are further excluded because the three human raters split three ways with no majority, leaving
$n=240$. Panel A compares within-jury agreement (Fleiss' $\kappa$, human raters and LLM models
separately) against cross-jury agreement (Cohen's $\kappa$ and accuracy, human consensus vs.\ LLM
majority vote). Panel B reports the confusion matrix underlying the Panel A cross-jury comparison.
Panel C reports pairwise agreement between every individual human rater and every individual LLM
model, without conditioning on human consensus.
\end{tablenotes}
\end{threeparttable}

Comparing the human majority label against the LLM majority label and re-weighing by population shares, given the stratification, the two agree on 81\% of articles, up from 78.8\% skipping re-weighting, and with Cohen's $\kappa$=0.68 (Panel A).\footnote{Cohen's $\kappa$ measures inter-rater agreement between two raters while correcting for agreement expected by chance. Fleiss' $\kappa$ generalizes this approach to assess agreement among more than two raters. Values of $\kappa$ range from $-1$ to $1$, with higher values indicating stronger agreement beyond chance.}

We treat the human-majority label as the best available proxy for the underlying true classification, not as ground truth. This distinction matters because within-jury agreement is only moderate (Fleiss' $\kappa$ = 0.50): individual raters often disagree, so no single rater's judgment is authoritative on its own. Two features nonetheless support its use as the comparison standard. First, the three human raters worked independently, using a structured codebook (Appendix \ref{subsec:human_validation}) and seeing the same event text as the LLM pipeline, with no additional context. Aggregating three independent judgments plausibly reduces idiosyncratic error relative to any single rater, so the reliability of the majority label exceeds what the pairwise $\kappa$ alone suggests. Second, this independence is precisely what distinguishes the human majority from the LLM's own three-model majority vote: the three LLM models see an identical prompt and codebook and may still share correlated errors from overlapping pretraining data, so their higher raw within-jury agreement (Fleiss' $\kappa$ = 0.74) need not reflect higher reliability. 

Hitherto, the comparison mirrors the design of the LLM's own majority-vote classification, but conditioning on articles where at least two human raters agree selects for the easier cases, mechanically inflating accuracy. We address this directly by also computing pairwise agreement between every individual rater and every individual model, without conditioning on consensus (Table \ref{tab:human_llm_validation},  Panel C); agreement ranges from 62.5\% to 75\%. On the same percentage-agreement scale, this is comparable to pairwise agreement among the human raters themselves (67.0\%--68.3\%; Fleiss' $\kappa$=0.50) and somewhat below pairwise agreement among the LLM models (81.7\%--83.5\%; Fleiss' $\kappa$=0.74). The headline number therefore survives this selection concern.

Moreover, the confusion matrix underlying this comparison (Panel B) shows that disagreement is concentrated at the relevance margin rather than the direction margin: of the 51 articles off the matrix diagonal, only one involves a directional confusion between DOVISH and HAWKISH — a human-labeled DOVISH article the LLM scores HAWKISH — and no HAWKISH article is ever mislabeled DOVISH. The remaining 45 errors instead consist of NOT\_RELEVANT articles that the LLM assigns a directional label (20 DOVISH and 25 HAWKISH), against only 5 cases of the reverse (a directional article the LLM scores NOT\_RELEVANT). In a nutshell, the LLM jury leans toward over-classifying relevance (Type I errors), not toward inverting the sign of pressure or toward missing genuinely relevant events (Type II errors). This is reassuring for our main results. First, directional confusion between DOVISH and HAWKISH is essentially absent, so misclassification cannot be driving the sign of our estimates. Second, the remaining errors are almost all false alarms: since this adds noise to the event count rather than flipping its sign, it should, if anything, bias our estimates toward zero — working against finding an effect.\footnote{Individual confusion matrices in Figure \ref{fig:individual_rater_model_matrix} in Appendix A.}\footnote{A related concern is that a Reuters article may report an event that occurred well before the article's publication date, which would misassign its timing. This is unlikely given that newswires cover breaking developments, but we nonetheless ask human reviewers to flag such cases: across the validation sample, raters flagged only between 2 and 5 articles (roughly 1\%).}

\subsection{Alternative Specifications and Samples}
\label{subsec:alt_specs}

We assess the robustness of the results to alternative specifications. The baseline model (equation \ref{eq:baseline}) has as variable of interest the count of dovish pressure events per day and includes a beginning-of-month dummy (or quarterly-roll dummy for treasury futures), an FOMC announcement-day dummy, and a set of controls capturing the macro-financial environment, following \citet{bauer_swanson_2023}.

We change the definition of our dependent variable to a dummy equal to 1 conditional on the presence of at least one daily dovish pressure event in the day, and then augment the control set along three dimensions, each addressing a
distinct channel through which political pressure and interest rate
expectations may be jointly determined.  The first is the macroeconomic
state.  Political pressure intensifies in recessions, in credit-market
stress, and in financial crises  We therefore add, in turn, an NBER recession indicator, substitute the inflation gap with a forward-looking indicator, introduce a daily corporate credit
spread, and re-estimate the baseline excluding crises windows. Additionally, we replace the inflation gap and unemployment gap with year × month fixed effects, which absorb all time variation at the monthly frequency — including slow-moving macroeconomic conditions, seasonal patterns, and any other confounders that are constant within a calendar month. One advantage of the fixed effects approach is that it avoids taking a stand on the precise construction of the gap measures, which are inherently model-dependent: both the inflation gap and the unemployment gap rely on reference values — the 2\% PCE target and the CBO estimate of the natural rate of unemployment — that are chosen by us. 

The
second is the flow of macroeconomic news.  Scheduled data releases move
rates for reasons unrelated to political pressure, and if the two are
correlated in time, the baseline coefficient would partly absorb the
release response.  We address this directly by conditioning on the most important release-day indicators- CPI, advance GDP, PPI, and retail sales- and, in
a separate specification, on the daily US Citi Economic Surprise
Index (CESI)---a continuous measure of how macro data has been landing
relative to consensus.\footnote{The CESI specification, which controls most directly for real-time data
surprises, is available only from January 2005, shortening the estimation sample by roughly
twenty years relative to the full 1987--2025 baseline.  We addressed this by
replacing the Citigroup Economic Surprise Index with direct dummies for the five major scheduled US
macroeconomic releases.}  The third is systematic day-of-week variation
in financial returns, which we absorb with day-of-week fixed effects.

\begin{table}[htbp]
  \centering
  \caption{Coefficient Stability of Dovish Pressure Across Alternative Specifications}
  \label{tab:coef_robustness_across_specs}
  \small
  \begin{tabular}{lrrrr}
  \hline\hline
  Specification & FFc1 & FFc2 & FFc3 & FFc4 \\
  \hline
  \multicolumn{5}{l}{\textit{Baseline}} \\
  Baseline & -0.91*** (0.33) & -0.62** (0.27) & -0.57** (0.27) & -0.50** (0.25) \\
  \\[-6pt]
  \multicolumn{5}{l}{\textit{Functional form}} \\
  Dummy pressure ($1[n_{\text{dovish}}>0]$) & -0.98*** (0.31) & -0.72*** (0.27) & -0.66** (0.28) & -0.57** (0.27) \\
  \\[-6pt]
  \multicolumn{5}{l}{\textit{Macro conditions}} \\
  NBER recession & -0.90*** (0.33) & -0.61** (0.27) & -0.56** (0.27) & -0.49* (0.25) \\
  Expected inflation cycle & -0.93*** (0.33) & -0.64** (0.27) & -0.59** (0.27) & -0.53** (0.25) \\
  Credit spread & -0.92*** (0.33) & -0.63** (0.27) & -0.57** (0.28) & -0.50* (0.26) \\
  Excl.\ 2009 \& 2020 crises & -0.82** (0.33) & -0.55** (0.25) & -0.52** (0.27) & -0.48* (0.27) \\
  Year $\times$ month FE & -0.85** (0.36) & -0.52* (0.28) & -0.47* (0.28) & -0.42 (0.31) \\
  \\[-6pt]
  \multicolumn{5}{l}{\textit{Macro news}} \\
  Macro releases & -0.90*** (0.33) & -0.62** (0.27) & -0.57** (0.27) & -0.50** (0.25) \\
  US Citigroup Index & -0.88** (0.43) & -0.52 (0.36) & -0.42 (0.36) & -0.39 (0.33) \\
  \\[-6pt]
  \multicolumn{5}{l}{\textit{Calendar effect}} \\
  Day-of-week FE & -0.91*** (0.33) & -0.62** (0.27) & -0.56** (0.27) & -0.49** (0.25) \\
  \hline\hline
  \end{tabular}
  \\[4pt]
  \parbox{0.95\linewidth}{\footnotesize \textbf{Note:} The baseline specification is equation \ref{eq:baseline}. Robustness variants are grouped by the potential concern they address.\textit{Functional form}: (i) substitutes the daily count of dovish pressure events with a dummy equal to 1 if at least one event \textit{Macro conditions}: (ii) adds the NBER recession indicator; (iii) replaces the inflation gap with the one-year Michigan Survey expected inflation gap; (iv) adds the daily change in the BAA$-$AAA corporate bond spread; (v) excludes calendar years 2009 (Global Financial Crisis) and 2020 (Covid-19); (vi) replaces the inflation and unemployment gaps with year $\times$ month fixed effects. \textit{Macro news}: (vii) adds announcement dummies for CPI, advance GDP, PPI, and retail sales releases from ALFRED; (viii) replaces the NFP releases dummy with the first difference of the US Citigroup Economic Surprise Index. \textit{Calendar}: (ix) adds day-of-week fixed effects. HAC (Newey-West) standard errors. *, **, *** = 10\%, 5\%, 1\% statistical significance.}
\end{table}

The baseline finding is robust to the alternative definition and choice of the different control sets. Table ~\ref{tab:coef_robustness_across_specs} reports the
impact of a political pressure event across the baseline specification  and the nine aforementioned ones for futures up to three-months ahead.
The point estimates are remarkably stable and significant at the 5-10\% level for almost every maturity-control pair. The only loss of significance occurs for the FFc2-FFc4 maturities when we add the CESI. Nevertheless, this should be taken with caution as we are removing half of the original sample, the coefficients do not change drastically, and the coefficient for the current-month future remains significant.\footnote{Additionally, for the specification with yearXmonth effects, the dovish estimate on FFc3 and FFc4 become borderline significant and insignificant, respectively. Moreover, the coefficients associated with the simple dummy are, if anything, higher than the baseline, in line with our diminishing returns to pressure story.}

The tightness of this range, across specifications that differ
materially in the confounds they absorb, is the central robustness finding: if the baseline coefficient reflected an omitted macroeconomic factor, it should load into at least one of these controls and move accordingly, however it does not.

\subsection{Placebo Test}
\label{subsec:placebo}

A further identification concern is that our variable of interest, $n_{\text{dovish},t}$ is a
count of news article/events. This means the coefficient could in principle load
on Fed-related news attention generally rather than on directional
pressure specifically---for instance, if markets respond to heightened
political scrutiny of the Federal Reserve regardless of the
scrutiny's content.  To address this concern, we construct a
within-sample placebo using the pool of articles that passed the
politician-and-Fed filter---a US elected official talking about
the Federal Reserve---but that the LLM classified as not containing
directional political pressure.  This pool shares the upstream
selection process with the treatment series (the same politicians,
the same Fed mentions, the same newswire, the same daily arrival
rate) and differs from it only in the classifier's judgment that the
article carries no pressure signal.  A considerably smaller coefficient on this
placebo compared to baseline would indicate that the baseline estimate reflects the
directional content captured by the classifier rather than mechanical
features of Fed-related news flow.

\begin{figure}[h!]
    \centering
    \caption{Placebo Test: Is the Content of Political Pressure Articles Informative?}
    \label{fig:placebo}
    \begin{subfigure}[b]{0.85\textwidth}
        \centering
        \includegraphics[width=\textwidth]{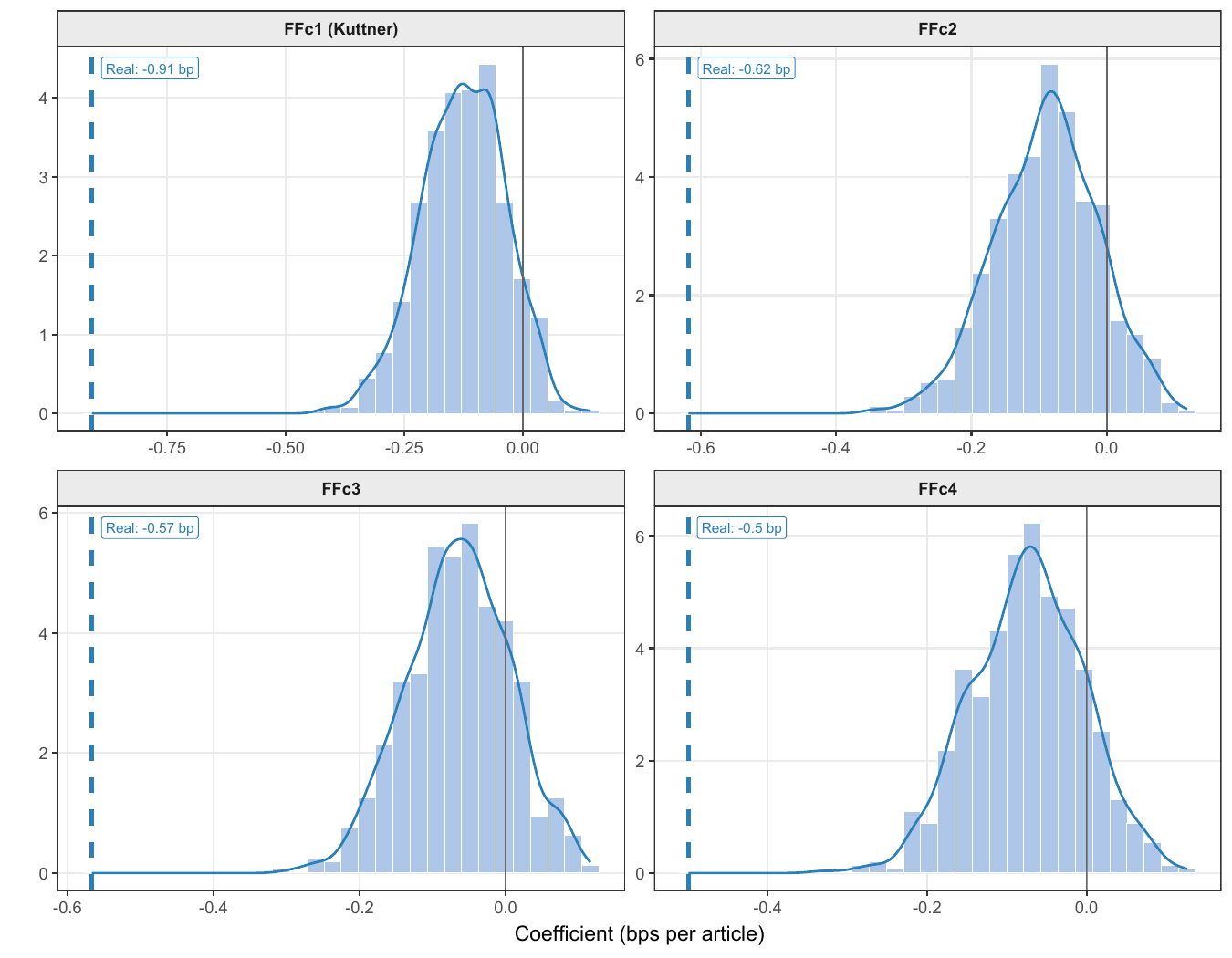}
    \end{subfigure}
    \vspace{0.2em}
     \footnotesize{\begin{flushleft}
    \textbf{Note}: Each panel shows the
  distribution of the coefficient on a placebo pressure variable across 678 simulation
      runs. In each run, a number of
      articles equal to the actual count of dovish pressure events is drawn at random from the pool of articles classified as 
      \textsc{not\_relevant} by the LLM majority     
  vote; these are
      aggregated to daily counts and entered in place
   of
      $n_{\mathrm{dovish},t}$ in equation \ref{eq:baseline}.
      The dashed vertical line marks the real        
  coefficient from the
      same specification.
    \end{flushleft}}
\end{figure}

Formally, we draw 678 placebo samples from the pool of
non-pressure articles. Each draw samples, with replacement, a number
of articles equal to 508, the total number of dovish pressure events
entering the baseline regression sample (i.e.,
$\sum_t n_{\text{dovish},t}$); the placebo pool is restricted to the
same trading-day calendar underlying the regression sample, so that
every sampled article can be aggregated onto a valid trading day.
The sampled articles are aggregated to daily counts, and the baseline
specification is re-estimated for FFc1 through FFc4 using this
placebo count in place of $n_{\text{dovish},t}$. The procedure
generates a distribution of 678 placebo coefficients under the null
that the baseline estimate reflects only the volume of Fed-related
news flow rather than its directional content. Figure~\ref{fig:placebo}
plots this distribution for each outcome alongside the baseline
coefficient, shown as a vertical line.

The placebo distribution is tightly centred to the left of zero, but only marginally so, approximately 0.1 for all four maturities, consistent
with the null hypothesis that randomly selected Fed-related political
articles have no systematic effect on rate expectations.  The
baseline coefficient lies in the extreme left tail of the placebo
distributions. Hence, we conclude that The
baseline effect is not a mechanical artefact of article
volume or of Fed-related news coverage more broadly; it is specific
to articles the classifier identifies as carrying dovish pressure
content, which in turn further validates the directional content of the
classification pipeline.

\pagebreak

\section{Conclusion}
\label{sec:conclusion}

% =============================================================================
% Section: Conclusion
% =============================================================================
This paper examines whether politically motivated pressure on the Federal Reserve by elected officials shifts near-term interest rate expectations as priced in financial markets. Using approximately forty years of Reuters newswire data from 1988 to 2025 and a majority-vote LLM classification pipeline to create a new political pressure measure, we document a robust negative association between dovish pressure and futures rates: the response of current-month futures to an additional dovish event is roughly 1bps coefficients, equivalent to a 4\% additional probability of a rate cut in the next FOMC. The effect is statistically significant up to the 3-months
ahead tenor, but decreases monotonically and disappears at the medium and long-term horizon. Hawkish pressure yields positive but uniformly insignificant coefficients across the same
specifications.

A natural concern is that the estimated association reflects the
unusually salient and sustained pressure campaigns of the two Trump
administrations rather than a structural feature of the
pressure--markets relationship.  The evidence does not support this
reading.  Re-estimating the baseline specification on the subsample
that excludes the windows of 20 January 2017 to 20 January 2021 and
20 January 2025 onward yields an effect on the current-month futures of 1.58 bps, greater than the full-sample specification. The monotonic relationship also remains intact. We interpret this pattern as evidence that
the mechanism is a general feature of the post-Volcker US monetary
framework rather than an artefact of an idiosyncratic episode.

The joint pattern across instruments---short-end rally, muted long-end
response, stable inflation compensation and market volatility, and dollar depreciation---is
consistent with markets pricing anticipated monetary accommodation rather
than a credibility tax in the sense of \citet{rogoff_1985} and
\citet{alesina_summers_1993}, which would require the nominal anchor to
move. Market participants appear to price the presidential
principal--agent relationship specifically---consistent with the
President's unique hold on the appointment and reappointment channel
over the Chair and Board of Governors---rather than a diffuse
political-pressure effect operating through rhetoric at large.  The
absence of a congressional response is consistent with
\citet{binder_2018}, who documents the limited traction of legislative
oversight on the conduct of monetary policy.

Three caveats qualify the interpretation.  First, the reduced-form
coefficient cannot separate completely a compliance channel, through which markets
update about the Fed's reaction function, from an information channel,
through which pressure co-occurs with a deteriorating macroeconomic outlook that would be priced independently.  We address this to the extent that daily identification through controls permits. Moreover, we show that the response to dovish pressure more than doubles when the President's party ``holds a majority'' at the Federal Reserve Board, a pattern the information channel cannot generate, since a weakening outlook calls for the same easing irrespective of Board composition. We read this as evidence that a compliance channel is operative, while acknowledging that a residual information component cannot be fully excluded within a reduced-form design. Second, the LLM classification is imperfect; the resulting
measurement error biases estimates toward zero, so the reported coefficients should be read as a lower bound on the true market response. Third, the regressor counts Reuters articles classified as exerting pressure, and therefore reflects both the underlying intensity of
political pressure and the newswire's editorial decision to cover it.
This is a feature shared with essentially all newspaper-based index
constructions in the macro-finance literature---from the
\citet{baker_bloom_davis_2016} policy uncertainty index onward---and
one the literature has generally treated as a second-order concern
relative to the informational content that media coverage conveys.

Three directions follow naturally.  First, the methodology is
portable: the same LLM classification pipeline, applied to newswire
corpora for the ECB, Bank of England, and Bank of Japan, would test
whether anticipated accommodation is a general feature of
advanced-economy central banking or specific to the US institutional
framework.  Second, a natural next step is to ask whether the realised
policy path tracks the expectational shift documented here.  We
document that federal funds futures track the settlement rate more
tightly around pressure events than on average, on the order of 25
percent of one standard deviation, consistent with markets correctly
anticipating the subsequent path.  A sharper and more complete answer
is nonetheless warranted.  Third, and conditional on compliance, the
estimates invite a welfare accounting.  Does sustained pressure move
realised inflation above target by a quantitatively meaningful
margin, and at what cost in terms of output variability?  That is ultimately the question
that decides whether political pressure on the Federal Reserve is a
first-order concern for monetary policy or a second-order feature of
the communication environment.

\pagebreak

\bibliographystyle{aer}
\bibliography{references}

\pagebreak

\appendix
\section*{Appendix}

\section{Political Pressure Measurement}

\subsection{Additional Figures}

\begin{figure}[H]
    \centering
    \caption{Reuters Corpus: Descriptive Statistics}
    \label{fig:corpus}
    \begin{subfigure}[b]{0.48\textwidth}
        \centering
        \includegraphics[width=\textwidth]{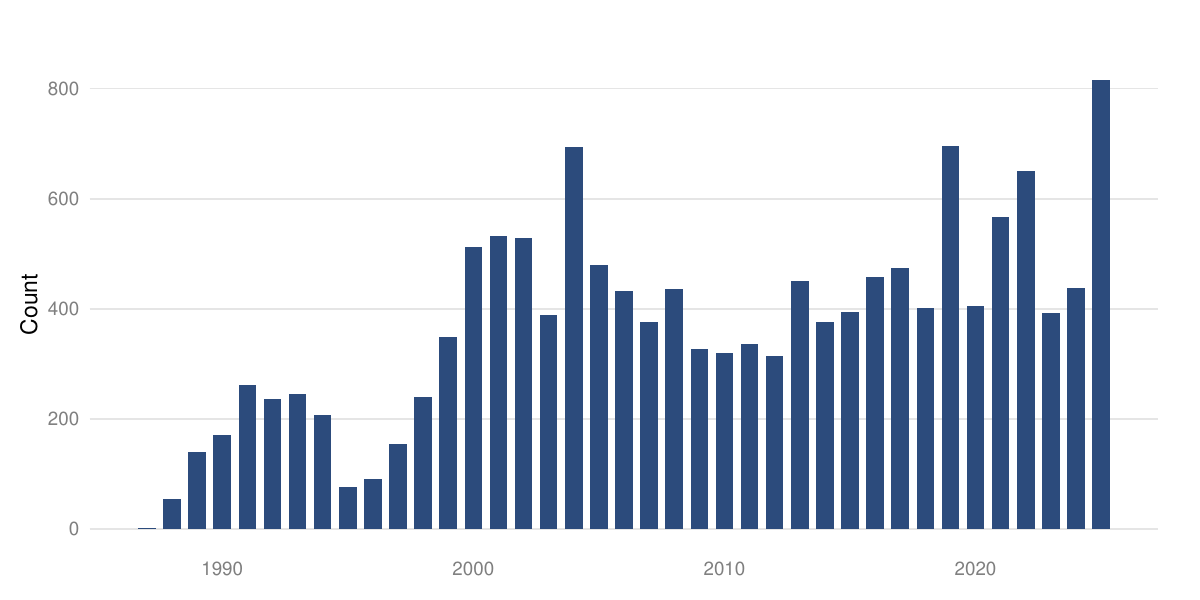}
        \caption{Articles by Year}
        \label{fig:corpus_a}
    \end{subfigure}
    \hfill
    \begin{subfigure}[b]{0.48\textwidth}
        \centering
        \includegraphics[width=\textwidth]{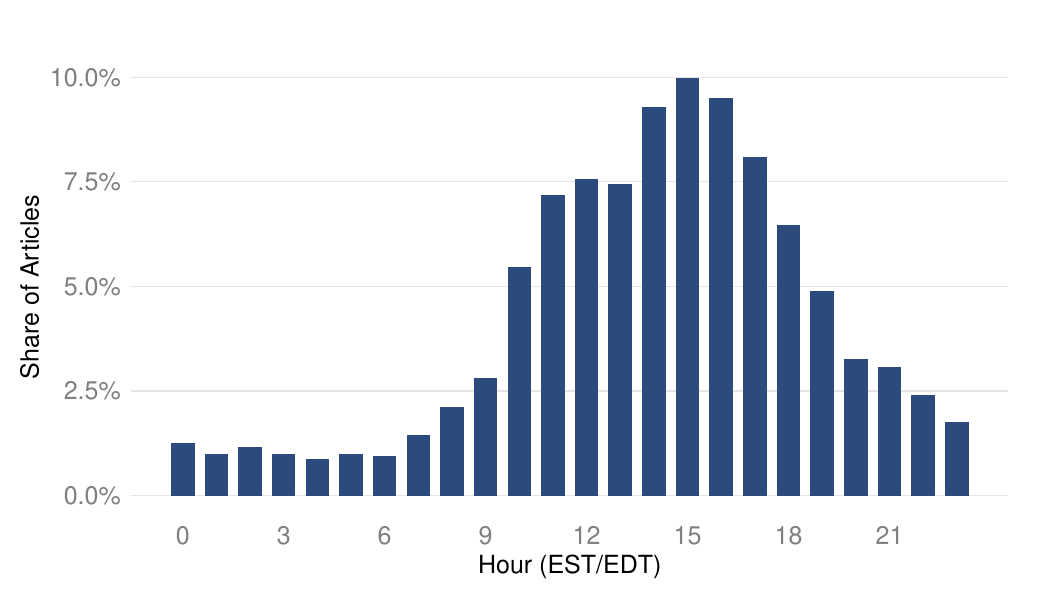}
        \caption{Publication Hour}
        \label{fig:corpus_b}
    \end{subfigure}
    
    \vspace{1em}
    
    \begin{subfigure}[b]{0.48\textwidth}
        \centering
        \includegraphics[width=\textwidth]{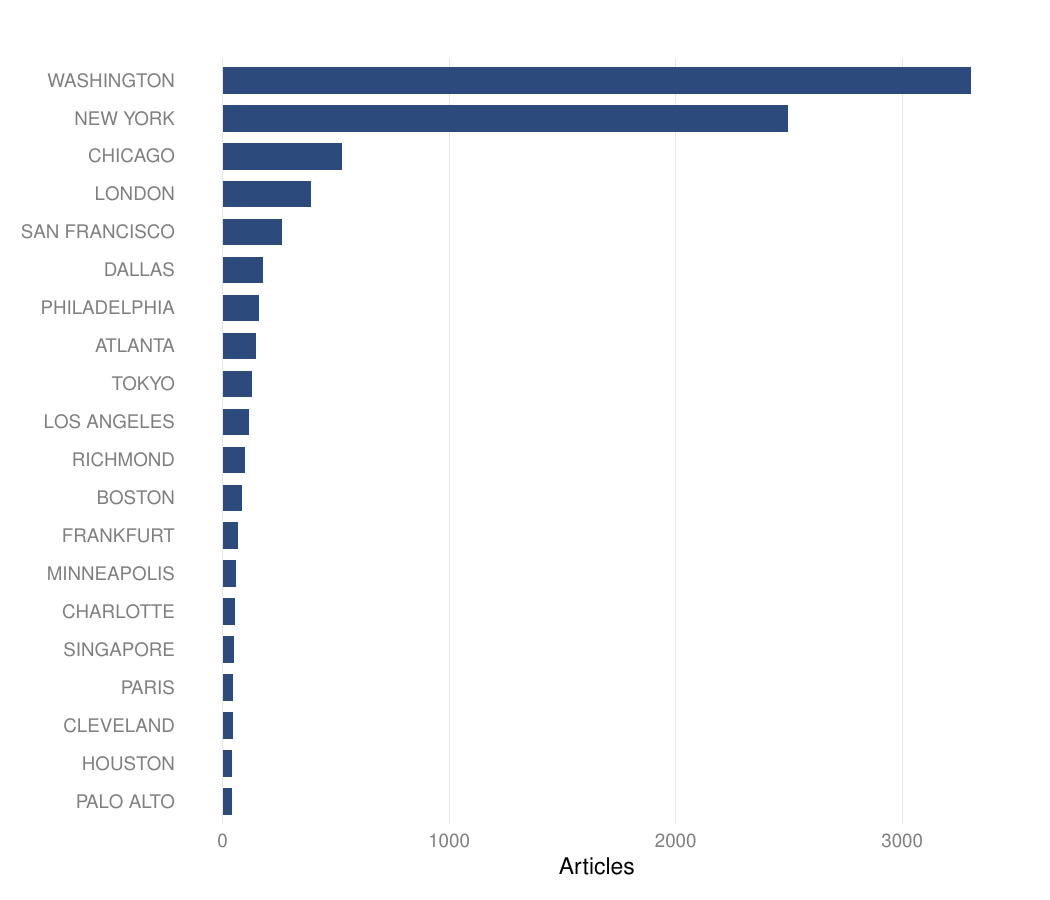}
        \caption{Top 20 Locations}
        \label{fig:corpus_c}
    \end{subfigure}
    \hfill
    \begin{subfigure}[b]{0.48\textwidth}
        \centering
        \includegraphics[width=\textwidth]{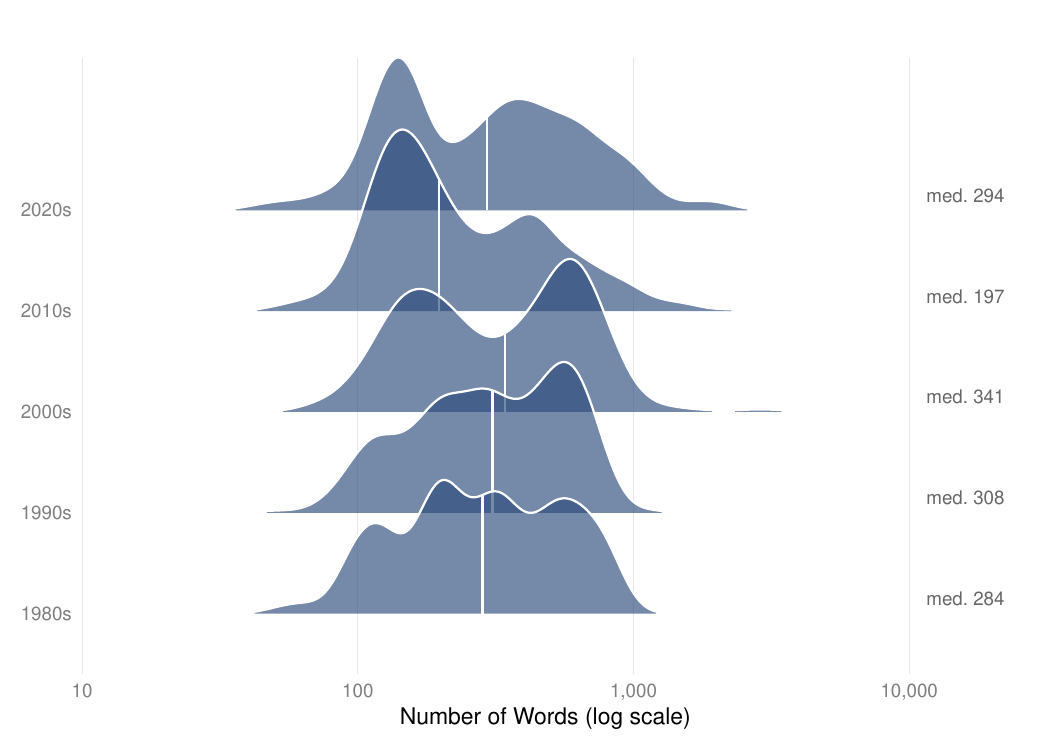}
        \caption{Article Length by Decade}
        \label{fig:corpus_d}
    \end{subfigure}
    \vspace{0.2em}
    \footnotesize{\begin{flushleft}
    \textbf{Note}: The figure reports descriptive statistics for the raw Reuters corpus retrieved from Factiva over the period 1988--2025. Panel (a) shows the number of articles per year. Panel (b) shows the distribution of publication hours reported in the header. Panel (c) reports the top 20 locations reported in the header. Panel (d) shows the evolution of article length by decade with the median number words noted on the right.
    \end{flushleft}}
\end{figure}

\begin{figure}[H]
    \centering
\caption{Inter-model agreement in the LLM classification ensemble}
    \label{fig:intermodel_agreement}
    \includegraphics[width=\linewidth]{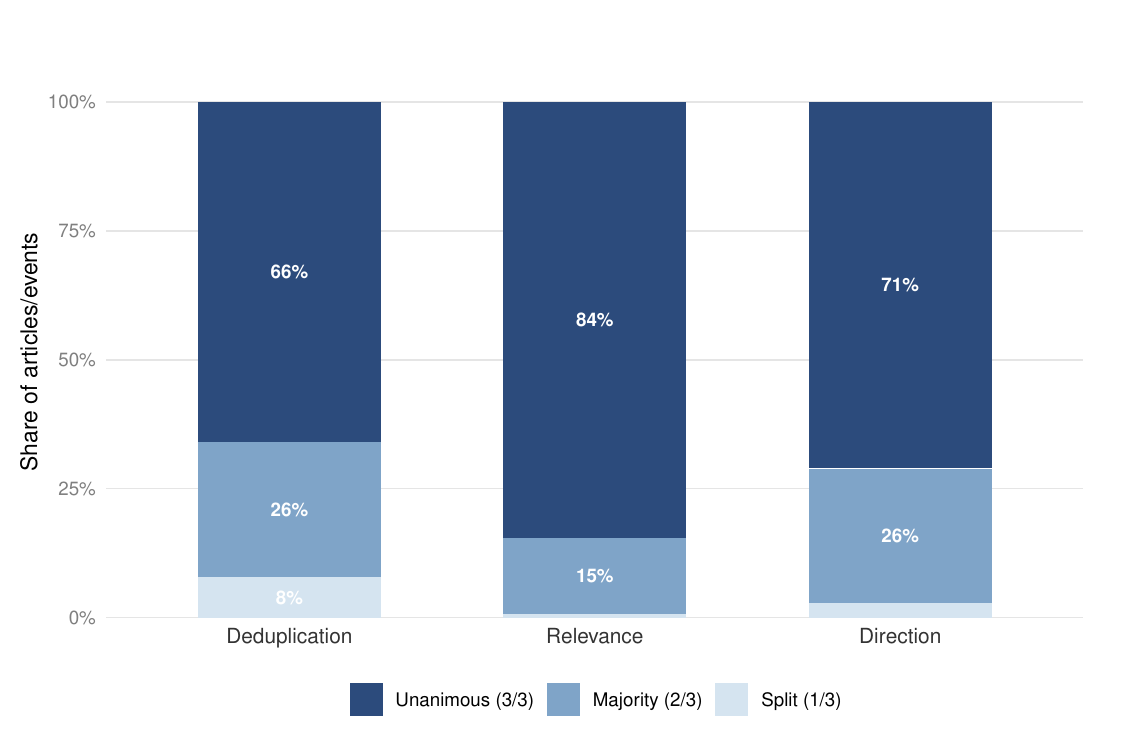}
    \footnotesize{\begin{flushleft}
    \textbf{Note:} The figure reports the degree of agreement across the three open-weight LLMs used in the classification ensemble---Qwen (\texttt{qwen3-235b-a22b-2507}), Meta Llama (\texttt{llama-4-maverick}), and DeepSeek (\texttt{deepseek-chat-v3-0324})---shown separately for the deduplication stage (left, whether a group of articles refer to the same underlying event), relevance judgment (centre, whether an article contains a US politician talking about the FED), and the direction judgment (right, whether the event is classified as dovish, hawkish, or not relevant).  \emph{Unanimous} denotes articles on which all three models agree; \emph{Majority} denotes articles on which exactly two of three agree; \emph{Split} denotes articles on which no majority is available, either because one of the three models returned a malformed response or because all three models assign a different label.
\end{flushleft}}
\end{figure}

\begin{figure}[H]
    \centering
    \caption{Characterizing pressure: from who to whom using what}
    \label{fig:sankey}
    \includegraphics[width=1\linewidth]{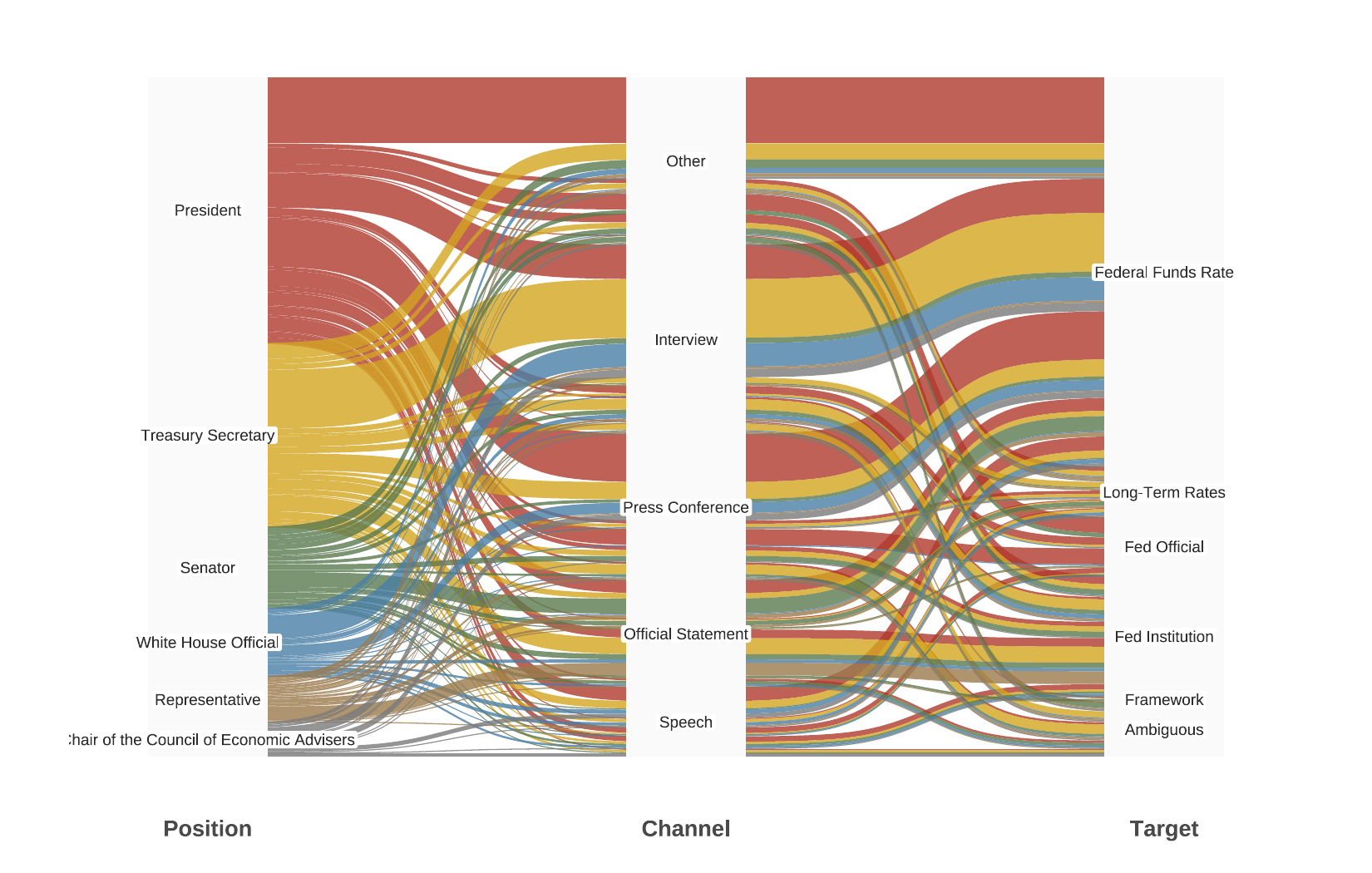}
    \footnotesize{\begin{flushleft}
    \textbf{Note:} The diagram traces each politics pressure event along three dimensions: the political actor's
  institutional role
      (left), the communication channel through which
   the statement was
      made (centre), and the target of the statement 
  (right). Flows are
      coloured by actor role. The six most frequent  
  actor roles are shown
      individually; remaining roles are excluded.
    \end{flushleft}}
\end{figure}

\begin{figure}[H]
    \centering
    \caption{Direction of Political Pressure by Target}
    \label{fig:direction_target}
    \includegraphics[width=0.72\linewidth]{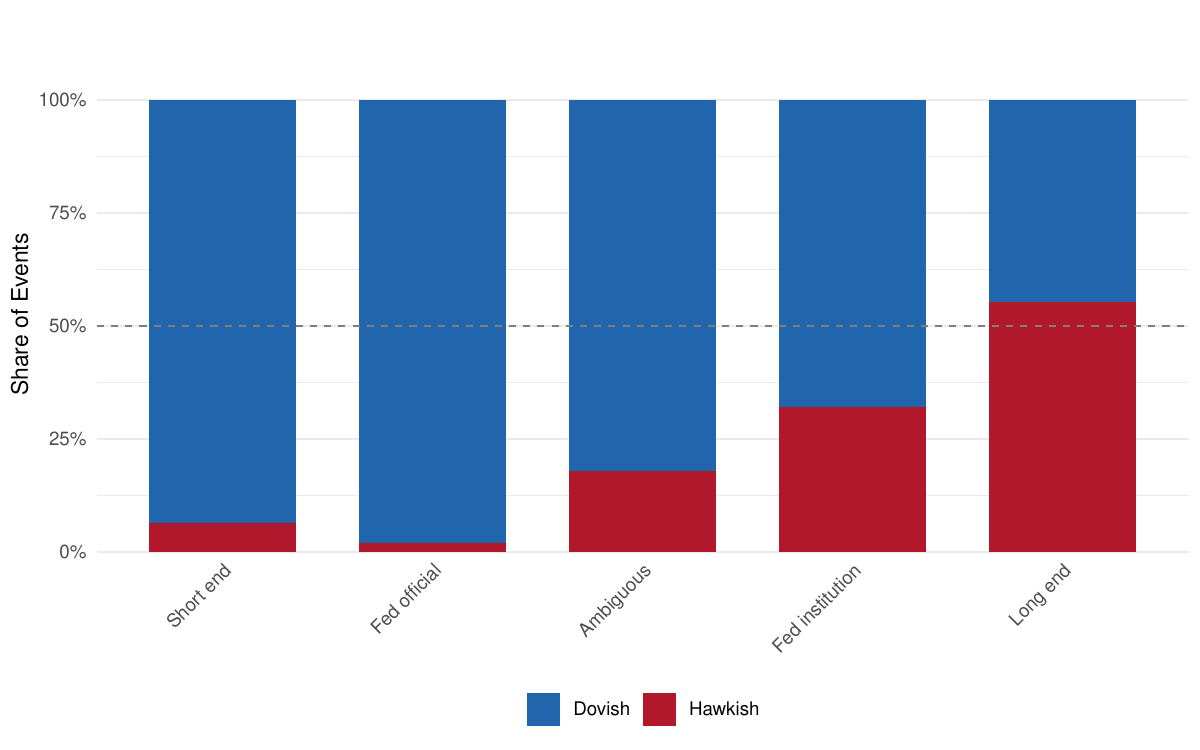}
\footnotesize{\begin{flushleft}
\textbf{Note:} The figure reports the share of dovish and hawkish events within each target category.  \emph{Short end} denotes pressure aimed at the near-term policy rate; \emph{Fed official} denotes pressure directed at a specific person on FOMC Board; \emph{Fed institution} denotes pressure aimed at the
Federal Reserve as an institution, including its mandate, or structure; \emph{Long end} denotes pressure aimed at
the longer-run policy stance or long-term interest rates;
\emph{Ambiguous} groups articles for which the target could not be
determined with confidence.  
\end{flushleft}}
\end{figure}

\pagebreak

\subsection{Human Validation of LLM Classification}
\label{subsec:human_validation}

To assess the reliability of the LLM-based classification, we validate it against independent human labels on a stratified subsample of 250 articles (100 dovish, 75 hawkish, 75 not-relevant), roughly one-third of the full classified sample. Random sampling would under-represent hawkish and not-relevant events given their lower base rate in the corpus, so we stratify by the LLM's own classification to ensure sufficient representation of each category.

\paragraph{Raters.} Three research assistants, each holding a bachelor's degree in economics and currently enrolled in an economics master's program, independently labeled every article in the validation sample. Raters were limited to two hours of annotation work per day to guard against attention decay over the course of the exercise; the full sample was completed over approximately one week.

\paragraph{Task design.} Raters received the identical set of 250 articles, each shown in full text, in a different randomized order per rater to avoid order effects. Raters were given the same classification prompt and category definitions (DOVISH / HAWKISH / NOT\_RELEVANT) used in the LLM pipeline, ensuring the human and machine tasks were as directly comparable as possible. Raters worked independently and were blind both to the LLM's classification and to each other's labels; no discussion or consensus step took place. To discourage reliance on AI tools in producing labels, raters were explicitly told that we valued their own judgment even at the cost of speed, and that there was no time pressure to complete the task quickly.

\paragraph{Aggregation.} For each article, we take the majority label across the three raters as the human consensus label. Articles without a majority (i.e., three-way splits) are dropped from the human-consensus-vs-LLM comparison (Panel A and Panel B of Table \ref{tab:human_llm_validation}) rather than adjudicated, since we have no principled basis for breaking such ties; this affects 10 of the 224 articles with complete ratings. These articles are retained, however, in the individual-level pairwise comparisons of Panel C, which do not require a consensus label.

\paragraph{LLM side.} For comparability, each of the three LLMs (Qwen-2-235B, LLaMA-4 Maverick, DeepSeek-V3) is queried once per article at temperature 0, so that each model's classification is a single deterministic vote rather than a majority over repeated calls; the LLM majority label used in Panel A is the majority across these three single-model votes, mirroring the construction of the human consensus label.

\begin{figure}[h!]
    \centering
    \caption{Individual rater x model confusion matrices}    
    \label{fig:individual_rater_model_matrix}
    \begin{subfigure}[b]{\textwidth}
        \centering
        \includegraphics[width=\textwidth]{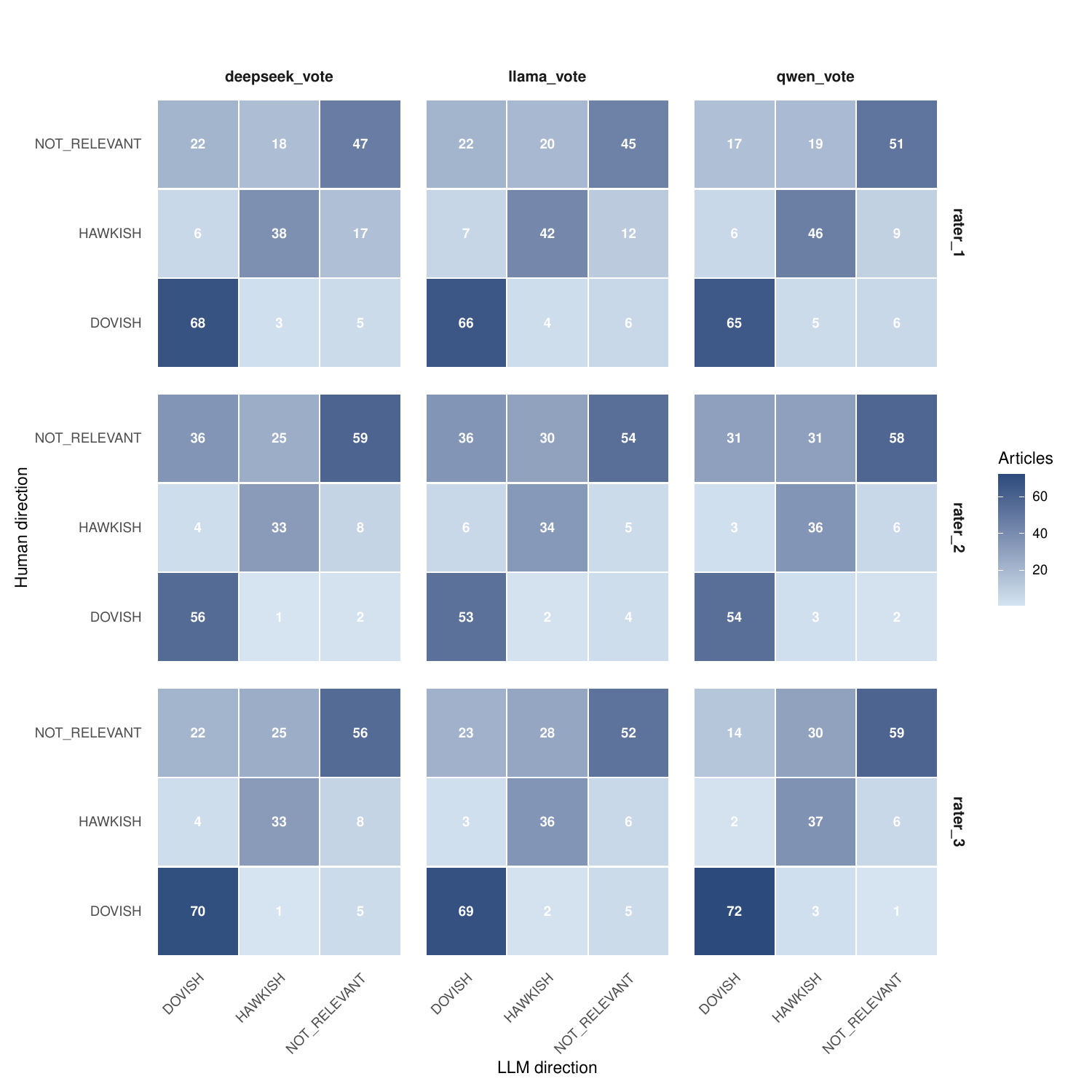}
    \end{subfigure}
    \vspace{0.2em}
    \footnotesize{\begin{flushleft}
    \textbf{Note}: Each panel reports the
  3$\times$3 confusion matrix between one human rater (rows: rater\_1--rater\_3)
  and one LLM model (columns: DeepSeek, Llama, Qwen), classifying articles as
  DOVISH, HAWKISH, or NOT\_RELEVANT. The sample is the $n = 224$ articles with
  complete ratings from all three human raters and all three LLM models; no
  conditioning is applied on human-rater consensus, so the same 224 articles
  enter every panel. Cell values are article counts; darker shading indicates a
  higher count.
    \end{flushleft}}
\end{figure}
\pagebreak

\subsection{Comparison with Existing Measures}

\begin{table}[H]
\centering
\caption{Confusion matrices across validation samples}
\label{tab:confusion_matrices}
\footnotesize
\begin{subtable}[t]{0.56\textwidth}
\centering
\caption{Ours vs.\ Bianchi et al.}
\label{tab_confusion_bianchi}
\small
\begin{tabular}{lcc}
\hline\hline
& \multicolumn{2}{c}{Bianchi et al. (Tweets + news)} \\
\cmidrule(lr){2-3}
Ours (Reuters) & Event & No event \\
\hline
Event      & 36 (2.6\%) & 70 (5.0\%) \\[4pt]
No event   & 21 (1.5\%) & 1284 (91.0\%) \\
\hline\hline
\end{tabular}

\end{subtable}
\hfill
\begin{subtable}[t]{0.56\textwidth}
\centering
\caption{Ours vs.\ Frankovic et al.}
\label{tab_confusion_frankovich}
\small
\begin{tabular}{lcc}
\hline\hline
& \multicolumn{2}{c}{Frankovic et al. (Truth Social posts)} \\
\cmidrule(lr){2-3}
Ours (Reuters) & Event & No event \\
\hline
Event      & 22 (10.0\%) & 48 (21.8\%) \\[4pt]
No event   & 9 (4.1\%) & 141 (64.1\%) \\
\hline\hline
\end{tabular}
\end{subtable}
\begin{flushleft}
\textbf{Note:} Each cell reports count of trading days and percentage of total in the mandate sample.  \emph{Event} = at least one dovish pressure observation on that day.  Mandate periods: Bianchi, 16 Jun 2015--20 Jan 2021; Frankovich, 20 Jan 2025--20 Jan 2026.
\end{flushleft}
\end{table}

\subsection{Robustness to an Expanded Set of Keywords}
\label{subsec:expandend_keywords}

A remaining concern is that the original set of keywords used to retrieve Reuters newswires from Factiva may have missed political pressure events mentioning post-2008 policy instruments (e.g., forward guidance, QE) or the underlying economic conditions motivating pressure and stemming from the Federal Reserve dual mandate (e.g., labor-market slack). To assess whether this narrower search misses a first-order component of the pressure signal, hence biasing our results, we re-run the retrieval and classification pipeline using an expanded set of keywords (Table \ref{tab:expanded_keywords}).

\begin{table}[htbp]
  \centering
  \caption{Additional Monetary Policy Keywords in the Factiva search}
  \label{tab:expanded_keywords}
  \small
  \begin{tabular}{llll}
  \hline\hline
  forward guidance & balance sheet & labor market & bond purchases \\
  quantitative easing & taper & jobs & asset-purchase programme \\
  QE & tapering & job creation & asset-purchase program \\
  quantitative tightening & balance-sheet runoff & economic growth & stimulus \\
  QT & maximum employment & recession & accommodation \\
  dual mandate & unemployment & full employment & tightening \\
  asset purchases & employment & credit conditions & easing \\
  asset purchase & labour market & liquidity & restrictive policy \\
  \hline\hline
  \end{tabular}
  \\[4pt]
  \parbox{\linewidth}{\footnotesize \textbf{Note:} Keywords added to the original search set (Table \ref{tab:keywords}).}
\end{table}

Table \ref{tab:expanded_pipeline_funnel} reports the resulting funnel. Of the 14,429 raw articles retrieved (compared to the original 11,496) deduplication and the relevance filter leave 1,024 events (rather than 923), of which the direction classifier labels 646 as dovish and 101 as hawkish (instead of 579 and 87, respectively).

\begin{table}[htbp]
  \centering
  \caption{Number of Articles/Events through the LLM pipeline ( with Expanded Keyword Set))}
  \label{tab:expanded_pipeline_funnel}
  \small
  \begin{tabular}{lr}
  \hline\hline
  Stage & Articles \\
  \hline
  Raw articles ingested & 14,429 \\
  Deduplicated & 10,129 \\
  Relevant (mentions Fed) &  1,024 \\
  Direction-classified -- Dovish &    646 \\
  Direction-classified -- Hawkish &    101 \\
  \hline\hline
  \end{tabular}
    \\[4pt]
  \parbox{\linewidth}{\footnotesize \textbf{Note:} \textbf{Note}:  Each LLM stage is run in 
parallel across Qwen (\texttt{qwen3-235b-a22b-2507}), Meta Llama 
(\texttt{llama-4-maverick}), and DeepSeek 
(\texttt{deepseek-chat-v3-0324}), and only observations for which at least two of the three models agree are retained.}
\end{table}

\begin{figure}[h!]
    \centering
    \caption{Political pressure coefficients: original vs. expanded keyword sample}    
    \label{fig:comparison_expandend_keywords_coeff}
    \begin{subfigure}[b]{0.88\textwidth}
        \centering
        \includegraphics[width=\textwidth]{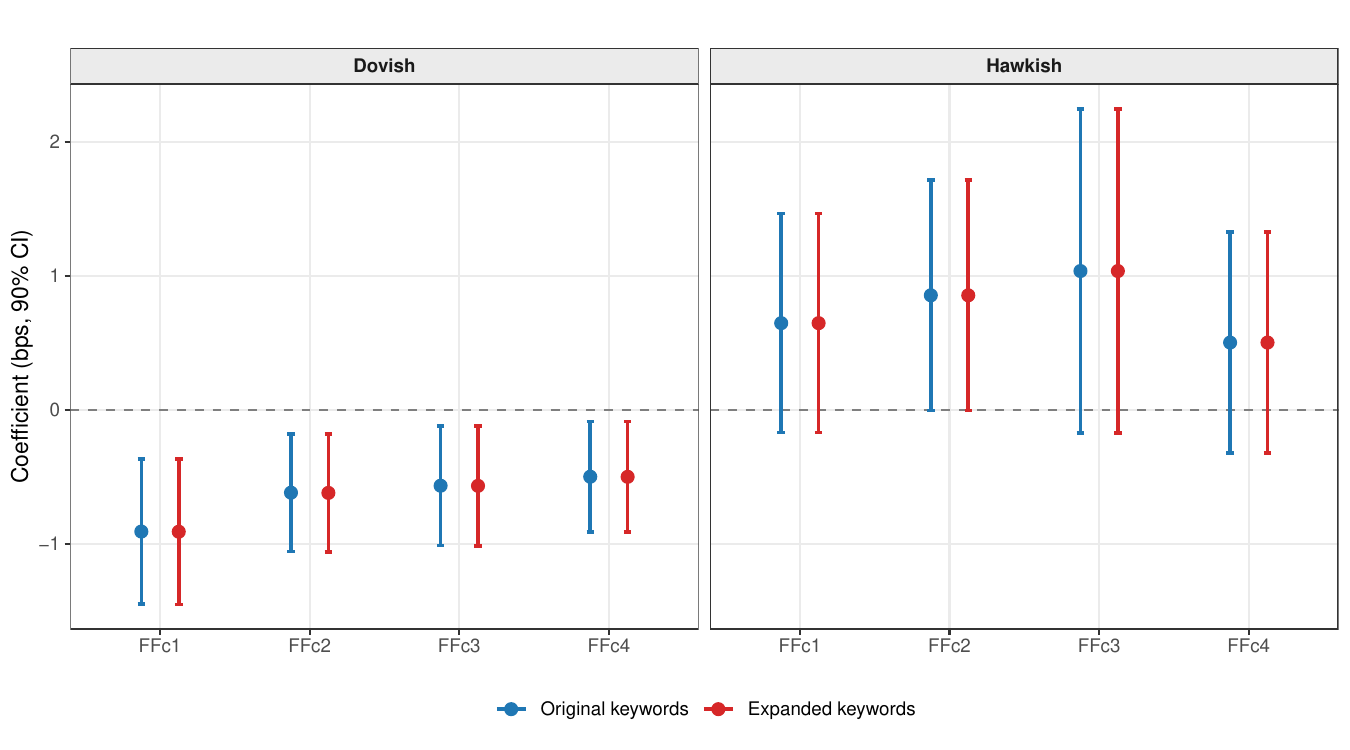}
    \end{subfigure}
    \vspace{0.2em}
    \footnotesize{\begin{flushleft}
    \textbf{Note}:  Each point reports the effect on daily political-pressure events count (Dovish or Hawkish) from the baseline model (equation \ref{eq:baseline}), estimated  under the original keyword sample (blue) and the expanded keyword sample (red).  Sample: 3 October 1988–3 December 2025. Whiskers represent 90\% confidence intervals, computed  with Newey-West (HAC) standard error, date-clustered.
    \end{flushleft}}
\end{figure}

Finally, Figure \ref{fig:comparison_expandend_keywords_coeff} compares the baseline coefficients from Figure \ref{fig:coef_drivers_ffc} with the coefficients estimated when the expanded-keyword events are added to the original sample. The point estimates are quantitatively close across both dovish and hawkish specifications and at every horizon, with overlapping 90\% confidence intervals throughout. Widening the search beyond direct policy language therefore neither changes the estimated effect nor meaningfully sharpens it, indicating that the baseline results are not an artefact of the original keyword choice.

\pagebreak

% ============================================================
%  Appendix: LLM Classification Prompts
% ============================================================

\lstdefinestyle{llmprompt}{
  basicstyle  = \small\ttfamily,
  breaklines  = true,
  breakatwhitespace = true,
  keepspaces  = true,
  frame       = single,
  framesep    = 8pt,
  xleftmargin = 0pt,
  xrightmargin= 0pt,
  columns     = flexible,
  numbers     = none,
  captionpos  = t,
  abovecaptionskip = 4pt,
  belowcaptionskip = 6pt,
  aboveskip   = 10pt,
  belowskip   = 10pt
}

\section{LLM Classification Prompts}
\label{sec:app:prompts}

% --------------------------------------------------------
%  Prompt 1 — Deduplication
% --------------------------------------------------------

\medskip
\noindent\textbf{Prompt 1 --- Event Deduplication}

\begin{singlespace}
\begin{lstlisting}[style=llmprompt]
You are an expert system for classifying whether news articles describe
the same real-world event.

Your task:
For each pair of articles below, determine if they refer to the SAME
underlying event or to DISTINCT events.

Definition of "same event":
- Both articles describe the same real-world occurrence.
- The actors, action, and timing match.
- One article is an update or expanded version of the other.
- Both texts reference the same speech, interview, press release,
  announcement, meeting, data release, incident, or physical occurrence.
- Differences in wording, detail, or angle do not matter if the underlying
  event is the same.

Definition of "different event":
- The actions occurred at different times.
- They describe different statements, decisions, speeches, or occurrences
  -- even if the topic is related.
- Commentary, reaction, or analysis triggered by another event counts as
  a separate event.
- Similar topic or subject matter does NOT imply the same event.

Comparative criteria -- evaluate alignment of:
- WHO acted or spoke
- WHAT was said or done
- WHEN the event occurred (explicitly or implicitly)
- WHERE it happened (if relevant)
- Whether one article directly updates or refers to the same occurrence

Rules:
1. Ignore generic topic similarity.
2. Focus strictly on whether both texts describe the same specific
   occurrence in reality.
3. Compare ALL pairs of articles ([N] comparisons needed for [N] articles).
4. If there is insufficient evidence that both articles describe the same
   specific real-world occurrence, classify as "different_event".
5. Multiple actions or statements by the same actor on different dates or
   occasions are separate events unless explicitly described as the same
   occurrence.

ARTICLES (doc_id: [doc_id]):
[formatted article texts with separators]

Return ONLY a JSON array with one object per pair:
[
  {
    "doc_id": "[doc_id]",
    "group_seq": 1,
    "article_1": 1,
    "article_2": 2,
    "classification": "same_event" or "different_event"
  },
  ...
]

Return ONLY the JSON array, no other text.
\end{lstlisting}
\end{singlespace}

% --------------------------------------------------------
%  Prompt 2 — Relevance
% --------------------------------------------------------

\medskip
\noindent\textbf{Prompt 2 --- Relevance Classification}

\begin{singlespace}
\begin{lstlisting}[style=llmprompt]
You are assisting me in analyzing news articles about US politicians and
the Federal Reserve.

## Definitions ##

A "politician commenting on Fed policy" means a US politician (President,
Vice President, Senator, Representative, Treasury Secretary, cabinet member,
White House official, or presidential candidate) makes a statement --
directly quoted or paraphrased -- about:
  - The Federal Reserve or Fed policy
  - Fed officials (Chair, Vice Chair, Governors, regional bank presidents)
  - Interest rates or monetary policy
  - Quantitative easing or the Fed balance sheet

## What DOES count ##
- Direct quotes from politicians about the Fed
- Paraphrased statements attributed to politicians about Fed policy
- Reported remarks from speeches, interviews, tweets, press conferences

## What does NOT count ##
- Journalist analysis or opinion about Fed-politician relations
- Articles that mention both a politician and the Fed without the
  politician commenting
- Historical background or context about past Fed decisions
- Fed officials commenting on politicians (reverse direction)
- General economic commentary not specifically about Fed policy
- Foreign politicians commenting on US monetary policy
- Private sector economists or analysts commenting

## Response Instructions ##

CRITICAL: Do NOT explain your reasoning. Do NOT use numbered steps.
Output ONLY these two blocks with NOTHING before or after:
<ANALYSIS>...</ANALYSIS>
<JSON>...</JSON>

Part 1: <ANALYSIS> (100 words max)
- Is a politician quoted or paraphrased making a statement about the Fed?
- If so: Who is the politician? What is the target? Through what channel?
- If not: Why does this article not qualify?

Part 2: <JSON>
{
  "unique_article_id": "[unique_article_id]",
  "politician_comments_on_fed": 0 or 1,
  "politician_name": "string or null",
  "politician_position": "string or null",
  "statement_type": "direct_quote" or "paraphrased" or "none",
  "communication_channel": "speech" or "press_conference" or "interview"
                           or "tweet" or "congressional_hearing"
                           or "official_statement" or "other" or "unknown",
  "target_type": "fed_official" or "fed_institution" or "policy" or "none",
  "target_name": "string or null"
}

## Article to analyze ##
[article_text]

Return ONLY the <ANALYSIS> and <JSON> blocks.
\end{lstlisting}
\end{singlespace}

% --------------------------------------------------------
%  Prompt 3 — Direction
% --------------------------------------------------------

\medskip
\noindent\textbf{Prompt 3 --- Direction Classification}

\begin{singlespace}
\begin{lstlisting}[style=llmprompt]
You are classifying the DIRECTION of political pressure on Federal Reserve
monetary policy.

## Context ##
This article has been identified as containing a US politician commenting
on the Federal Reserve.
Politician identified: [politician_name]

## Definitions ##

DOVISH -- The politician wants EASIER monetary policy:
- Lower interest rates / rate cuts
- More economic stimulus / looser monetary conditions
- Concerns about: jobs, growth, recession, credit availability,
  strong dollar hurting exports
- Criticism that current policy is "too tight" or rates are "too high"

HAWKISH -- The politician wants TIGHTER monetary policy:
- Higher interest rates / rate hikes
- Fighting inflation / tighter monetary conditions
- Concerns about: inflation, rising prices, currency debasement,
  asset bubbles
- Criticism that current policy is "too loose" or rates are "too low"

NOT_RELEVANT -- No pressure on monetary policy direction:
- General economic commentary without Fed pressure
- Factual reporting without politician advocacy
- Discussion of Fed appointments/personnel without policy implications
- Praise for Fed actions without directional preference

## Classification Rules ##

1. Explicit statements override context: if the politician explicitly
   mentions rates or policy direction, use that.
2. Use context for implicit pressure: if the politician criticizes the Fed
   without stating a direction, infer from the economic context.
3. Default logic when context is unclear:
   - Criticism of rate HIKES              -> dovish
   - Criticism of rate CUTS               -> hawkish
   - Criticism during HIGH INFLATION      -> hawkish
   - Criticism during RECESSION/HIGH UNEMPLOYMENT -> dovish
   - Calls for "action" during inflation  -> hawkish
   - Calls for "action" during recession  -> dovish
4. Personnel attacks without policy content: use economic context to infer
   direction; if truly impossible to determine, classify as not_relevant.

## Response Format ##

Part 1: <ANALYSIS> (100 words max)
- What is the politician's statement?
- Is the direction explicit or inferred from context?
- What economic context helped determine direction (if applicable)?

Part 2: <JSON>
{
  "unique_article_id": "[unique_article_id]",
  "pressure_direction": "dovish" or "hawkish" or "not_relevant"
}

## Article to analyze ##
[article_text]
\end{lstlisting}
\end{singlespace}

\section{Additional Statistics}

\begin{table}[H]
  \centering
  \caption{Summary Statistics}
  \label{tab:summary_statistics}
  \footnotesize
  \setlength{\tabcolsep}{4pt}
  \renewcommand{\arraystretch}{0.95}
  \begin{tabular}{llrrrrrr}
  \hline\hline
  & Variable & Unit & $N$ & Mean & SD & Min & Max \\
  \hline
  \multicolumn{8}{l}{\textit{Pressure measure}} \\
  & Dovish pressure ($n_{\text{dovish}}$) & count & 9,361 & 0.054 & 0.262 & 0.000 & 4.000 \\
  & Hawkish pressure ($n_{\text{hawkish}}$) & count & 9,361 & 0.008 & 0.090 & 0.000 & 3.000 \\
  \\[-6pt]
  \multicolumn{8}{l}{\textit{Panel A: FF futures}} \\
  & FFc1 (Kuttner surprise) & bps & 9,338 & -0.043 & 4.963 & -98.500 & 86.800 \\
  & FFc2 rate change & bps & 9,352 & -0.050 & 3.539 & -52.000 & 63.000 \\
  & FFc3 rate change & bps & 9,349 & -0.055 & 3.624 & -52.000 & 46.500 \\
  & FFc4 rate change & bps & 9,348 & -0.055 & 3.893 & -52.500 & 40.000 \\
  \\[-6pt]
  \multicolumn{8}{l}{\textit{Panel B: Treasury futures}} \\
  & TU (2Y Treasury) & bps & 8,882 & 0.151 & 9.780 & -33.323 & 29.947 \\
  & FV (5Y Treasury) & bps & 9,346 & 0.281 & 24.244 & -75.332 & 67.437 \\
  & TY (10Y Treasury) & bps & 9,354 & 0.404 & 36.425 & -107.479 & 99.110 \\
  & US (30Y Treasury) & bps & 9,352 & 0.702 & 60.319 & -165.595 & 158.369 \\
  \\[-6pt]
  \multicolumn{8}{l}{\textit{Panel C: Volatility}} \\
  & VIX (CBOE) & pts & 5,441 & -0.000 & 1.346 & -12.075 & 19.200 \\
  & MOVE Index & pts & 5,616 & -0.021 & 4.315 & -40.500 & 41.500 \\
  \\[-6pt]
  \multicolumn{8}{l}{\textit{Panel D: FX \& commodities}} \\
  & DXY (USD Index) & \% & 9,353 & 0.000 & 0.505 & -3.056 & 2.853 \\
  & EUR/USD & \% & 9,358 & 0.001 & 0.607 & -3.669 & 3.719 \\
  \\[-6pt]
  \multicolumn{8}{l}{\textit{Panel E: Inflation expectations}} \\
  & 2Y inflation swap & bps & 4,651 & -0.000 & 7.601 & -153.100 & 121.000 \\
  & 10Y inflation swap & bps & 4,651 & -0.009 & 4.225 & -63.000 & 29.000 \\
  \hline\hline
  \end{tabular}
  \\[4pt]
  \parbox{0.95\linewidth}{\footnotesize \textit{Notes:} FFc1 is the Kuttner (2001) surprise scaled to basis points; its extreme values ($-$98.5 and $+$86.8 bps) correspond to the 2008 crisis and COVID-19 emergency cuts. FFc2--FFc4 are daily changes in implied Fed Funds rates (pp $\times$ 100 = bps). TU, FV, TY, and US are daily price returns on 2Y, 5Y, 10Y, and 30Y Treasury futures (price return $\times$ 100 = bps), winsorised at the 1st and 99th percentiles to remove contract-roll and data-error artifacts. FX and commodity series are daily log returns ($\times$ 100, \%). Inflation swap rates are daily first differences ($\times$ 100 = bps); coverage begins in 2004, yielding approximately half the full sample. VIX and MOVE are daily first differences (index points); both series begin in the mid-1990s, accounting for their lower observation counts. $n_{\text{dovish}}$ and $n_{\text{hawkish}}$ are daily counts of classified news events.}
\end{table}

\begin{figure}[h!]
    \centering
    \caption{Correlation Structure of Dependent Variables}    \label{fig:correlation_dep_var}
    \begin{subfigure}[b]{0.7\textwidth}
        \centering
        \includegraphics[width=\textwidth]{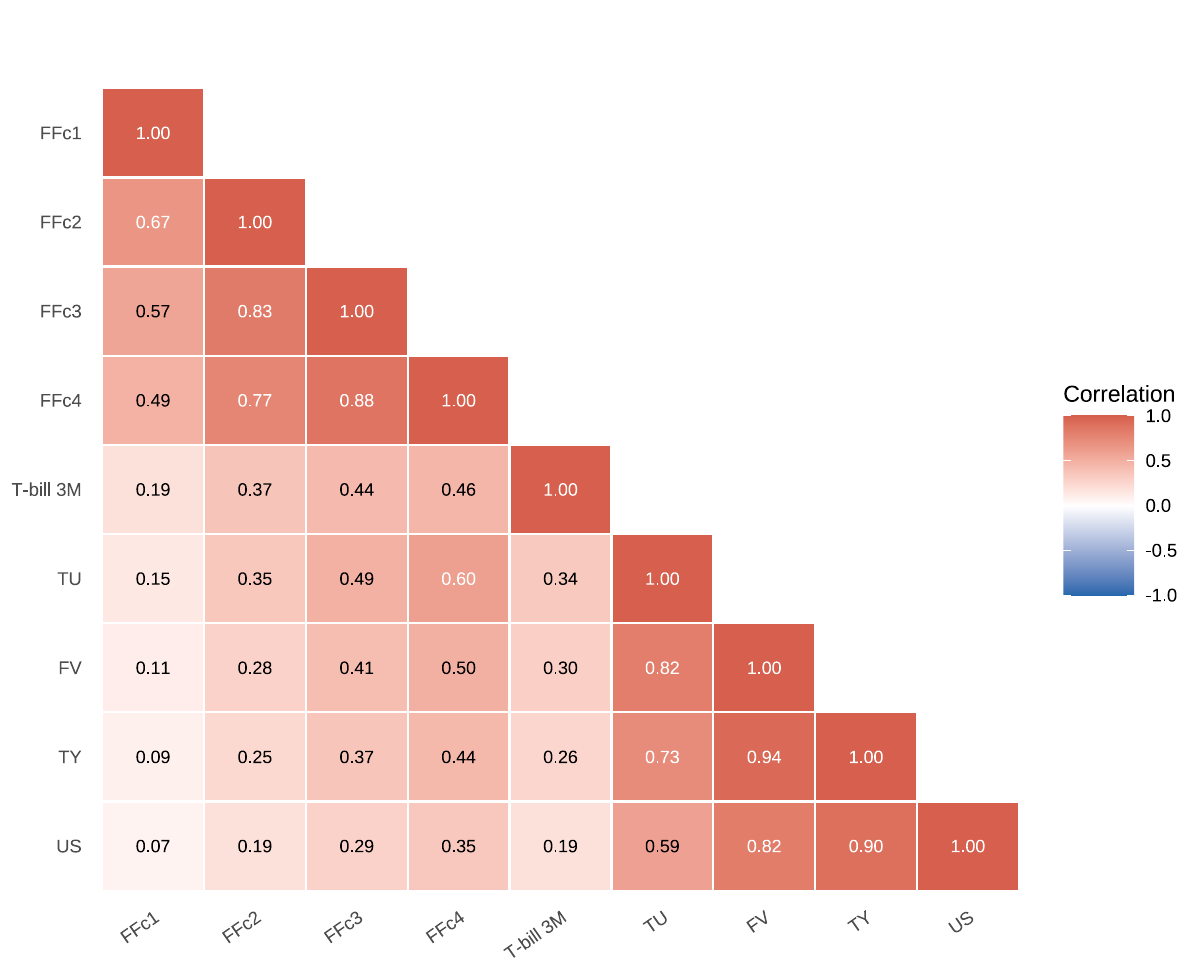}
    \end{subfigure}
    \vspace{0.2em}
    \footnotesize{\begin{flushleft}
    \textbf{Note}: FFc1 refers to the current-month contract.
    \end{flushleft}}
\end{figure}

\begin{figure}[H]
    \centering
    \caption{Dependent variables and day-of-week seasonality}
    \label{fig:fff_dow}
    \includegraphics[width=\linewidth]{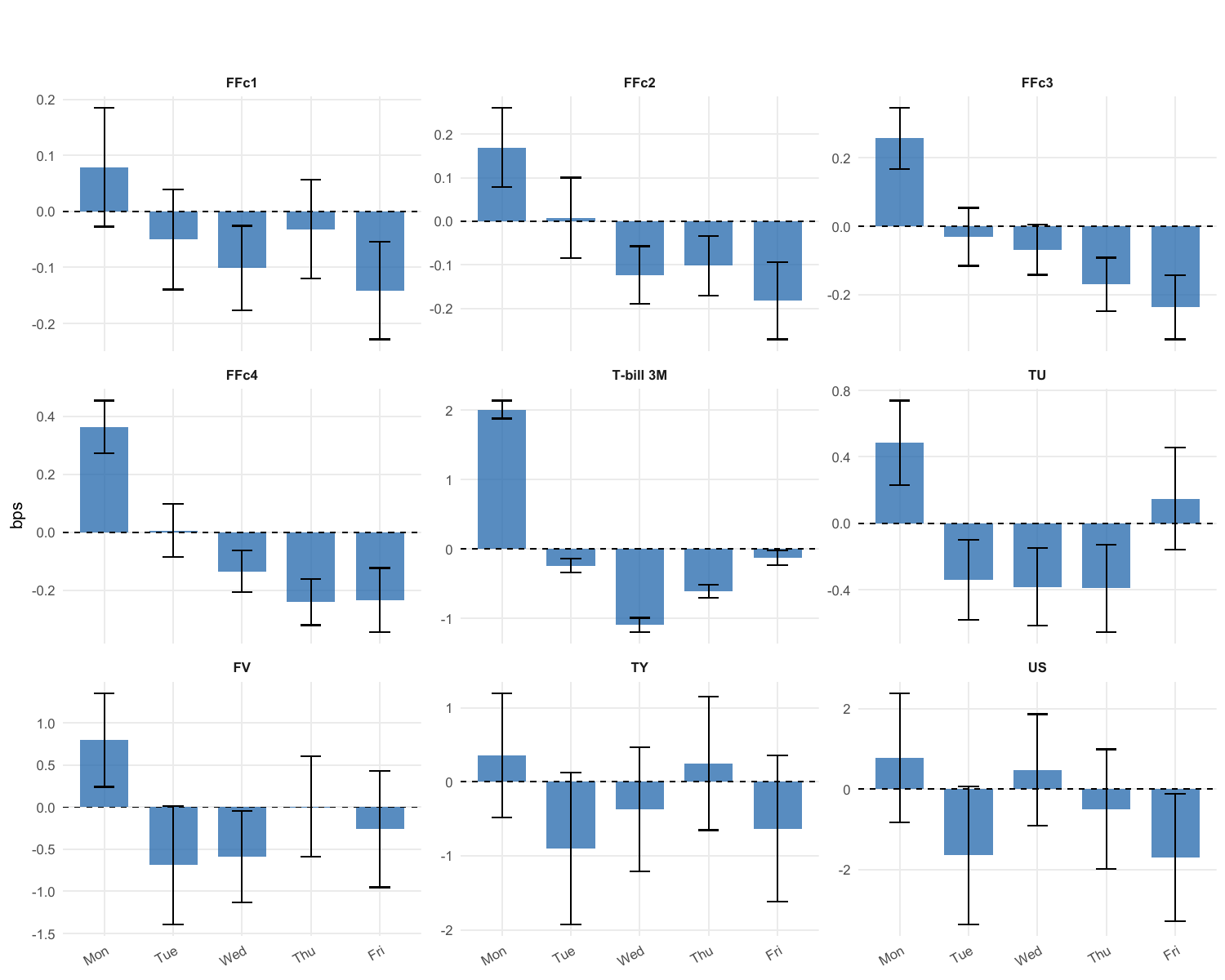}
    \footnotesize{\begin{flushleft}
    \textbf{Note:} 
    \end{flushleft}}
\end{figure}

\begin{figure}[H]
    \centering
    \caption{Treasury Futures: Rolling Volume by Decade}
    \label{fig:roll_vol_decade}
    \begin{subfigure}[b]{0.48\textwidth}
        \centering
        \includegraphics[width=\textwidth]{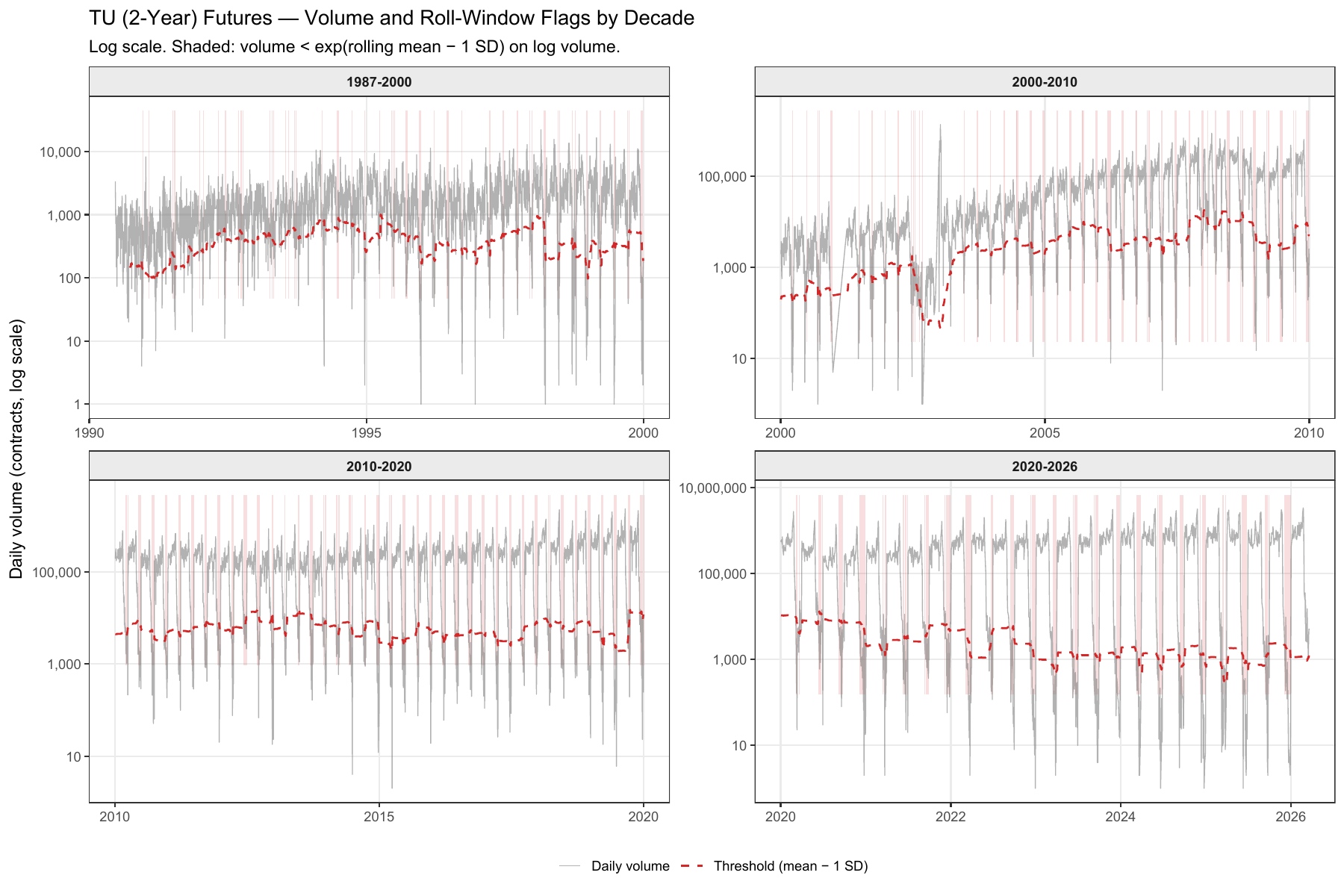}
        \caption{2-Year (TU)}
        \label{fig:roll_vol_tu}
    \end{subfigure}
    \hfill
    \begin{subfigure}[b]{0.48\textwidth}
        \centering
        \includegraphics[width=\textwidth]{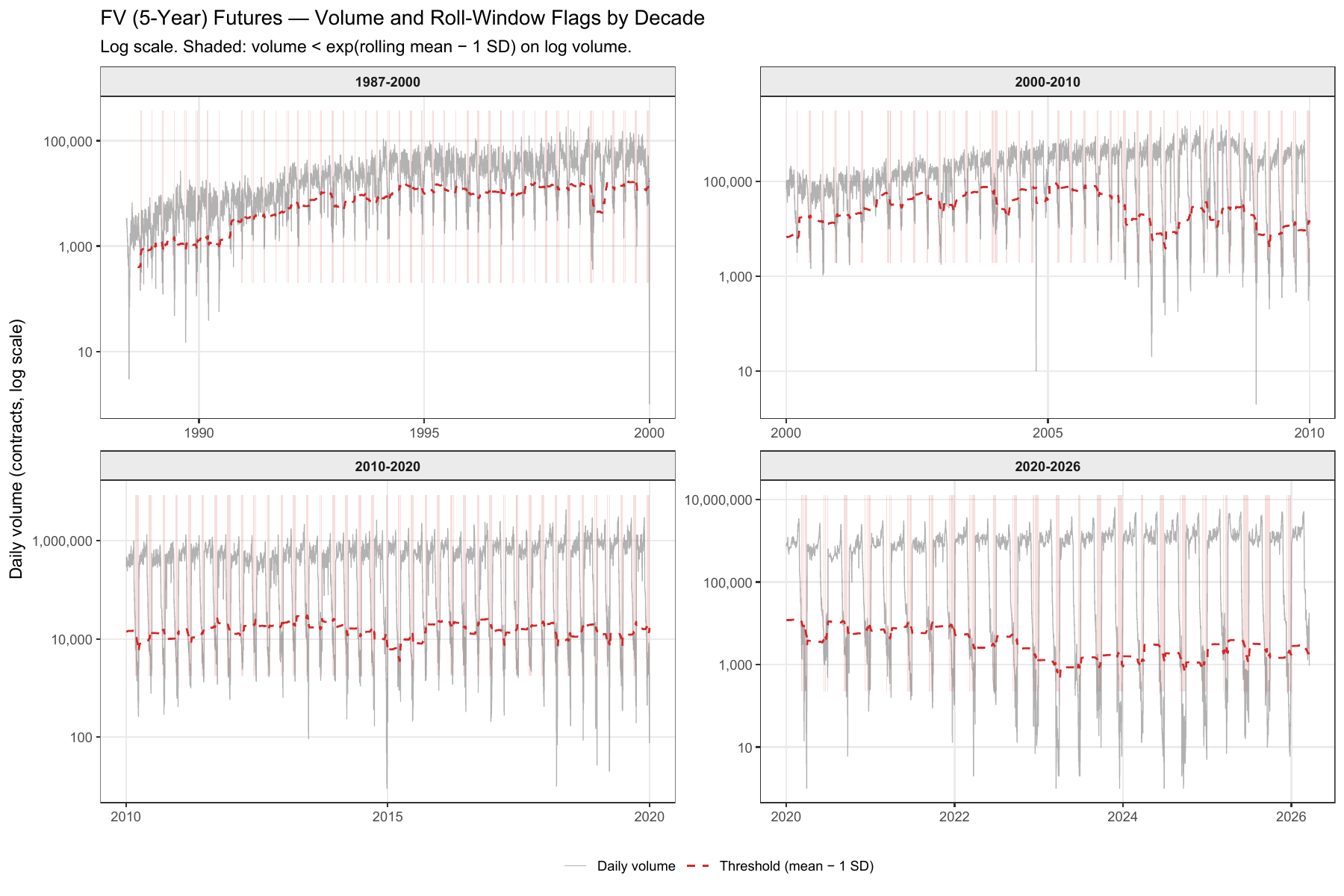}
        \caption{5-Year (FV)}
        \label{fig:roll_vol_fv}
    \end{subfigure}
    
    \vspace{1em}
    
    \begin{subfigure}[b]{0.48\textwidth}
        \centering
        \includegraphics[width=\textwidth]{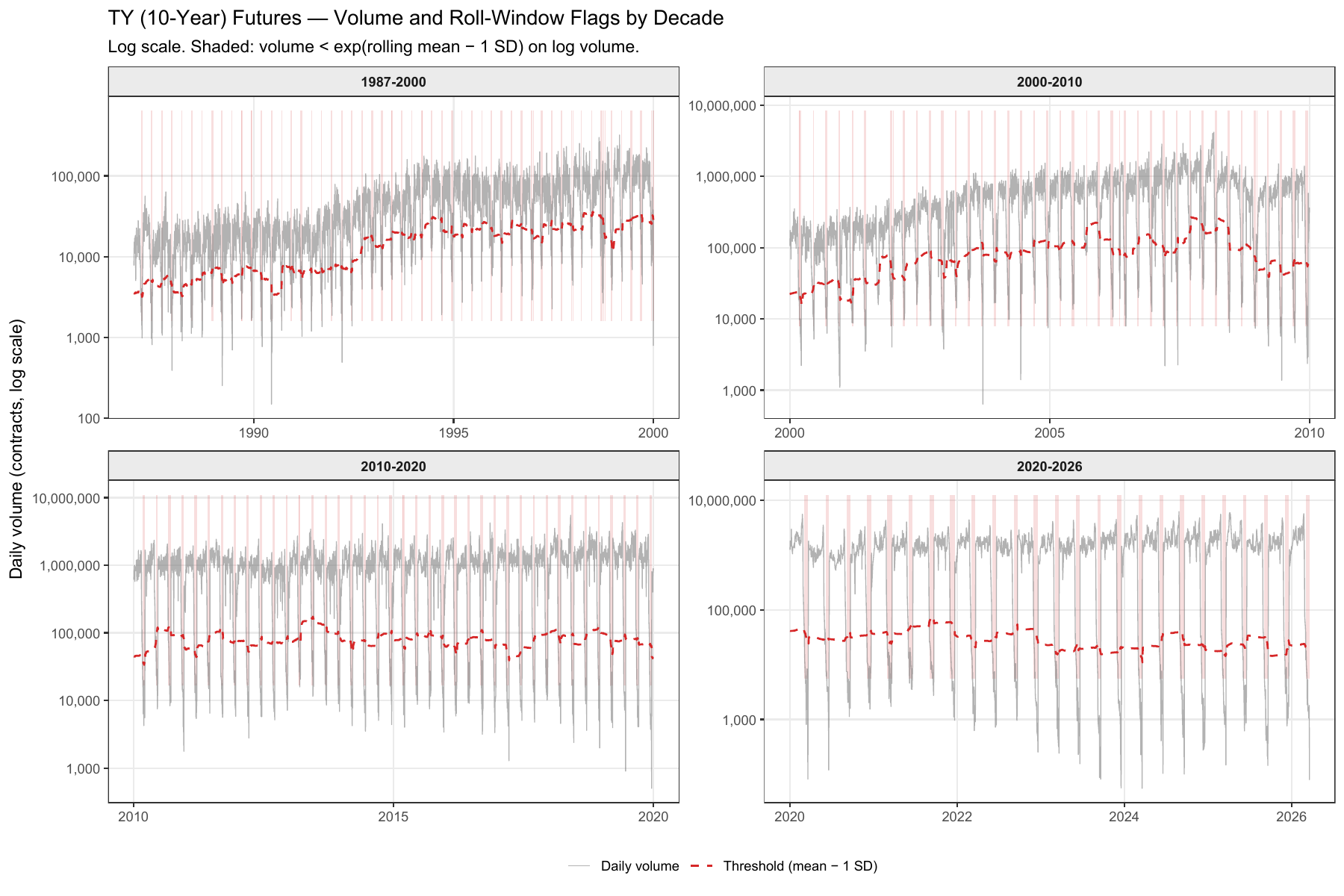}
        \caption{10-Year (TY)}
        \label{fig:roll_vol_ty}
    \end{subfigure}
    \hfill
    \begin{subfigure}[b]{0.48\textwidth}
        \centering
        \includegraphics[width=\textwidth]{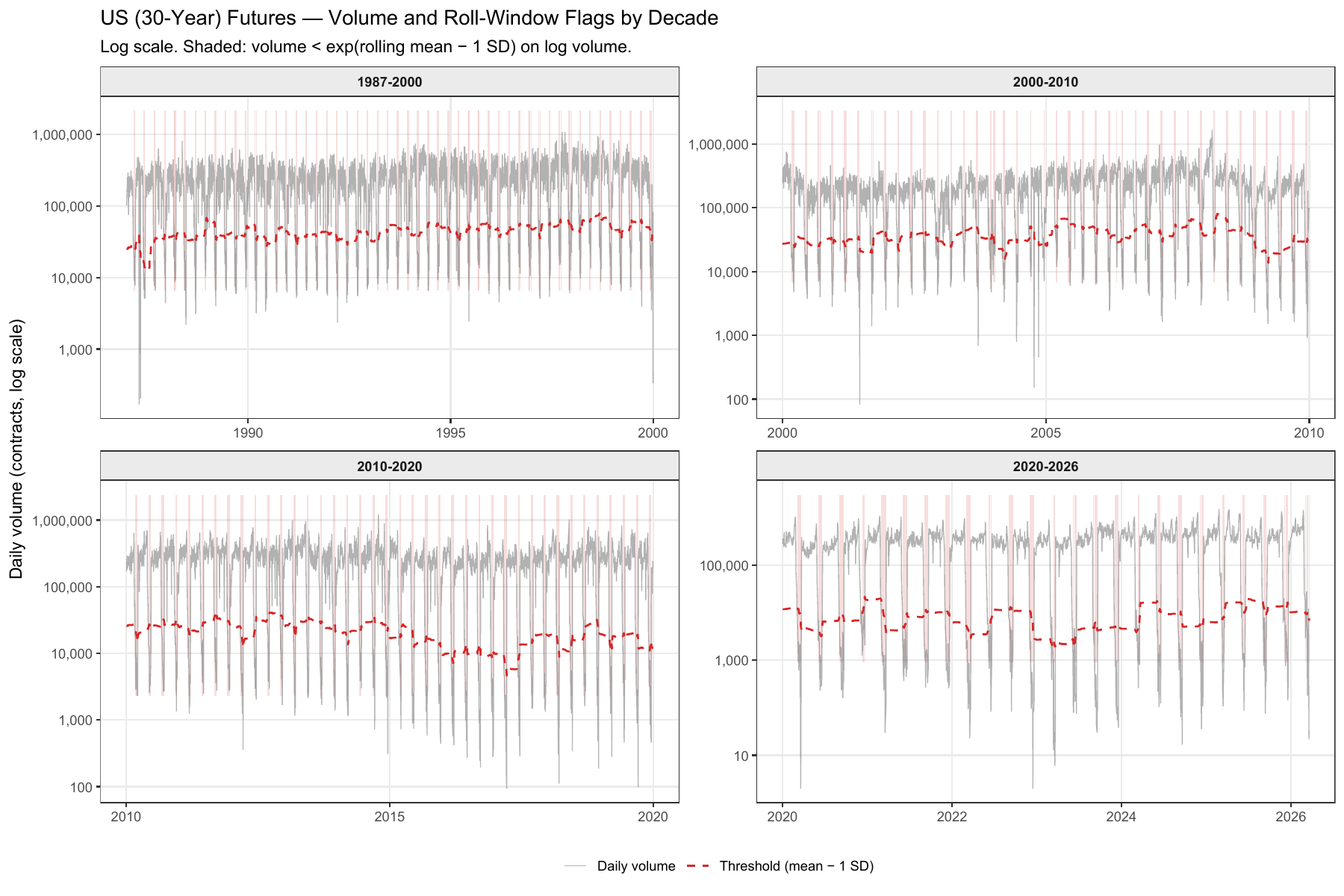}
        \caption{30-Year (US)}
        \label{fig:roll_vol_us}
    \end{subfigure}
    \vspace{0.2em}
    \footnotesize{\begin{flushleft}
    \textbf{Note}: Each panel shows the distribution of daily 
    log-volume for the corresponding Treasury futures contract, disaggregated by decade. Observations flagged as low-volume near contract expiry and excluded from the analysis are shown in a lighter shade.
    \end{flushleft}}
\end{figure}

\section{Additional Results}

\subsection{Dovish Pressure: Inflation Expectations, Financial Market Volatility, and Spot Exchange Rates}

\begin{table}[H]
   \caption{\label{tab_drivers_inflexp_dovish} Dovish Political Pressure and Inflation Expectations}
   \centering
   \footnotesize
   \setlength{\tabcolsep}{4pt}
   \renewcommand{\arraystretch}{0.95}
   \begin{tabular}{lcccc}
      \tabularnewline \midrule \midrule
      Dependent Variables: & \multicolumn{2}{c}{2Y Infl.\ Swap ($\Delta$, pp)} & \multicolumn{2}{c}{10Y Infl.\ Swap ($\Delta$, pp)}\\
                                             & Full                    & No Trump       & Full                         & No Trump \\   
      Model:                                 & (1)                     & (2)            & (3)                          & (4)\\  
      \midrule
      \emph{Variables}\\
      Constant                               & -0.0032$^{*}$           & -0.0013        & -0.0010                      & -0.0006\\   
                                             & (0.0018)                & (0.0023)       & (0.0008)                     & (0.0012)\\   
      Dovish pressure ($n_{\text{dovish}}$)  & -0.0022                 & -0.0120        & -0.0009                      & -0.0051\\   
                                             & (0.0027)                & (0.0107)       & (0.0015)                     & (0.0069)\\   
      Oil price (65-day \% return, LOCF)            & 0.0002$^{**}$           & 0.0003$^{***}$ & $6.86\times 10^{-5}$$^{**}$  & 0.0001$^{**}$\\   
                                             & ($7.98\times 10^{-5}$)  & (0.0001)       & ($3.43\times 10^{-5}$)       & ($5.23\times 10^{-5}$)\\    
      S\&P~500 (65-day \% return, LOCF)             & 0.0006$^{**}$           & 0.0006$^{*}$   & 0.0002$^{**}$                & 0.0003\\   
                                             & (0.0003)                & (0.0003)       & (0.0001)                     & (0.0002)\\   
      NFP announcement                       & 0.0120$^{**}$           & 0.0163$^{**}$  & 0.0080$^{***}$               & 0.0099$^{***}$\\   
                                             & (0.0060)                & (0.0078)       & (0.0029)                     & (0.0036)\\   
      FOMC announcement                      & -0.0008                 & -0.0021        & $5.28\times 10^{-6}$         & 0.0027\\   
                                             & (0.0061)                & (0.0071)       & (0.0034)                     & (0.0040)\\   
      Yield skewness (ISK)                   & 0.0056                  & 0.0055         & 0.0068$^{***}$               & 0.0068$^{***}$\\   
                                             & (0.0038)                & (0.0048)       & (0.0021)                     & (0.0024)\\   
      Inflation gap                          & -0.0008                 & -0.0015        & -0.0004                      & -0.0007\\   
                                             & (0.0010)                & (0.0011)       & (0.0004)                     & (0.0005)\\   
      Unemployment gap                       & 0.0004                  & -0.0012        & -0.0004                      & -0.0011$^{**}$\\   
                                             & (0.0010)                & (0.0010)       & (0.0005)                     & (0.0005)\\   
      \midrule
      \emph{Fit statistics}\\
      Observations                           & 4,636                   & 3,411          & 4,636                        & 3,411\\  
      R$^2$                                  & 0.044                   & 0.054          & 0.019                        & 0.019\\    
      \midrule \midrule
      \multicolumn{5}{l}{\emph{}}\\
   \end{tabular}
   
   \par \raggedright 
   \footnotesize \textbf{Note:}
   Dependent variables: daily change in 2-year and 10-year USD inflation swap rates (percentage points; $\times 100$ = basis points). Drivers specification: inflation gap, unemployment gap, oil price 65-days percentage return (LOCF), S\&P~500 65-days percentage return (LOCF), NFP announcement dummy, FOMC announcement dummy, and treasury yield skewness (ISK). Beginning-of-month dummy and two lags of the dependent variable included but omitted from display. ``No Trump'' drops 20 Jan 2017--20 Jan 2021 and post-20 Jan 2025. HAC (Newey-West) standard errors in parentheses. *, **, *** denote significance at 10\%, 5\%, 1\%.
\end{table}

\begin{table}[htbp]
   \caption{\label{tab_drivers_volatility_dovish} Dovish Political Pressure and Financial Volatility}
   \centering
   \footnotesize
   \setlength{\tabcolsep}{4pt}
   \renewcommand{\arraystretch}{0.95}
   \begin{tabular}{lcccc}
      \tabularnewline \midrule \midrule
      Dependent Variables: & \multicolumn{2}{c}{VIX ($\Delta$, pts)} & \multicolumn{2}{c}{MOVE ($\Delta$, pts)}\\
                                             & Full           & No Trump       & Full            & No Trump \\   
      Model:                                 & (1)            & (2)            & (3)             & (4)\\  
      \midrule
      \emph{Variables}\\
      Constant                               & 0.0308         & 0.0226         & 0.2361$^{***}$  & 0.2634$^{***}$\\   
                                             & (0.0253)       & (0.0289)       & (0.0757)        & (0.0890)\\   
      Dovish pressure ($n_{\text{dovish}}$)  & 0.0921         & -0.0071        & -0.2341         & 1.172\\   
                                             & (0.0843)       & (0.2589)       & (0.2140)        & (0.8650)\\   
      Oil price (65-day \% return, LOCF)            & 0.0009         & -0.0002        & 0.0049          & 0.0040\\   
                                             & (0.0008)       & (0.0012)       & (0.0032)        & (0.0046)\\   
      S\&P~500 (65-day \% return, LOCF)             & -0.0088$^{**}$ & -0.0085$^{*}$  & -0.0117         & -0.0134\\   
                                             & (0.0042)       & (0.0044)       & (0.0113)        & (0.0132)\\   
      NFP announcement                       & -0.1030        & -0.0661        & -2.720$^{***}$  & -3.317$^{***}$\\   
                                             & (0.0758)       & (0.0744)       & (0.3194)        & (0.3442)\\   
      FOMC announcement                      & -0.2084$^{*}$  & -0.2536$^{**}$ & -2.768$^{***}$  & -3.095$^{***}$\\   
                                             & (0.1067)       & (0.1256)       & (0.3611)        & (0.4544)\\   
      Yield skewness (ISK)                   & -0.2169$^{**}$ & -0.1897$^{**}$ & -0.5657$^{***}$ & -0.5230$^{*}$\\   
                                             & (0.0949)       & (0.0760)       & (0.2177)        & (0.2680)\\   
      Inflation gap                          & 0.0091         & 0.0148         & 0.0598          & 0.0614\\   
                                             & (0.0129)       & (0.0153)       & (0.0483)        & (0.0501)\\   
      Unemployment gap                       & 0.0084         & 0.0226         & 0.0241          & 0.0489\\   
                                             & (0.0166)       & (0.0157)       & (0.0323)        & (0.0444)\\   
      \midrule
      \emph{Fit statistics}\\
      Observations                           & 5,408          & 4,183          & 5,429           & 4,287\\  
      R$^2$                                  & 0.020          & 0.017          & 0.044           & 0.050\\   
      \midrule \midrule
      \multicolumn{5}{l}{\emph{}}\\
   \end{tabular}
   
   \par \raggedright 
   \footnotesize
   \textbf{Note:} Dependent variables: first difference of VIX and MOVE index levels (index points). Drivers specification: inflation gap, unemployment gap, oil price 65-day \% return (LOCF), S\&P~500 65-day \% return (LOCF), NFP announcement dummy, FOMC announcement dummy, and treasury yield skewness (ISK). Beginning-of-month dummy and two lags of the dependent variable included but omitted from display. ``No Trump'' drops 20 Jan 2017--20 Jan 2021 and post-20 Jan 2025. HAC (Newey-West) standard errors in parentheses. *, **, *** denote significance at 10\%, 5\%, 1\%.
\end{table}

\begin{table}[H]
   \caption{\label{tab_drivers_fx_dovish} Dovish Political Pressure and Spot Exchange Rates}
   \centering
   \footnotesize
   \setlength{\tabcolsep}{4pt}
   \renewcommand{\arraystretch}{0.95}
   \begin{tabular}{lcccc}
      \tabularnewline \midrule \midrule
      Dependent Variables:                   & \multicolumn{2}{c}{EUR/USD (\%)} & \multicolumn{2}{c}{DXY (\%)}\\
                                             & Full                   & No Trump       & Full            & No Trump\\
      Model:                                 & (1)                    & (2)            & (3)             & (4)\\  
      \midrule
      \emph{Variables}\\
      Constant                               & -0.0025                & -0.0076        & 0.0050          & 0.0095\\   
                                             & (0.0091)               & (0.0104)       & (0.0075)        & (0.0086)\\   
      Dovish pressure ($n_{\text{dovish}}$)  & 0.0451$^{*}$           & 0.1087$^{**}$  & -0.0256         & -0.0748$^{*}$\\   
                                             & (0.0268)               & (0.0473)       & (0.0233)        & (0.0403)\\   
      Oil price (65-days return, \%)             & 0.0010$^{***}$         & 0.0012$^{**}$  & -0.0010$^{***}$ & -0.0012$^{***}$\\   
                                             & (0.0004)               & (0.0005)       & (0.0003)        & (0.0004)\\   
      S\&P~500 (65-days return, \%)              & -0.0017                & -0.0018        & 0.0010          & 0.0010\\   
                                             & (0.0012)               & (0.0013)       & (0.0010)        & (0.0011)\\   
      NFP announcement                       & -0.0186                & -0.0094        & 0.0222          & 0.0187\\   
                                             & (0.0360)               & (0.0406)       & (0.0309)        & (0.0349)\\   
      FOMC announcement                      & 0.1140$^{***}$         & 0.1251$^{***}$ & -0.1202$^{***}$ & -0.1317$^{***}$\\   
                                             & (0.0434)               & (0.0485)       & (0.0338)        & (0.0378)\\   
      Yield skewness (ISK)                   & $-2.73\times 10^{-6}$  & 0.0146         & 0.0095          & -0.0012\\   
                                             & (0.0200)               & (0.0288)       & (0.0172)        & (0.0243)\\   
      Inflation gap                          & -0.0104$^{*}$          & -0.0110$^{*}$  & 0.0090$^{*}$    & 0.0094$^{*}$\\   
                                             & (0.0058)               & (0.0060)       & (0.0051)        & (0.0052)\\   
      Unemployment gap                       & 0.0011                 & -0.0014        & -0.0023         & -0.0006\\   
                                             & (0.0048)               & (0.0063)       & (0.0038)        & (0.0050)\\   
      $\Delta$ Bund 2Y (pp)                  & 0.5341$^{**}$          & 0.5153$^{**}$  & -0.1300         & -0.1273\\   
                                             & (0.2419)               & (0.2496)       & (0.1828)        & (0.2048)\\   
      \midrule
      \emph{Fit statistics}\\
      Observations                           & 8,454                  & 7,256          & 8,447           & 7,249\\  
      R$^2$                                  & 0.005                  & 0.006          & 0.004           & 0.005\\  
      \midrule \midrule
      \multicolumn{5}{l}{\emph{}}\\
   \end{tabular}
   
   \par \raggedright 
   \footnotesize \textbf{Note:}
   Dependent variables are log returns multiplied by 100 and expressed in percentage points. Oil price growth and S\&P~500 returns are 65-trading-day percentage returns, labelled as QoQ. All regressions include two lags of the dependent variable, omitted from display. ``No Trump'' drops 20 Jan 2017--20 Jan 2021 and post-20 Jan 2025. HAC (Newey-West) standard errors in parentheses. *, **, *** denote significance at the 10\%, 5\%, 1\% level.
\end{table}

\subsection{Hawkish Pressure}

\begin{table}[H]
   \caption{\label{tab_drivers_hawkish_ffc} Hawkish Political Pressure and Fed Funds Futures}
   \centering
   \footnotesize
   \setlength{\tabcolsep}{4pt}
   \renewcommand{\arraystretch}{0.95}
   \begin{tabular}{lcccc}
      \tabularnewline \midrule \midrule
      Dependent Variables:                     & FFc1 (Kuttner)          & FFc2                    & FFc3                    & FFc4\\  
      Model:                                   & (1)                     & (2)                     & (3)                     & (4)\\  
      \midrule
      \emph{Variables}\\
      Constant                                 & -0.0006                 & -0.0010$^{**}$          & -0.0010$^{**}$          & -0.0010$^{**}$\\   
                                               & (0.0006)                & (0.0004)                & (0.0005)                & (0.0005)\\   
      Hawkish pressure ($n_{\text{hawkish}}$)  & 0.0064                  & 0.0085                  & 0.0103                  & 0.0049\\   
                                               & (0.0056)                & (0.0052)                & (0.0070)                & (0.0049)\\   
      Oil price growth (QoQ, \%)               & $-2.04\times 10^{-5}$   & $-2.89\times 10^{-5}$   & $-3.95\times 10^{-5}$   & $-4.43\times 10^{-5}$$^{*}$\\    
                                               & ($2.56\times 10^{-5}$)  & ($2.35\times 10^{-5}$)  & ($2.48\times 10^{-5}$)  & ($2.33\times 10^{-5}$)\\    
      S\&P~500 return (QoQ, \%)                & 0.0001$^{*}$            & $4.73\times 10^{-5}$    & $4.53\times 10^{-5}$    & $4.13\times 10^{-5}$\\    
                                               & ($7.18\times 10^{-5}$)  & ($5.8\times 10^{-5}$)   & ($5.99\times 10^{-5}$)  & ($5.81\times 10^{-5}$)\\    
      NFP announcement                         & -0.0090$^{***}$         & -0.0055$^{**}$          & -0.0066$^{**}$          & -0.0050$^{*}$\\   
                                               & (0.0033)                & (0.0023)                & (0.0026)                & (0.0029)\\   
      Inflation gap                            & 0.0010$^{**}$           & 0.0007$^{*}$            & 0.0007                  & 0.0008\\   
                                               & (0.0005)                & (0.0004)                & (0.0005)                & (0.0005)\\   
      Unemployment gap                         & -0.0009$^{**}$          & -0.0006$^{***}$         & -0.0006$^{***}$         & -0.0007$^{***}$\\   
                                               & (0.0004)                & (0.0002)                & (0.0003)                & (0.0003)\\   
      FOMC announcement                        & -0.0113$^{**}$          & -0.0120$^{***}$         & -0.0125$^{***}$         & -0.0130$^{***}$\\   
                                               & (0.0050)                & (0.0040)                & (0.0042)                & (0.0039)\\   
      Yield skewness (ISK)                     & 0.0057$^{***}$          & 0.0065$^{***}$          & 0.0078$^{***}$          & 0.0090$^{***}$\\   
                                               & (0.0019)                & (0.0018)                & (0.0017)                & (0.0018)\\   
      \midrule
      \emph{Fit statistics}\\
      Observations                             & 9,004                   & 9,019                   & 9,015                   & 9,015\\  
      R$^2$                                    & 0.014                   & 0.031                   & 0.038                   & 0.031\\  
      \midrule \midrule
      \multicolumn{5}{l}{\emph{}}\\
   \end{tabular}
   
   \par \raggedright 
   \footnotesize textbf{Note:}
   All columns use the full sample. Coefficients on FF futures variables multiplied by 100 are expressed in basis points. All regressions include: beginning-of-month dummy and two lags of the dependent variable (omitted from display). HAC (Newey-West) standard errors in parentheses. *, **, *** denote significance at the 10\%, 5\%, 1\% level.
\end{table}

\begin{table}[htbp]
   \caption{\label{tab_drivers_fx_hawkish} Hawkish Political Pressure and Spot Exchange Rates}
   \centering
    \setlength{\tabcolsep}{4pt}
   \renewcommand{\arraystretch}{0.95}
   \begin{tabular}{lcccc}
      \tabularnewline \midrule \midrule
      Dependent Variables: & \multicolumn{2}{c}{EUR/USD (\%)} & \multicolumn{2}{c}{DXY (\%)}\\
                                               & Full           & No Trump       & Full            & No Trump \\   
      Model:                                   & (1)            & (2)            & (3)             & (4)\\  
      \midrule
      \emph{Variables}\\
      Constant                                 & -0.0004        & -0.0046        & 0.0038          & 0.0074\\   
                                               & (0.0090)       & (0.0104)       & (0.0074)        & (0.0085)\\   
      Hawkish pressure ($n_{\text{hawkish}}$)  & 0.0095         & -0.0787        & -0.0018         & 0.0534\\   
                                               & (0.0730)       & (0.0753)       & (0.0590)        & (0.0638)\\   
      Oil price growth (65-day \%, LOCF)       & 0.0010$^{***}$ & 0.0011$^{**}$  & -0.0010$^{***}$ & -0.0012$^{***}$\\   
                                               & (0.0004)       & (0.0005)       & (0.0003)        & (0.0004)\\   
      S\&P~500 return (65-day \%, LOCF)        & -0.0016        & -0.0018        & 0.0010          & 0.0011\\   
                                               & (0.0012)       & (0.0013)       & (0.0010)        & (0.0011)\\   
      NFP announcement                         & -0.0158        & -0.0059        & 0.0206          & 0.0163\\   
                                               & (0.0359)       & (0.0406)       & (0.0308)        & (0.0349)\\   
      FOMC announcement                        & 0.1188$^{***}$ & 0.1330$^{***}$ & -0.1229$^{***}$ & -0.1371$^{***}$\\   
                                               & (0.0433)       & (0.0483)       & (0.0338)        & (0.0378)\\   
      Yield skewness (ISK)                     & -0.0047        & 0.0137         & 0.0122          & -0.0006\\   
                                               & (0.0199)       & (0.0290)       & (0.0172)        & (0.0245)\\   
      Inflation gap                            & -0.0103$^{*}$  & -0.0101$^{*}$  & 0.0089$^{*}$    & 0.0087$^{*}$\\   
                                               & (0.0058)       & (0.0061)       & (0.0051)        & (0.0052)\\   
      Unemployment gap                         & 0.0010         & -0.0010        & -0.0023         & -0.0009\\   
                                               & (0.0048)       & (0.0063)       & (0.0038)        & (0.0050)\\   
      $\Delta$ Bund 2Y (pp)                    & 0.5315$^{**}$  & 0.5072$^{**}$  & -0.1286         & -0.1221\\   
                                               & (0.2419)       & (0.2461)       & (0.1791)        & (0.2048)\\   
      \midrule
      \emph{Fit statistics}\\
      Observations                             & 8,454          & 7,256          & 8,447           & 7,249\\  
      R$^2$                                    & 0.005          & 0.005          & 0.004           & 0.004\\  
      \midrule \midrule
      \multicolumn{5}{l}{\emph{}}\\
   \end{tabular}
   
   \par \raggedright  \footnotesize \textbf{Note:}
   Dependent variables: log return $\times$ 100 (percentage points). Canonical specification: 2 lags of dep.\ var.; oil price (65-day \% return, LOCF); S\&P~500 (65-day \% return, LOCF); NFP announcement dummy; FOMC announcement dummy; treasury yield skewness (ISK); inflation gap; unemployment gap. FF futures specifications additionally include a beginning-of-month dummy; Treasury futures specifications include a low-volume indicator. $\Delta$~Bund~2Y (daily change in the German 2-year Bund yield, pp) is used as the instrument-type control to partial out euro-area rate news (UIP logic). Beginning-of-month dummy and two lags of the dependent variable included but omitted from display. ``No Trump'' drops 20 Jan 2017--20 Jan 2021 and post-20 Jan 2025. HAC (Newey-West) standard errors in parentheses. *, **, *** denote significance at 10\%, 5\%, 1\%.
\end{table}

\begin{table}[H]
   \caption{\label{tab_drivers_inflexp_hawkish} Hawkish Political Pressure and Inflation Expectations}
   \centering
   \footnotesize
   \setlength{\tabcolsep}{4pt}
   \renewcommand{\arraystretch}{0.95}
   \begin{tabular}{lcccc}
      \tabularnewline \midrule \midrule
      Dependent Variables: & \multicolumn{2}{c}{2Y Infl.\ Swap ($\Delta$, pp)} & \multicolumn{2}{c}{10Y Infl.\ Swap ($\Delta$, pp)}\\
                                               & Full                    & No Trump       & Full                      & No Trump \\   
      Model:                                   & (1)                     & (2)            & (3)                       & (4)\\  
      \midrule
      \emph{Variables}\\
      Constant                                 & -0.0031$^{*}$           & -0.0011        & -0.0010                   & -0.0005\\   
                                               & (0.0019)                & (0.0023)       & (0.0008)                  & (0.0012)\\   
      Hawkish pressure ($n_{\text{hawkish}}$)  & -0.0198                 & -0.0322$^{**}$ & -0.0065                   & -0.0104\\   
                                               & (0.0142)                & (0.0160)       & (0.0066)                  & (0.0081)\\   
      Oil price (65-day \%, LOCF)              & 0.0002$^{**}$           & 0.0003$^{***}$ & $7\times 10^{-5}$$^{**}$  & 0.0001$^{**}$\\   
                                               & ($8.05\times 10^{-5}$)  & (0.0001)       & ($3.42\times 10^{-5}$)    & ($5.24\times 10^{-5}$)\\    
      S\&P~500 (65-day \%, LOCF)               & 0.0006$^{**}$           & 0.0006$^{*}$   & 0.0002$^{**}$             & 0.0003\\   
                                               & (0.0003)                & (0.0003)       & (0.0001)                  & (0.0002)\\   
      NFP announcement                         & 0.0118$^{**}$           & 0.0161$^{**}$  & 0.0079$^{***}$            & 0.0098$^{***}$\\   
                                               & (0.0060)                & (0.0078)       & (0.0028)                  & (0.0036)\\   
      FOMC announcement                        & -0.0012                 & -0.0029        & -0.0001                   & 0.0024\\   
                                               & (0.0061)                & (0.0073)       & (0.0034)                  & (0.0040)\\   
      Yield skewness (ISK)                     & 0.0057                  & 0.0056         & 0.0069$^{***}$            & 0.0069$^{***}$\\   
                                               & (0.0039)                & (0.0048)       & (0.0021)                  & (0.0024)\\   
      Inflation gap                            & -0.0006                 & -0.0013        & -0.0004                   & -0.0006\\   
                                               & (0.0010)                & (0.0011)       & (0.0004)                  & (0.0005)\\   
      Unemployment gap                         & 0.0005                  & -0.0012        & -0.0004                   & -0.0011$^{**}$\\   
                                               & (0.0010)                & (0.0010)       & (0.0005)                  & (0.0005)\\   
      \midrule
      \emph{Fit statistics}\\
      Observations                             & 4,636                   & 3,411          & 4,636                     & 3,411\\  
      R$^2$                                    & 0.045                   & 0.056          & 0.019                     & 0.020\\  
      \midrule \midrule
      \multicolumn{5}{l}{\emph{}}\\
   \end{tabular}
   
   \par \raggedright 
   \footnotesize \textbf{Note:}
   Dependent variables: daily change in 2-year and 10-year USD inflation swap rates (percentage points; $\times 100$ = basis points). Drivers specification: inflation gap, unemployment gap, QoQ oil price growth (LOCF), QoQ S\&P~500 return (LOCF), NFP announcement dummy, FOMC announcement dummy, and treasury yield skewness (ISK). Beginning-of-month dummy and two lags of the dependent variable included but omitted from display. ``No Trump'' drops 20 Jan 2017--20 Jan 2021 and post-20 Jan 2025. HAC (Newey-West) standard errors in parentheses. *, **, *** denote significance at 10\%, 5\%, 1\%.
\end{table}

\begin{table}[H]
   \caption{\label{tab_drivers_volatility_hawkish} Hawkish Political Pressure and Financial Volatility}
   \centering
   \footnotesize
   \setlength{\tabcolsep}{4pt}
   \renewcommand{\arraystretch}{0.95}
   \begin{tabular}{lcccc}
      \tabularnewline \midrule \midrule
      Dependent Variables: & \multicolumn{2}{c}{VIX ($\Delta$, pts)} & \multicolumn{2}{c}{MOVE ($\Delta$, pts)}\\
                                               & Full           & No Trump       & Full           & No Trump \\   
      Model:                                   & (1)            & (2)            & (3)            & (4)\\  
      \midrule
      \emph{Variables}\\
      Constant                                 & 0.0358         & 0.0237         & 0.2305$^{***}$ & 0.2812$^{***}$\\   
                                               & (0.0260)       & (0.0293)       & (0.0767)       & (0.0904)\\   
      Hawkish pressure ($n_{\text{hawkish}}$)  & -0.0247        & -0.2006        & -0.8982        & -0.9984\\   
                                               & (0.1999)       & (0.1899)       & (0.7839)       & (0.9238)\\   
      Oil price (65-day \%, LOCF)              & 0.0009         & -0.0001        & 0.0050         & 0.0042\\   
                                               & (0.0008)       & (0.0012)       & (0.0033)       & (0.0046)\\   
      S\&P~500 (65-day \%, LOCF)               & -0.0086$^{**}$ & -0.0085$^{*}$  & -0.0125        & -0.0139\\   
                                               & (0.0042)       & (0.0045)       & (0.0112)       & (0.0132)\\   
      NFP announcement                         & -0.0974        & -0.0669        & -2.737$^{***}$ & -3.321$^{***}$\\   
                                               & (0.0752)       & (0.0742)       & (0.3164)       & (0.3448)\\   
      FOMC announcement                        & -0.1986$^{*}$  & -0.2543$^{**}$ & -2.797$^{***}$ & -3.041$^{***}$\\   
                                               & (0.1074)       & (0.1276)       & (0.3598)       & (0.4571)\\   
      Yield skewness (ISK)                     & -0.2269$^{**}$ & -0.1898$^{**}$ & -0.5434$^{**}$ & -0.5342$^{**}$\\   
                                               & (0.0979)       & (0.0772)       & (0.2126)       & (0.2678)\\   
      Inflation gap                            & 0.0088         & 0.0160         & 0.0671         & 0.0688\\   
                                               & (0.0131)       & (0.0155)       & (0.0488)       & (0.0506)\\   
      Unemployment gap                         & 0.0076         & 0.0229         & 0.0273         & 0.0496\\   
                                               & (0.0167)       & (0.0158)       & (0.0323)       & (0.0444)\\   
      \midrule
      \emph{Fit statistics}\\
      Observations                             & 5,408          & 4,183          & 5,429          & 4,287\\  
      R$^2$                                    & 0.019          & 0.017          & 0.044          & 0.050\\  
      \midrule \midrule
      \multicolumn{5}{l}{\emph{Signif. Codes: ***: 0.01, **: 0.05, *: 0.1}}\\
   \end{tabular}
   
   \par \raggedright 
   \footnotesize  \textbf{Note:}
   Dependent variables: first difference of VIX and MOVE index levels (index points). Drivers specification: inflation gap, unemployment gap, QoQ oil price growth (LOCF), QoQ S\&P~500 return (LOCF), NFP announcement dummy, FOMC announcement dummy, and treasury yield skewness (ISK). Beginning-of-month dummy and two lags of the dependent variable included but omitted from display. ``No Trump'' drops 20 Jan 2017--20 Jan 2021 and post-20 Jan 2025. HAC (Newey-West) standard errors in parentheses. *, **, *** denote significance at 10\%, 5\%, 1\%.
\end{table}

\pagebreak

\subsection{Pressure Intensity and the Trump Period}
\label{subsec:pressure_intensity}

The result that the dovish coefficient is larger in the non-Trump subsample than in the full sample---documented in
Section~\ref{subsec:a_trump_story}---invites a natural question.  If Trump-era pressure is by any conventional measure \emph{more} intense than pressure under previous administrations, why does the per-event coefficient attenuate when the Trump windows are included?  This subsection offers a characterisation of the difference in pressure intensity between the two periods and uses it to rationalise the
coefficient pattern.

\begin{figure}[H]
    \centering
    \caption{Political pressure streaks: pre-Trump vs.\ Trump periods}
    \label{fig:pressure_streak_density_trump}
    \includegraphics[width=\linewidth]{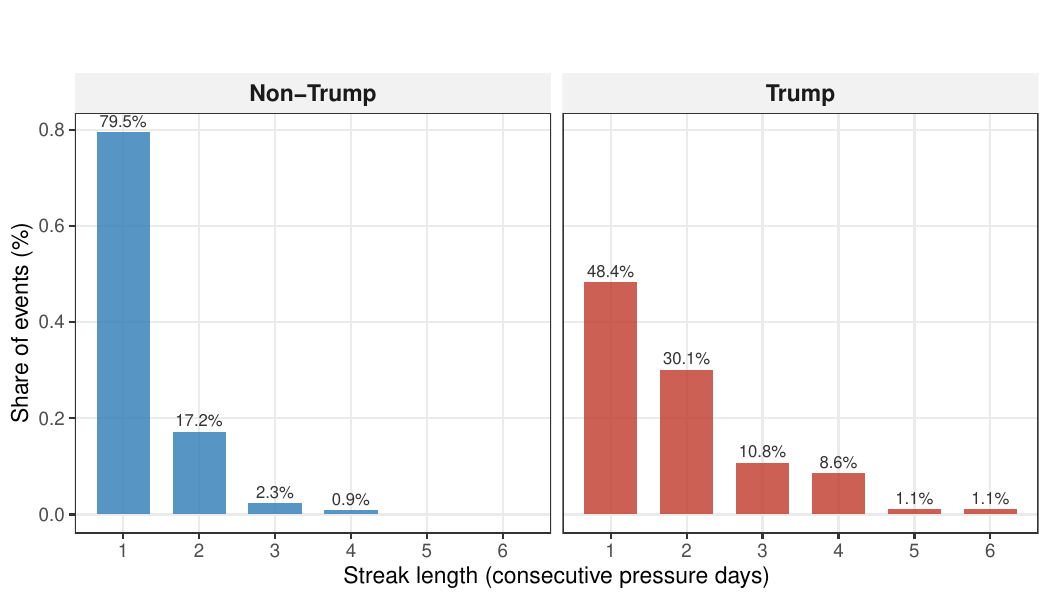}
    \footnotesize{\begin{flushleft}
    \textbf{Note:} The figure shows the distribution of pressure streak lengths separately for the pre-Trump and Trump periods. A streak is defined as a sequence of consecutive days with at least one dovish pressure article.
    \end{flushleft}}
\end{figure}

Figure~\ref{fig:pressure_streak_density_trump} documents the first piece of evidence.  We define a pressure streak as a sequence of consecutive days on which at least one dovish pressure article appears, and plot the distribution of streak lengths separately for the pre-Trump and Trump periods.  The two distributions differ markedly.  Pre-Trump streaks are overwhelmingly short, with isolated single-day events dominating the mass of the distribution.  Trump-period
streaks exhibit a substantially heavier right tail, reflecting the sustained multi-day and multi-week campaigns of public pressure against the Chair and the FOMC that characterise both Trump
administrations.

The streak distribution suggests that the two periods differ not only
in the overall frequency of pressure events but in their temporal
clustering.  A coefficient estimated on the pooled sample weighs each
event equally, implicitly assuming a constant marginal response
regardless of the intensity of recent pressure flow---an assumption
the Trump-era streak distribution makes difficult to sustain.  We test
it directly by augmenting the baseline specification with an
interaction between $n_{\text{dovish},t}$ and the rolling sum of dovish
articles over the preceding three days.

\begin{figure}[H]
    \centering
    \caption{Marginal effect of dovish pressure as a function of recent pressure intensity}
    \label{fig:me_interaction_dovish_w3}
    \includegraphics[width=\linewidth]{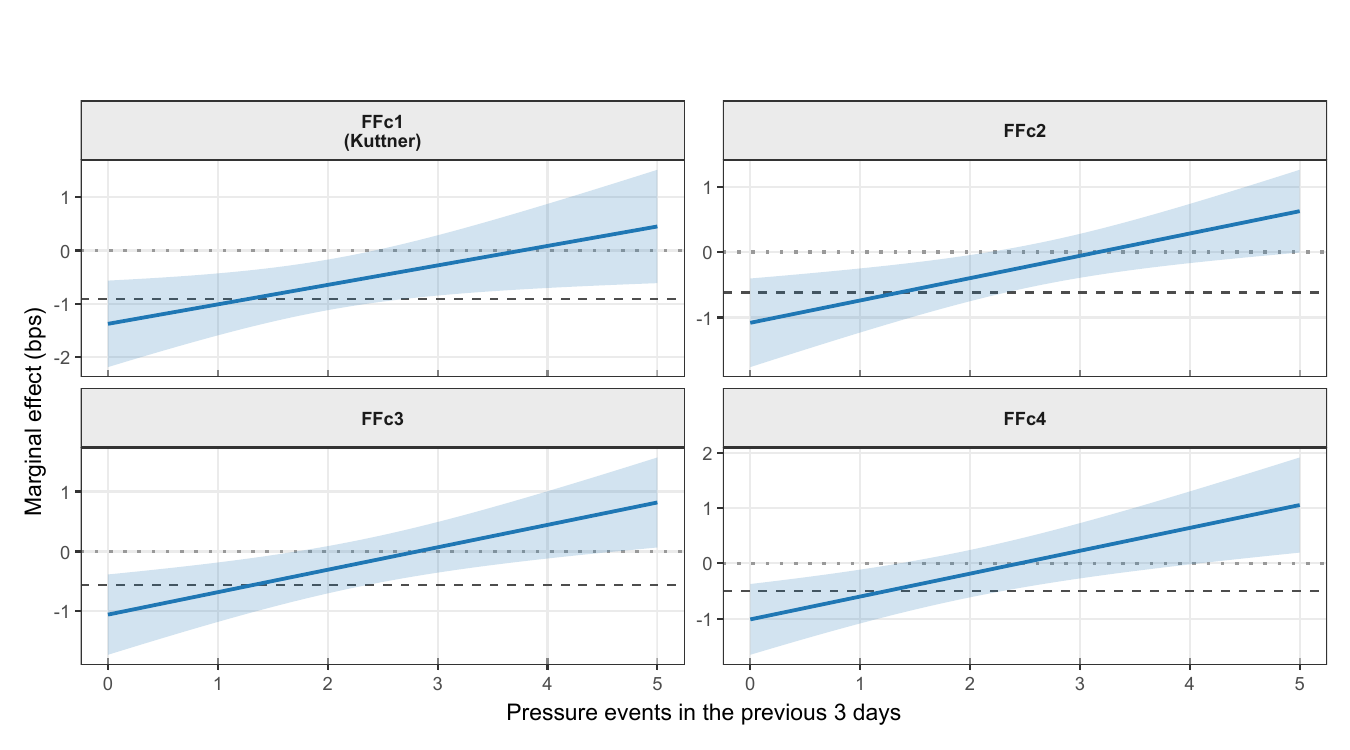}
    \footnotesize{\begin{flushleft}
    \textbf{Note:} The figure plots the marginal effect of political pressure events on the daily change in federal funds futures implied rates (in basis points) as a function of recent pressure intensity, measured as the sum of dovish articles over the preceding three days.  Shaded bands denote 90\% confidence intervals constructed via the delta method. The dashed horizontal line indicates the unconditional baseline estimate. 
    \end{flushleft}}
\end{figure}

Formally, for each maturity $m \in \{1, 2, 3, 4\}$ we estimate
\begin{equation}
\label{eq:interaction_dovish_w3}
\Delta r^{m}_{t}
  = \alpha_m
  + \beta_{1,m} \, n_{\text{dovish},t}
  + \beta_{2,m} \, S_{t-1}^{(3)}
  + \beta_{3,m} \, \bigl(n_{\text{dovish},t} \times S_{t-1}^{(3)}\bigr)
  + \sum_{l=1}^{2} \delta_{l,m} \, \Delta r^{m}_{t-l}
  + \boldsymbol{\gamma}_m' \mathbf{X}_t
  + \varepsilon_{m,t},
\end{equation}
where $\Delta y_{m,t}$ is the daily change in the FFc$m$ implied rate,
$n_{\text{dovish},t}$ is the count of dovish pressure articles on
day~$t$, and $S_{t-1}^{(3)} = \sum_{k=1}^{3} n_{\text{dovish},t-k}$ is
the rolling three-day sum of dovish articles over the preceding
window.  The vector $\mathbf{X}_t$ collects the baseline controls;
$\varepsilon_{m,t}$ is an error term and standard errors are HAC
(Newey--West).  The marginal effect of an additional dovish article
on day~$t$, evaluated at recent pressure intensity $s$, is
\begin{equation}
\label{eq:marginal_effect}
\frac{\partial \, \mathbb{E}[\Delta r^{m}_{t} \mid \cdot]}{\partial \, n_{\text{dovish},t}}
  = \beta_{1,m} + \beta_{3,m} \, s,
\end{equation}
which is the object plotted in
Figure~\ref{fig:me_interaction_dovish_w3} for each maturity $m$.

The marginal effect declines monotonically in recent pressure
intensity.  An isolated dovish event---one that arrives after a
quiet three-day window---moves near-term rate expectations by
substantially more than the unconditional baseline.  As recent
intensity rises, the marginal effect attenuates, and at the highest
observed intensities it is statistically indistinguishable from zero.
The pattern is consistent with market habituation: the first dovish
article in a campaign carries information about the executive's
stance and the likely path of the Fed's reaction function, but
successive articles in a sustained campaign convey progressively less
incremental information, because the stance has already been priced.

This interaction rationalises the non-Trump coefficient pattern.  The
Trump periods are disproportionately populated by high-intensity
streaks, and within those streaks the marginal event carries a
smaller-than-average effect.  Pooling across the full sample
therefore weights the estimate toward the low-marginal-effect region
of the distribution, attenuating the pooled coefficient relative to
the non-Trump subsample---where pressure events are more often
isolated and closer to the high-marginal-effect region.  The
non-Trump estimate of $-1.58$ basis points is not, on this reading,
evidence that markets responded more to Bush or Obama-era pressure
than to Trump-era pressure; rather, it reflects the fact that
Bush and Obama-era pressure arrived in the information-rich,
isolated form that markets price most sharply. Habituation at the
per-event margin does not, however, imply a small cumulative effect:
a sustained campaign with many events can move expectations
substantially even as each additional event contributes less, and the
welfare-relevant object is the cumulative shift in the expected
policy path rather than the marginal response to any single article.
The substantive conclusion is unchanged---political pressure shifts
rate expectations throughout the post-1987 sample---but the mechanism
is sharpened: what markets price is new information about the
executive's stance, and sustained repetition of the same stance
delivers diminishing returns.

\subsection{Heterogeneity by Actor}
\label{app:heterogeneity_actor}

The baseline specification pools all political pressure events,
implicitly treating a presidential statement and a congressional
speech as equivalent signals.  We relax this assumption by
estimating the baseline separately for five actor categories:
President, Treasury Secretary, Senator, Representative, and other
executive officials.  Presidential pressure dominates the Reuters
sample in frequency, so cross-actor comparisons are affected by
differential precision: non-presidential categories have smaller
samples and wider confidence intervals.

\begin{figure}[h!]
    \centering
    \caption{Heterogeneity by source of political pressure}
    \label{fig:heterogeneity_actor}
    \includegraphics[width=\linewidth]{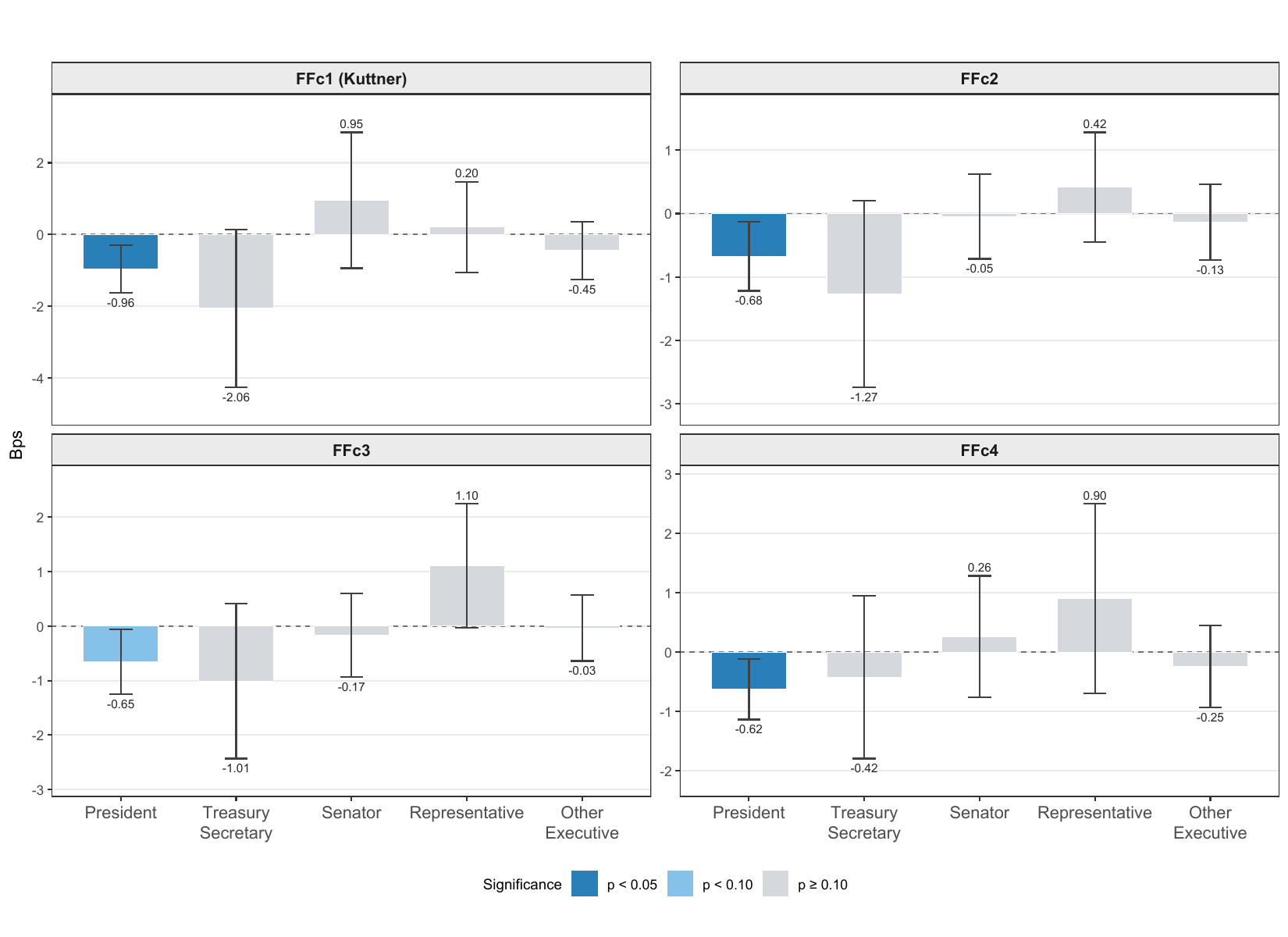}
    \footnotesize{\begin{flushleft}
    \textbf{Note:} The figure reports the coefficient on
    $n_{\text{dovish},t}$ from the baseline specification,
    estimated separately for five actor categories, across FFc1
    through FFc4.  Bars denote point estimates; whiskers denote
    95\% confidence intervals.  Dark blue bars indicate significance at
    the 5\% level; light blue bars indicate significance at
    the 10\% level; grey bars are not significant.
    \end{flushleft}}
\end{figure}

Figure~\ref{fig:heterogeneity_actor} shows that the baseline effect
is driven entirely by presidential pressure.  The presidential
coefficient is negative and significant across all
four maturities, ranging from $-0.96$ basis points at FFc1 to
$-0.62$ at FFc4.  All four remaining actor categories---Treasury
Secretary, Senator, Representative, and other executive
officials---enter with coefficients that are statistically
indistinguishable from zero at every maturity.  Market
participants appear to price the unique institutional authority of
the President specifically, rather than political rhetoric more
broadly; neither opposition statements nor pressure from within
the governing party have detectable market effects unless the
signal originates from the President himself.  This concentration
of the effect, however, can only be documented \emph{ex post} and
with a pressure measure that preserves the identity of each
actor---prior aggregate indices, which collapse all political
signalling into a single count, would not have revealed it.  The
finding thus also illustrates the value of actor-level resolution
in the construction of the pressure measure itself.

\subsection{Persistence of the Exchange Rate Response}
\label{app:lp_fx}

\begin{figure}[H]
    \centering
    \caption{Cumulative Exchange-Rate Responses to Dovish Political Pressure}
    \label{fig:lp_fx_appendix}
    \includegraphics[width=0.9\linewidth]{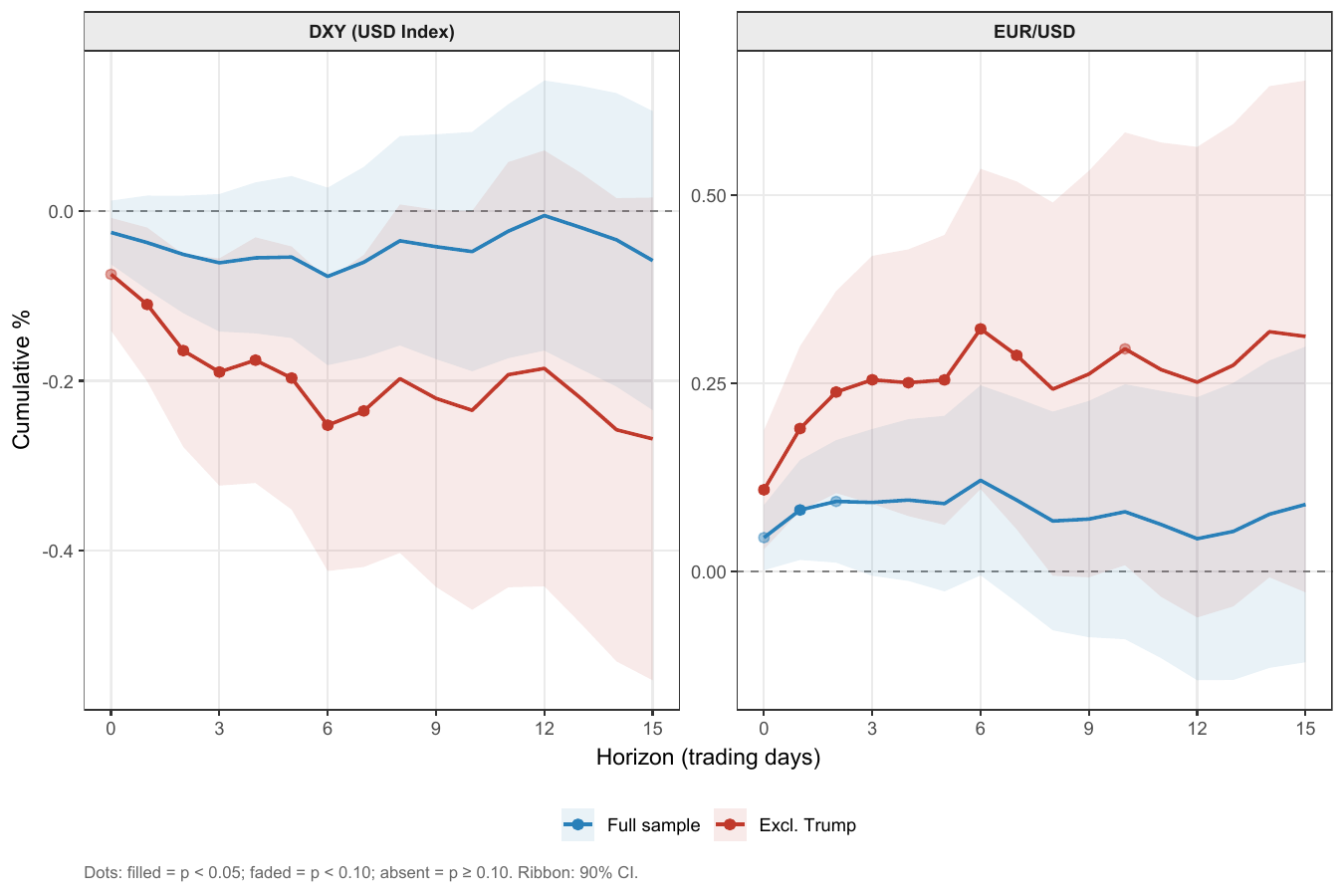}
    \footnotesize{
    \begin{flushleft}
    \textbf{Note:} The figure reports cumulative impulse-response
    functions estimated using local projections for the Dollar Index
    (DXY) and the EUR/USD exchange rate. Blue lines correspond to the
    full sample, while red lines correspond to the sample excluding the
    Trump presidencies. Shaded areas denote 90\% confidence intervals.
    Filled markers indicate estimates significant at the 5\% level,
    faded markers indicate estimates significant at the 10\% level, and
    the absence of a marker indicates estimates that are not statistically
    significant at the 10\% level. A negative DXY response and a positive
    EUR/USD response both indicate a depreciation of the U.S. dollar.
    \end{flushleft}
    }
\end{figure}

Figure~\ref{fig:lp_fx_appendix} reports the cumulative impulse-response
functions of the DXY and EUR/USD exchange rates following a dovish
political-pressure shock. The responses are estimated using the local
projection framework described in the Section \ref{sec:strategy} and control variables as in Paragraph \ref{par:fx}.
As in the baseline local projections, the coefficient at horizon
$h=0$ coincides with the corresponding contemporaneous OLS estimate,
while the coefficients at subsequent horizons measure the cumulative
exchange-rate response through trading day $h$.

The results reveal substantial heterogeneity across samples. In the full sample, we find no statistically significant cumulative response of the broad dollar index. By contrast, the cumulative response of the EUR/USD is positive and statistically significant over the first few trading days, indicating a short-run appreciation of the euro relative
to the dollar following dovish political pressure. The response becomes
statistically insignificant at longer horizons as the confidence
interval widens.

The responses are stronger and more persistent when the Trump
presidencies are excluded. In this restricted sample, dovish political
pressure is followed by a statistically significant cumulative decline
in the DXY and a statistically significant cumulative increase in the
EUR/USD. Both responses are consistent with a depreciation of the U.S.
dollar. The contrast between the two samples mirrors the baseline OLS
results and is consistent with the interpretation that political
pressure has a larger marginal effect when pressure events are less
frequent and therefore potentially more informative to financial
markets.

\end{document}